\documentclass{caps}
\usepackage{natbib}
\usepackage{epsfig}

\def\lsim{\lower 2pt \hbox{$\, \buildrel {\scriptstyle <}\over{\scriptstyle \sim}\,$}}
\def\gsim{\lower 2pt \hbox{$\, \buildrel {\scriptstyle >}\over{\scriptstyle \sim}\,$}}
\def\lapp{\ifmmode\stackrel{<}{_{\sim}}\else$\stackrel{<}{_{\sim}}$\fi}
\def\gapp{\ifmmode\stackrel{>}{_{\sim}}\else$\stackrel{>}{_{\sim}}$\fi}
\def\aproxpropto{\ifmmode\stackrel{\propto}{_{\sim}}\else$\stackrel{\propto}{_{\sim}}$\fi}

\begin{document}

\setcounter{chapter}{6}

\author{Victoria M. Kaspi and Mallory S. E. Roberts\\Physics Department, McGill University, Rutherford Physics Building,\\ 3600 University Street, Montreal, QC, Canada, H3A 2T8 \and
Alice K. Harding\\NASA Goddard Space Flight Center, Laboratory for High Energy Astrophysics,\\ Code 661.0, Greenbelt, MD, USA, 20771}

\chapter{Isolated Neutron Stars}

\section{Introduction}

This chapter deals with X-ray emission from isolated neutron stars for
which the energy for the observed X-rays is thought to originate from
the rotation of the neutron star, or from an internal heat reservoir
following formation.  Rotation power can manifest itself as pulsed
emission, or as nebular radiation produced by a relativistic wind of
particles emitted by the neutron star.  Residual heat of formation is
observed as soft X-ray emission from young neutron stars.  Such thermal
radiation, however, can also be produced as a result of reheating from
internal or external sources.  Rotation-powered pulsed and nebular
X-ray emission, as well as thermal emission, can often be observed in a
single object simultaneously; this is both fascinating and annoying, as
one invariably contaminates the study of the other.  There are also a
handful of neutron stars for which the origin of the observed X-ray
emission is unclear but may be related to the above processes; we will
discuss those as well.

Rotation-powered neutron stars are generally referred to as ``radio
pulsars'' since it is at radio wavelengths that the vast majority of
the catalogued\footnote{http://www.atnf.csiro.au/research/pulsar/psrcat/}
population (currently numbering $\sim$1400) is observed.  However,
the radio emission is energetically unimportant, and we now know of several
rotation-powered neutron stars that are not detected as radio sources
in spite of deep searches \citep[e.g.][]{ckm+98,mcd+01}.  We therefore
use the more physically motivated term ``rotation-powered.''

The total available spin luminosity in a rotation-powered pulsar is
given by the rate of loss of rotational kinetic energy, $\dot{E} \equiv
I \omega \dot{\omega} \equiv 4 \pi^2 I \dot{P} / P^3$, where $I$ is the
stellar moment of inertia and $\omega \equiv 2\pi/P$ is the angular
frequency with $P$ the spin period. Thus a simple measurement of $P$
and $\dot{P}$ for an isolated neutron star determines the available
rotational power in a model-independent way, assuming a value for $I$,
typically taken to be $10^{45}$~g~cm$^2$.  Also generally inferred for
these sources are their surface di-polar magnetic fields, $B = 3.2
\times 10^{19} ( P \dot{P})^{1/2}$~G (for $B$ on the equator), 
and their characteristic
spin-down ages, via $\tau_c = P/2\dot{P}$.  Although these latter two
inferences are model-dependent ($B$ because it assumes simple magnetic
dipole braking in a vacuum -- which is almost certainly not the case -- and
$\tau_c$ because it represents the true physical age only for the same
assumption and also for a negligible initial spin period), the
important point is that $P$ and $\dot{P}$ provide, in addition to
$\dot{E}$, at least estimates of other important physical information.

\begin{figure}
\centering
\epsfig{file=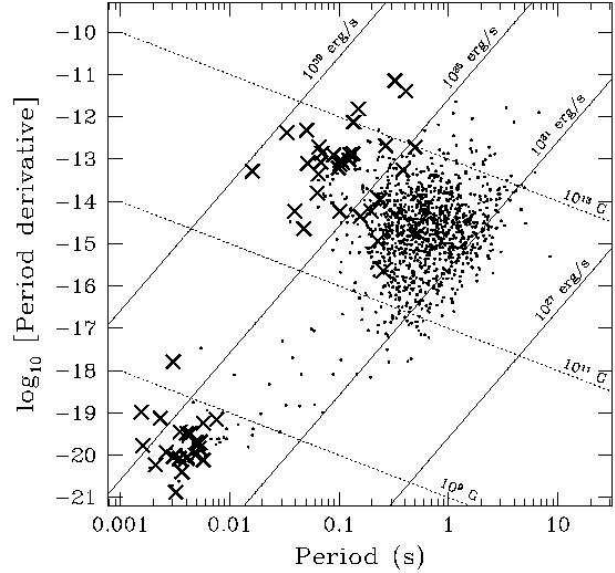, width=9cm}
\vspace{-1in}
\caption{$P$--$\dot{P}$ diagram for the 1403 currently catalogued 
rotation-powered pulsars.  The 66 X-ray detected sources are indicated with an
``X.''  Solid lines show constant $\dot{E}$, and dotted
lines show constant inferred surface di-polar magnetic field.
``X-ray-detected'' means pulsed, unpulsed or nebular emission, be it thermal or
non-thermal.  X-ray detected sources are generally those with the
greatest spin-down luminosity, although the correlation is not perfect
because of the wide range of source distances and observational selection
effects.  The indicated sources
are summarized in Tables~\ref{ta:nonthermal}, \ref{ta:thermal}, 
and \ref{pwn_table}.
Note that the X-ray-detected millisecond pulsars in the globular
cluster 47 Tucanae do not have observable intrinsic $\dot{P}$'s due to
contamination by acceleration in the cluster potential; for those
sources (numbering 15), we have used inferred $\dot{P}$'s
\protect\citep[see][]{fkl+01,gch+02}.}
\label{fig:ppdot}
\end{figure}

It is for this reason that $P$--$\dot{P}$ diagrams are so
useful.  Figure~\ref{fig:ppdot} shows the $P$--$\dot{P}$ phase space,
with all catalogued rotation-powered neutron stars with measured $\dot{P}$s 
indicated as dots.  Lines
of constant $\dot{E}$ and $B$ are indicated.  Note the two different
populations of rotation-powered neutron stars:  those with periods
between $\sim$16~ms and $\sim$8~s, having $B \gapp 
10^{11}$~G, and the lower magnetic field ($B \lapp 10^{10}$~G),
very rapidly rotating class ($P \lapp 100$~ms), the
millisecond pulsars.  The division is made particularly obvious when
binarity is considered, as the latter sources are mostly in
binaries, while the former are only rarely in binaries.  The
millisecond pulsars are often referred to as ``recycled'' pulsars
as they are thought to be formed in a past episode of mass
transfer as part of an accreting X-ray binary phase (see chapter
1 by Psaltis).  Finally, note the absence of sources having 
$B \gapp 10^{14}$~G.  Neutron stars having fields higher than
this are thought to be ``magnetars,'' discussed in chapter 14 by
Woods et al.  

%In those sources, which appear to be radio-quiet, the decay of the
%ultra-high magnetic field is thought to be the origin of very luminous
%X-rays.  Interestingly, although on the $P$--$\dot{P}$ diagram
%magnetars and apparently purely rotation-powered neutron stars overlap
%slightly \citep{msk+03}, the emission properties have not been seen to
%overlap at all, for reasons that are unclear.

All sources which have been detected in X-rays are identified with an
``X'' in Figure~\ref{fig:ppdot}.  For the purposes of this Figure, any
X-ray detection, be it pulsed, unpulsed or nebular, thermal or non-thermal, is
counted.  Note the clear correlation with $\dot{E}$ -- most X-ray
identified rotation-powered neutron stars have large $\dot{E}$ and vice
versa.  The correlation is not perfect; the main reason for this is
that the distances to these sources vary by over two orders of
magnitude.  Distant high-$\dot{E}$ sources are harder to detect, and
nearby low-$\dot{E}$ sources may be detected through their thermal
emission.  Note also the large number of millisecond pulsar detections
is due to many of them being in X-ray studied globular clusters having
larger scale heights, hence smaller line-of-sight absorbing column densities.

Tables~\ref{ta:nonthermal}, \ref{ta:thermal} and \ref{pwn_table}
provide a summary list of all X-ray-detected rotation-powered neutron
stars, along with their most important properties.  Note that the
number of X-ray detected rotation-powered pulsars has more than doubled
in the past 5 years (from 27 in the 1997 review by Becker \&
Tr\"umper\nocite{bt97}, to 66 today).  This is due to improvements in
X-ray sensitivity, in particular from the {\it Rossi X-ray Timing
Explorer (RXTE)}, the {\it Chandra X-ray Observatory} and the {\it XMM-Newton
Telescope}, and also improvements in radio telescope sensitivity, most
notably at the Parkes, Arecibo, and Green Bank radio telescopes, which
have permitted discoveries of sources that could later be followed up
with X-ray observations.

%This Chapter summarizes the most recent developments in X-ray observations 
%of rotation-powered and other isolated neutron stars.  
%For previous reviews on 
%the same subject, see \cite{sw88,bt97,pccm02}.
%We do not, however
%cover soft gamma repeaters or anomalous X-ray pulsars, also probably isolated
%neutron stars, but which are covered in Chapter XXX.  

\section{Magnetospheric Emission}
\label{sec:magnetospheric}

\subsection{Motivation}

After the discovery \citep{sr68,ccl+69} of the Crab pulsar in 1968 and the realization 
 \citep{og69,fw69} that the spin-down energy loss of the neutron star, as estimated from its 
measured $P$ and $\dot P$, could power all the visible radiation of the Crab 
nebula, it was evident that neutron stars are energetically important sources of particle 
production and acceleration.  It is now believed that the radiating particles 
in pulsar wind nebulae (see \S\ref{sec:pwn}) are electron-positron pairs produced
by electromagnetic cascades within the pulsar magnetosphere.  
Since the pulsar magnetosphere is also believed to be the origin of the pulsed non-thermal 
emission, understanding this emission will lead us to an understanding of the energetics of 
these sources and ultimately how the pulsar converts its spin-down power into both the pulsed 
and unpulsed observable radiation.  However, the pulsed radio emission, which first enabled 
the discovery of pulsars, makes up a tiny ($\sim 10^{-6}$) portion of the energy budget.  Pulsed 
emission at high energies, although harder to detect, is energetically much more significant 
(making up $\sim 10^{-3}$ of the spin-down power in the Crab and up to $0.1$ in other 
pulsars) and thus is a more direct probe of the particle acceleration in the magnetosphere.  

Aside from the important issue of energetics, we also want to study the non-thermal emission 
at high energies in order to understand the physical processes in the pair cascades which 
generate the electron-positron pairs.  It is widely believed that generation of a pair plasma is 
required for the instabilities that produce coherent radio emission in pulsars \citep{mel00a}.  
Why radio pulsars die after they have spun down to periods of several seconds may possibly be 
understood as their inability to produce pair plasmas when particle acceleration becomes too 
feeble \citep{rs75,ha01,hmz02}.  Studying the radiation and pair cascades at high energy is 
also a study in the physical processes, such as one-photon pair creation and photon 
splitting, that occur only in the strong magnetic fields of neutron stars.  Finally, the 
generation of non-thermal emission is intimately linked to the polar-cap heating which 
contributes to the thermal emission, particularly in older pulsars.

\subsection{Summary of Observations}

At X-ray energies, 66 rotation-powered pulsars have been detected, nearly half of 
which have periods $P < 6$ ms.  Many of these detections were first made by {\it 
ROSAT}  \citep{bt97} and {\it ASCA}  \citep{sai98}, but the number has increased rapidly in recent 
years with new observations by {\it RXTE}, {\it Chandra} and {\it XMM-Newton}  \citep{ba02}.  Table 1.1 
lists the pulsars thought to show non-thermal emission components in their X-ray spectra with 
observed parameters, including detected emission at other wavelengths.  The vast majority are 
also radio pulsars; many of the target sources were chosen from radio pulsar catalogs.  
However, there have been some detections of X-ray pulsars, such as PSR J0537$-$6910 
\citep{mgz+98}, PSR J1846$-$0258 \citep{gvbt00}, PSR J1811$-$1926 \citep{ttd+97}, and PSR 
J0205+6449 \citep{mss+02}, without a radio ephemeris.  Radio pulses have since been detected 
from only one of these \citep[PSR J0205+6449;][]{csl+02}.  Clearly, short period pulsars seem to have 
the highest levels of non-thermal X-ray emission.  Most of the field millisecond pulsars have 
purely non-thermal spectra.  On the other hand, the millisecond radio pulsars in the globular cluster
47 Tucanae that 
have been detected as point sources by {\it Chandra} \citep{gch+02} have spectra that are quite 
soft, and most likely thermal.  The longer-period pulsars with detected X-ray emission have 
spectra that are dominated by thermal emission.  

\begin{table*}[t]
\caption{Non-Thermal X-ray Detected Rotation-Powered Pulsars}
\label{ta:nonthermal}
\vspace*{41pc}
\end{table*}

\nocite{tkt+98,hsg+02,bwt+04,wob+04,ncl+04,bt99,khvb98}

\subsubsection{Spectra}

Generally, the X-ray spectra of spin-powered pulsars show a mix of
thermal and non-thermal components.  Often several thermal components
(discussed in \S\ref{sec:thermal}) and sometimes even several
non-thermal components can be identified.  Cleanly separating the
non-thermal, or power-law, components from the thermal components can
be difficult, especially in sources where the thermal components
dominate \citep[e.g. Geminga, PSRs 0656+14 and B1055$-$52;][]{hw97,gcf+96}.
Detectors with
sensitivity at energies up to at least 10 keV, such as {\it RXTE,
Chandra} and {\it XMM}, have been able to make the best measurements of
non-thermal emission components.  This non-thermal emission probably
originates from the radiation of particles accelerated in the pulsar
magnetosphere. Models for this emission will be discussed below.

\begin{figure}[t]
\begin{center}
\epsfig{file=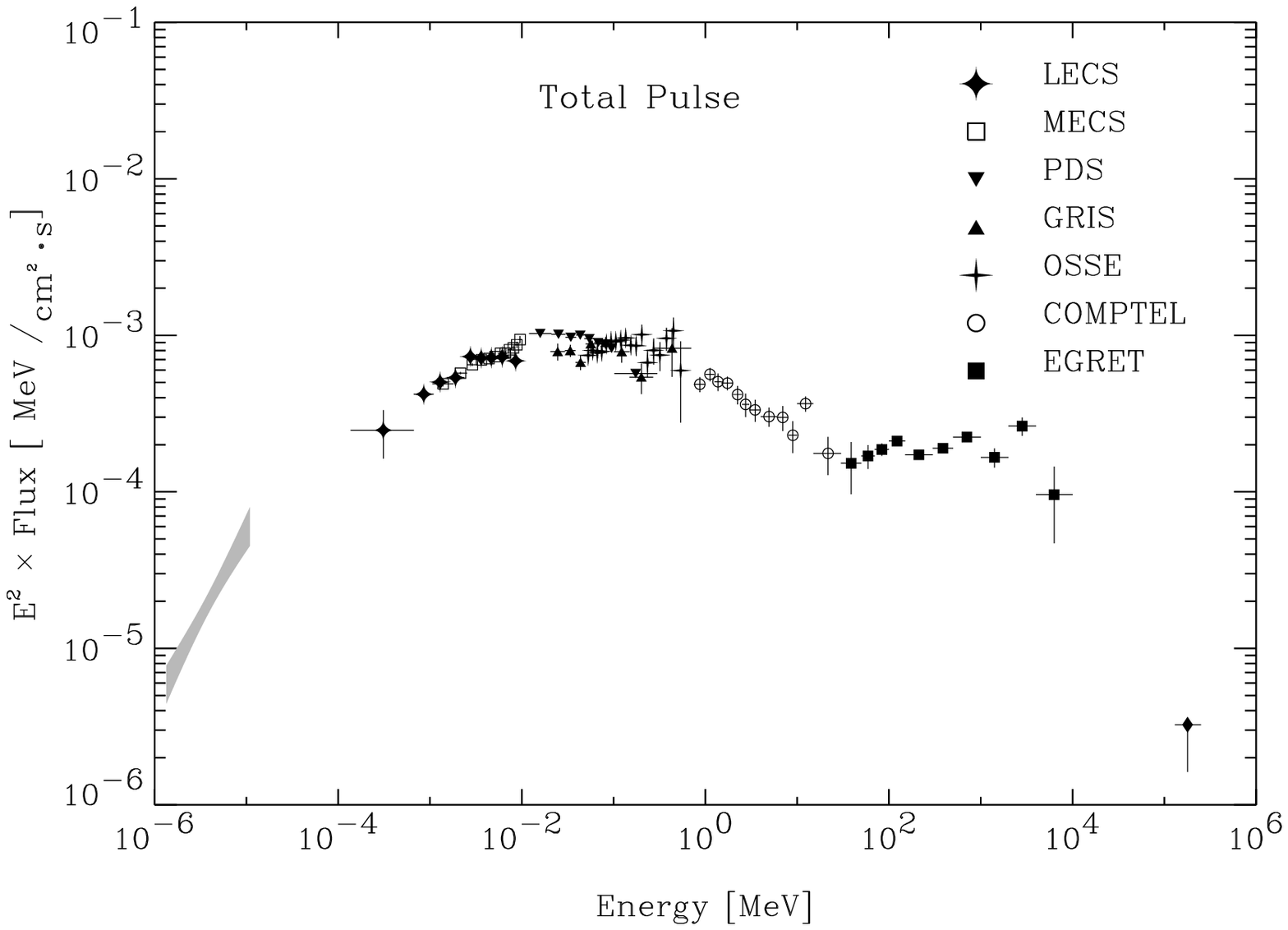, width=10cm}
\caption{Broadband pulsed spectrum  \protect\citep{khc+01} of the Crab,
as an example of emission characteristics of a young pulsar.}
\label{fig:crabspectrum}
\end{center}
\end{figure}

\begin{figure}[t]
\epsfig{file=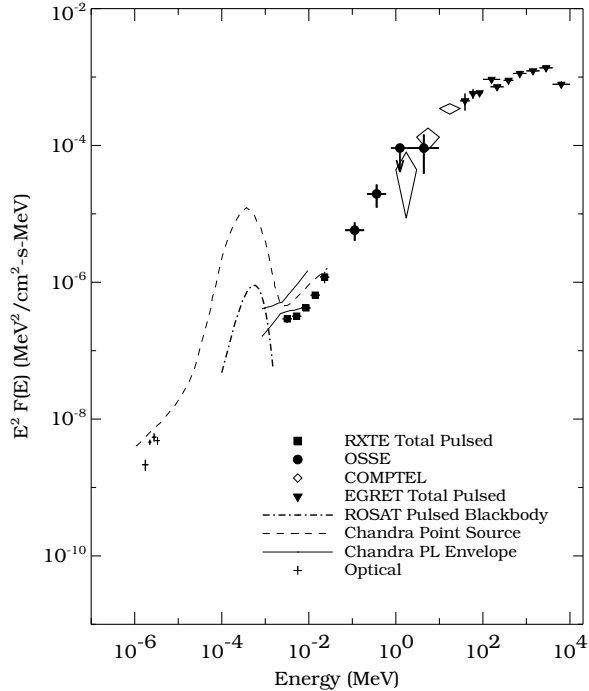, width=5cm}
\caption{Broadband pulsed spectrum \protect\citep{hsg+02} of the Vela pulsar,
as an example of emission characteristics of a middle-aged pulsar.}
\label{fig:velaspectrum}
\end{figure}

All of the $\gamma$-ray pulsars that have been detected by the 
{\it Compton Gamma-Ray Observatory (CGRO)}
are also X-ray pulsars with measured non-thermal emission components.
One interesting question is whether the non-thermal X-ray spectrum is
part of the same emission as that detected at $\gamma$-ray energies.
For the youngest of these pulsars (Crab, PSR B1509$-$58), the total
power in pulsed emission peaks in the hard X-ray band (Figure
\ref{fig:crabspectrum}).  There is strong emission through the entire
X-ray and $\gamma$-ray bands and a smooth connection between the two.
However, in the case of the Crab the smoothness of the spectrum is
misleading as the situation may be more complicated.  \citet{khc+01}
have argued for several separate emission components in the Crab
optical to $\gamma$-ray spectrum, evidenced by the strong frequency
dependence of the interpeak emission and the peak 2 to peak 1 ratio,
both of which have a maximum around 1 MeV.  The middle-aged pulsars
have comparatively weak non-thermal emission in the X-ray band since
their power peaks at GeV energies, and there is a gap in the detected
spectrum between the X-ray and $\gamma$-ray bands (Figure
\ref{fig:velaspectrum}).  In several cases (Vela, PSR B1055$-$52) an
extrapolation between the two is plausible, but in others
\citep[Geminga, PSR~B1706$-$44;][]{ghd02} a connection is not clear.  In
the case of Vela (Figure \ref{fig:velaspectrum}), there is strong
evidence for two separate non-thermal X-ray components \citep{hsg+02},
one connecting to the $\gamma$-ray (OSSE) spectrum and the other
possibly connected with the optical spectrum.

Millisecond pulsars seem to be bright X-ray emitters, as a relatively
large number of the known millisecond radio pulsars are X-ray pulsars.
Most have hard power-law spectra with photon indices 1.5--2.0
\citep{ba02}, with the exception of a few \citep[e.g. PSRs J0437$-$4715,
J2124$-$3358;][]{zps+02,skt+01} which also have thermal components.
Since they are too old to have detectable cooling emission, the
presence of pulsed thermal emission from millisecond pulsars probably
requires polar-cap heating.

There are a number of newly discovered X-ray pulsars \citep[PSR J1420$-$6048;][]{rrj01}, 
\citep[PSR J1930+1852;][]{clb+02}, \citep[PSR J2229+6114;][]{hcg+01}, \citep[PSR J1105$-$6107;]
[]{kbm+97} that lie 
in or near the error circles of EGRET unidentified $\gamma$-ray sources \citep{hbb+99}.  Since these pulsars 
were discovered after the end of the {\it CGRO} mission and EGRET detected too few photons to 
search for pulsations, it is not known whether they are also $\gamma$-ray pulsars.  There are 
also some X-ray pulsars discovered earlier (PSR B1823$-$13, PSR B1800$-$21) that are associated 
with EGRET sources and are candidate $\gamma$-ray pulsars \citep{kan02}.

\begin{figure}[t] 
\epsfig{file=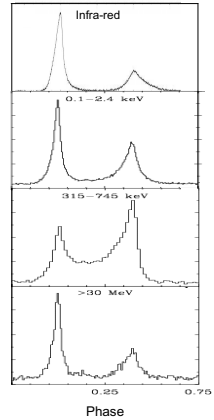, width=3cm}
\caption{Multiwavelength pulse profiles \citep{kan02} of the Crab,
as an example of emission characteristics of a young pulsar.}
\label{fig:crablc}
\end{figure}

\subsubsection{Pulse Profiles}

The pulse profiles of the non-thermal X-ray emission usually show much more modulation and 
narrower pulses than does the thermal emission.  The pulsed fraction for many pulsars with mixed 
emission rises from around 10\% -- 20\% in the (0.1 - 1 keV) band where the thermal emission 
dominates, to 80\% -- 90\% where the non-thermal emission dominates \citep{ba02}.  There is 
often an accompanying phase shift between the thermal and non-thermal pulses of the middle-aged 
pulsars.  

\begin{figure}[t] 
\epsfig{file=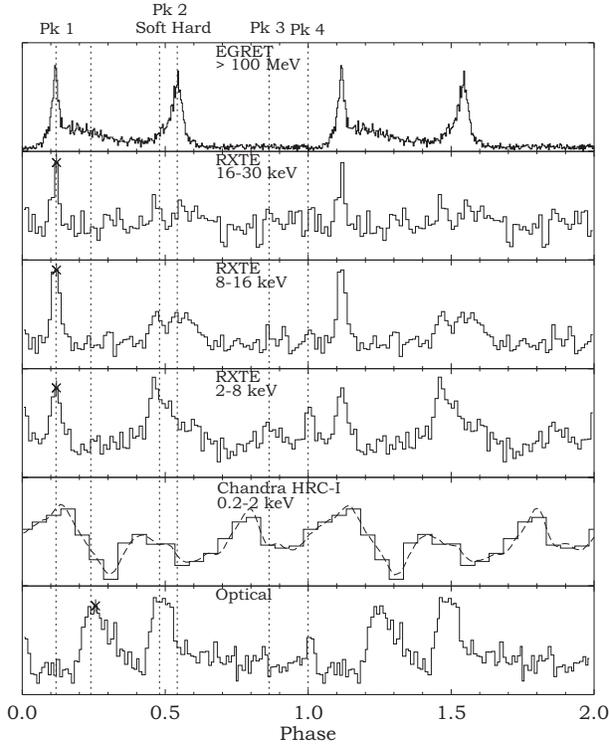, width=8cm}
\caption{Multiwavelength pulse profiles \citep{hsg+02} of the Vela pulsar,
as an example of emission characteristics of a middle-aged pulsar.}
\label{fig:velalc}
\end{figure}

There seems to be no pattern in the relative phases of the non-thermal
X-ray, radio and gamma-ray pulses.  The non-thermal X-ray and radio pulses of many
pulsars that are young or have very short periods, including the Crab
(with the exception of the radio pre-cursor), PSRs J1617$-$5055,
J0205+6114, and J1930+1852 and most of the millisecond pulsars, are in
phase and/or the profiles look very similar.  
Somehow for these pulsars, the radio and
high-energy emission must originate from the same region in the
magnetosphere.  However, the profiles of these pulsars vary from
double-peaked to both broad and narrow single-peaked shapes, and a few
young X-ray pulsars (PSRs J1811$-$1926, J0537$-$6910 and
J1846$-$0258) have no detected radio counterparts.  The non-thermal
X-ray pulses of middle-aged pulsars, on the other hand, are not in
phase with their radio pulses, indicating a different location in the
magnetosphere of the two emission components (e.g.  Figure
\ref{fig:velalc}).  The same trend seems to be true for the X-ray and
$\gamma$-ray pulses of the $\gamma$-ray pulsars, with the profiles of
the fastest, including millisecond \citep[PSR~J0218+4232,][]{khv+00b}, pulsars
being alike and in phase while the profiles of the slower pulsars
are not in phase \citep{kan02}.  Thus, a fast rotation rate seems to be
instrumental in causing alignment of the pulses across the spectrum.

\subsection{Emission Models}

\begin{figure} 
\epsfig{file=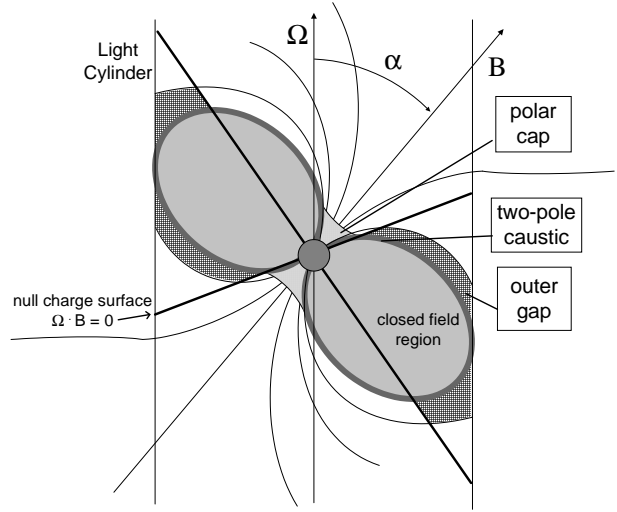, width=10cm}
\caption{Schematic 2D view of the high-energy emission geometry of several proposed models.}
\label{psrmods}
\end{figure}

Particle acceleration inside the pulsar magnetosphere gives rise to pulsed 
non-thermal radiation and possibly also thermal emission from backflowing particles that heat 
the neutron-star surface.  Rotating, magnetized neutron stars are natural unipolar inductors, 
generating huge vacuum electric fields.  However, as was first noted by 
\citet{gj69} in their classic paper, a rotating neutron star will not be 
surrounded by vacuum, since a large surface charge will build up on the star.  
This surface charge is unstable, because the induced vacuum electric field has a component 
parallel to the 
magnetic field at the stellar surface that exceeds the gravitational force by many orders of 
magnitude and is capable of pulling charges out of the star.  If charges are not trapped in 
the surface by binding forces (i.e. if the crust is solid) a charge density, known as the Goldreich-Julian or 
corotation charge density, $\rho_{_{GJ}} \simeq - {\bf {\omega} \cdot B}/(2\pi c)$,
builds up in the neutron-star magnetosphere.  If the magnetospheric charge reaches the 
Goldreich-Julian value everywhere, it is able to short out the electric field 
parallel to the magnetic field ($E_{\parallel}$), and the dipole magnetic field will corotate with 
the star.  Corotation of the pulsar magnetosphere must break down at large distances from the 
neutron star due to particle inertia.  Where and how this happens is not yet understood, but 
it is certain that corotation cannot persist past the speed-of-light cylinder radius, $R_{LC} = 
c/\omega$, which is the distance at which the corotation velocity reaches $c$ (see Figure \ref{psrmods}).  
It is believed that in the outer parts of the magnetosphere there must be a transition to the 
wind zone, where the energy density of the particles is large enough to distort the poloidal 
dipole field into a toroidal relativistic wind flow that carries the spin-down energy of the 
pulsar, in the form of magnetic and particle energy, into a surrounding nebula where (at 
least in the case of the Crab pulsar) it is dissipated as synchrotron 
radiation \citep{rg74,kc84}.  Although the standard picture of a pulsar magnetosphere described 
above is usually accepted on faith, global magnetospheric simulations \citep{km85,phb02,sa02} have not 
yet been able to show whether and how a pulsar magnetosphere reaches the nearly force-free 
(ideal MHD) state envisioned by Goldreich \& Julian.  However, if it is assumed that a force-free, 
corotating magnetosphere has been achieved, then \citet{ckf99} have found a solution 
which smoothly connects the corotating magnetosphere to a relativistic wind.  Inside the
magnetosphere and at the light cylinder, the ratio (known as $\sigma$) of the magnetic energy density 
to the particle energy density is large, primarily because according to the acceleration models 
to be discussed below, the particles receive only a fraction of the full potential drop across 
the open field lines.  Therefore, $\sigma$ is large at the start of the wind flow, but must
drop to 1 or below to accelerate the particles at the wind-termination shock.
The pulsed emission is presumed to originate 
inside the corotating magnetosphere, and strong $E_{\parallel}$ may develop to accelerate particles 
at two possible sites where ${\bf E \cdot B} \neq 0$.  
These sites have given rise to two classes of high energy emission models: polar-cap 
models, where the acceleration and radiation occur near the magnetic poles in the inner 
magnetosphere, and outer-gap models, where these processes occur in the outer magnetosphere.

\subsubsection{Polar-Cap Model}

Polar-cap models \citep{dh82,um95,dh96} advocate that particle acceleration occurs near the 
neutron star surface and that high-energy emission results from a curvature radiation or 
inverse Compton induced pair cascade in a strong magnetic field. There is some variation 
among 
polar-cap models, with the primary division being whether or not there is free emission of 
particles from the neutron-star surface. The subclass of polar-cap models based on free 
emission of particles of either sign, called space-charge limited flow models, assumes that 
the surface temperature of the neutron star (many of which have now been measured to be in the 
range $T \sim 10^5 - 10^6$~K, see \S\ref{sec:thermal}) exceeds the ion and electron
thermal emission temperatures.  Although $E_{\parallel} = 0$ at the neutron-star 
surface in these models, the space charge along open field lines above the surface falls 
short of the corotation charge, due to the curvature of the field \citep{as79} and to general 
relativistic inertial frame dragging \citep{mt92}.  The $E_{\parallel}$ generated by the charge 
deficit accelerates particles, which radiate inverse Compton (IC) photons (at particle 
Lorentz factors $\gamma \sim 10^2 - 10^6$) and curvature (CR) photons (at Lorentz factors 
$\gamma \gsim 10^6$).  Because lower Lorentz factors are required to produce pairs through IC 
emission, an IC pair formation front (PFF) will form first, close to
the surface.  However, it is found \citep{hm02} 
that the IC pair formation fronts do not produce sufficient pairs to 
screen the $E_{\parallel}$ completely, thus allowing acceleration to the Lorentz
factors $\sim 10^7$ sufficient to produce a CR PFF, where there are sufficient pairs to 
completely screen the $E_{\parallel}$.  The CR pair front will therefore limit the particle 
acceleration voltage and determine the high-energy emission luminosity.  The CR photons can 
produce secondary electrons and positrons through one-photon pair production in the strong 
magnetic field near the neutron star surface.  These pairs are produced in excited Landau 
states and radiate synchrotron photons which are energetic enough to produce more pairs.  The 
combined curvature and synchrotron radiation from such pair cascades has been shown to 
produce high-energy phase-resolved spectra similar to observed spectra of X-ray and $\gamma$-
ray Vela-like pulsars \citep{dh96}.  The $\gamma$-ray spectra cut off very sharply (as
a ``super-exponential") due to one-photon pair production attenuation, at the 
pair escape energy, i.e. the highest energy at 
which photons emitted at a given location can escape the magnetosphere without pair producing.
A rough estimate of this cutoff energy, assuming emission
along the polar-cap outer rim, $\theta_{_{\rm PC}} \simeq (2\pi r / cP)^{1/2}$, 
at radius $r$, is 
\begin{equation} \label{Ec}
E_c \sim 2\,\,{\rm GeV}\,P^{1/2}\,\left({r\over R}\right)^{1/2}\, {\rm max}
\left\{0.1, \,B_{0,12}^{-1}\,\left({r\over R}\right)^3\right\}
\end{equation}
where $B_{0,12}$ is the surface magnetic field in units of $10^{12}$ G and $R$ is the
stellar radius.  
Although many high-energy photons are attenuated well below 1 GeV near the surface,
in polar-cap cascades much of the radiation occurs at least 1-2 stellar
radii above the surface, where the photon escape energy for pair
production is several GeV (causing a spectral cutoff there).
At all but the highest fields
there is a prediction that the spectral cutoff energy should be inversely
proportional to surface field strength, or $B_0 = 6.4 \times 10^{19}\,(P\dot P)^{1/2}$~G 
for a dipole field at the pole.  In fields above $\sim 2 \times 10^{13}$~G, photon splitting, 
in which a single photon splits into two lower energy photons, becomes the
dominant attenuation process and lowers the photon escape energy \citep{bh01}. 
The cascade radiation produces a hollow cone of emission 
around the magnetic pole, with opening angle determined by the polar cap half-angle, 
$\theta_{_{PC}}$, at the radius of emission $r$.  The polar cap 
opening angle is very small (a few degrees) unless the emission occurs more than a few 
stellar radii above the surface.  But the pair cascades over most of the polar cap occur 
within several stellar radii of the stellar surface.  Therefore, the wide double-peaked pulses of 
observed $\gamma$-ray pulsars like the Crab, Vela and Geminga, which require beam opening 
angles of the order of the magnetic inclination angle in this model, cannot be produced 
unless the pulsar is nearly aligned.

Recently, this seemingly impossible requirement of polar-cap models has been eased somewhat 
by the realization that emission at the rim of the polar cap will occur at higher altitudes.  
The height of the CR pair front is within a stellar radius of the neutron-star surface over 
most of the polar cap.  However, near the edge of the polar cap, at the last open field line 
which is assumed to be a conducting boundary, the parallel electric field drops to zero.  
Particles therefore need a longer distance to reach the Lorentz factors necessary to produce 
pairs, causing the PFF to rise to higher altitude.  A slot gap \citep{aro83b} is formed as the 
PFF turns up to asymptotically approach the last open field line.  The pair cascades that 
form on the edge of the slot gap at altitudes of several stellar radii above the neutron-star 
surface form a wide hollow cone of high-energy emission \citep{mh03}.  Even so, the inclination 
angles required in this model to reproduce the Crab and Vela profiles are less than 
$10^{\circ}-20^{\circ}$.  This may be at odds with the $\sim 60^{\circ}$ viewing angle 
inferred from the {\it Chandra} image of the Crab nebula \citep{wht+00}.  However, acceleration
of primary electrons in the slot gap can continue to high altitudes \citep{mh04}, 
where the emission may be viewed at larger angles due to the flaring of field lines.

In polar-cap models, surface heating will occur as a result of the screening of the parallel 
electric field by pairs.  Above the pair front, positrons will decelerate and some fraction 
will turn around and accelerate downward to the neutron-star surface.  Because polar-cap heating 
emission is intimately tied to the pulsar 
acceleration mechanism, thermal X-ray emission provides very strong constraints on
pulsar models.  In a space-charge 
limited flow model, the trapped positrons needed to screen the $E_{\parallel}$ are only a 
small fraction of the number of primary electrons because this field was created by a small 
charge imbalance 
in the first place.  
The polar-cap heating luminosity is predicted by space-charge limited flow models to be only 
a few percent of the luminosity of the non-thermal magnetospheric emission, and will fall in 
the soft X-ray band \citep{aro81}.  Recent self-consistent models of polar-cap heating in a 
space-charge limited flow model \citep{hm01,hm02} have predicted heating fluxes from 
positrons trapped at CR and IC pair fronts.  
The heating from CR pair fronts predicts surface X-ray luminosity 
\begin{equation}
L_+^{(CR)} \simeq 10 ^{31}~{\rm erg\, s^{-1}}~  
\left\{ \begin{array}{ll}
    0.4~P^{-6/7}\tau _6^{-1/7} & {\rm if}\: P\lsim 0.1~B_{0,12}^{4/9}, \\
    1.0~P^{-1/2} & {\rm if}~P\gsim 0.1~B_{0,12}^{4/9},
\end{array} 
\right.
\label{L+CR}
\end{equation}
where $\tau_6$ is the pulsar characteristic age in units of $10^6$ yrs. Polar-cap heating
may make the dominant contribution to thermal emission from old and middle-aged pulsars.  The 
heating from IC pair fronts, with predicted X-ray luminosity
\begin{equation}
\label{L+ICS}
L_+^{(IC)} \simeq 2.5 \times 10 ^{27}~{\rm erg\, s^{-1}}~P^{-3/2}
\end{equation}
can account for thermal emission components detected in the spectrum of some millisecond 
pulsars.  

Polar-cap models predict that emission from the open field line region is visible at all 
pulse phases.  This is because the particles that are accelerated and initiate pair cascades 
at low altitudes radiate curvature (and possibly inverse Compton) emission at higher altitude 
on field lines extending to the light cylinder.  Such off-pulse emission may have been 
detected at $\gamma$-ray energies by EGRET \citep{fmnt98} and at X-ray energies by {\it 
Chandra} \citep{tbj+01}. 

\subsubsection{Outer-Gap Model}

Outer-gap models \citep{chr86a,rom96a} assume that acceleration occurs in vacuum gaps that 
develop in the outer magnetosphere, along the last open field line above the null charge surfaces, 
where the Goldreich-Julian charge density changes sign (see Figure~\ref{psrmods}), and that 
high-energy emission results from photon-photon pair production-induced cascades.  The gaps arise 
because charges escaping through the light cylinder along open field lines above the null 
charge surface cannot be replenished from below.  Pairs from the polar-cap cascades, which 
flow out along the open field lines, may pollute the outer gaps to some extent (and vice 
versa), but this effect has yet to be investigated.
The electron-positron pairs needed to provide the current in the outer gaps are 
produced by photon-photon pair production.  In young Crab-like pulsars, the pairs are 
produced by CR photons from the primary particles interacting with non-thermal synchrotron 
X-rays from the same pairs.  In older Vela-like pulsars, where non-thermal X-ray emission is 
much lower, the pairs are assumed to come from the interaction of primary particles 
with thermal X-rays from the neutron-star surface.  
Some of the accelerated pairs flow downward to heat the 
surface and maintain the required thermal X-ray emission.  The modern outer-gap 
Vela-type models \citep{rom96a,zc97} all adopt this picture.  

Although there seems to be agreement on the radiation processes
involved in the outer gap, the full geometry of the gap is still not
solved.  Two approaches to such a solution are currently underway, but
neither is near to defining the complete three-dimensional gap
geometry.  One group \citep{zc97,crz00} solves the 1D Poisson equation
perpendicular to the magnetic field lines, resulting in a gap geometry
for young pulsars that is a long, thin sheet bounded below by the last
open field line and above by production of pairs.  The other group
\citep{hs01,hhs03} obtains solutions to the 1D Poisson equation along
the magnetic field (assuming a gap width across field lines) and finds
that the gap is limited parallel to the field by pair creation.  The
actual gap geometry is probably somewhere in between.  The long narrow
outer-gap geometry of \citet{ry95} and \citet{crz00}, has been
successful in reproducing the observed double-peaked pulse profiles for
large inclination angles.   In these models, the electrons/positrons
flowing outward/inward along the last open field line radiate
outward/inward from the null surface.  The emission forms a wide fan
beam which produces in general a wide double-peaked profile, much like
the observed Crab profile.  The narrow peaks are the result of caustics
formed in the observer's frame by the cancelation of the phase shifts
from relativistic aberration and light travel time, so that emission
from a large range of altitudes is compressed into a narrow phase
range.  However, due to the fact that no outward emission originates
below the null surface, the profile falls off very abruptly at the
outer edges (see Figure \ref{Crabpol}) with no leading and trailing or
off-pulse emission.  The inability to produce trailing and off-pulse 
emission, as is seen in high-energy pulsar profiles, is thus a serious 
problem for current outer-gap models.  The
models of \citet{hs01} and \citet{hhs03} allow external currents to
flow through the outer gaps, thus producing a possible extension of the
gap below the null surface.  In the case of the Crab however, the
extent and magnitude of the emission below the null surface is small
and would not produce much off-pulse emission.  The formation of the leading
peak in double-peaked profiles is also problematic since it requires
emission very close to the light cylinder, where the structure of the 
magnetic field has not been determined.  

The spectrum of radiation from the outer gap is a combination of synchro-curvature radiation 
from pairs inside and outside of the gap and inverse Compton scattering of the synchrotron 
emission by the pairs.  Unlike in polar-cap models, pair production in outer-gap models is 
essential to the production of the high-energy emission: it allows the current to flow and 
particle acceleration to take place in the gap. Beyond a death line in period-magnetic field 
space, and well before the traditional radio-pulsar death line, pairs cannot close the outer 
gap and the pulsar cannot emit high-energy radiation.  This outer-gap death line for 
high-energy pulsars \citep{cr93} falls around $P = 0.3$ s for $B \sim 10^{12}$ G.  Geminga is 
very close to the outer-gap death line and recent self-consistent models \citep{hhs03} have 
difficulty accounting for GeV $\gamma$ rays from pulsars of this age.  Polar-cap models, on 
the other hand, predict that all pulsars are capable of high-energy emission at some level, 
so that detection as a radio-loud high-energy 
pulsar is thus a matter of sensitivity.  

Half of all the particles produced in the outer gaps are accelerated back toward the 
neutron-star surface and radiate curvature photons that can produce pairs in the strong magnetic
field near the surface, initiating pair cascades \citep{zc97,wrhz98}.  
But the pairs still have enough residual energy to heat the surface at the footpoints
of open field lines that thread the outer gaps.  The resulting thermal emission has
$T \sim 1$ keV, which is much higher than the  
thermal emission temperatures observed in pulsars.  According to the model, 
this emission is not observed directly (except right along the poles) but only through 
the blanket of pairs produced by the downward-going particle cascades.  The 1-keV 
photons are reflected back to the surface by the pair blanket through cyclotron resonance 
scattering \citep{hr93}, 
and are re-radiated from the entire surface at a 
temperature around 0.1 keV.  Thus, these outer-gap models predict three X-ray emission
components: hard thermal emission from
direct heating of the polar caps (seen only along the poles), soft thermal emission
reflected from the pair blanket, and non-thermal emission from the downward pair 
cascades \citep{cz99}.  The components actually observed from a particular 
pulsar depend on inclination and viewing angle.  The predicted X-ray luminosities for
pulsars in the {\it ROSAT} band can account for the observed $L_x = 10^{-3} \dot{E}$ relation.

\subsection{It's The Geometry...}

Theoreticians have spent years building pulsar emission models by starting from the 
fundamental electrodynamics of the particle acceleration and radiation processes, predicting 
observable pulsar characteristics from the bottom up.  This has had some success, but both 
polar-cap and outer-gap models as currently formulated also have some basic, unresolved 
problems.  Ultimately, it is the emission geometry required by the observations, which tells 
us the distribution of radiating particles, that will drive us to the correct understanding of 
pulsar high-energy emission.  

Putting together the results of a number of attempts, in different types of models, to 
reproduce the narrow and often double-peaked pulse profiles of high-energy pulsars, one common 
thread seems to emerge.  Nearly all successful geometrical models of pulsar high-energy 
emission have assumed enhanced particle acceleration and radiation along the last open field
lines of a magnetic dipole.  In their polar-cap emission model, \citet{dh96} were led to 
assume strongly enhanced particle flow on the rim of the polar 
cap.  Acceleration in the slot gap along the polar-cap rim later provided a physical basis for 
this idea \citep{mh03}.  The outer-gap models of \citet{ry95} and 
\citet{crz00} assumed acceleration and emission essentially only along 
the last open field line between the null charge surface and the light cylinder in a thin 
outer gap.  But these models cannot produce the non-thermal off-pulse emission seen through 
the whole pulse phase of the Crab optical and Vela and Geminga $\gamma$-ray profiles.  Very 
recently, \citet{dr03} have explored a purely geometrical emission model, known as 
the ``two-pole caustic" model in which they assume that particles radiate all along the last 
open field lines of both poles, from the neutron-star surface to the light cylinder 
(see Figure~\ref{psrmods}).  Taking 
into account relativistic effects of aberration and light travel time, the phase shifts of 
photons emitted at different altitudes along the trailing edge of the polar open field line 
region cancel, so that the emission in the observers frame is compressed into a small range 
of phase, forming a caustic peak in the profile.  The caustic peaks in the two-pole caustic model 
have the same origin as the caustic peaks that form from outward emission 
above the null charge surface in the outer-gap models.  What makes the 
two-pole caustic model fundamentally 
different is the addition of outward emission below the null surface, that allows an observer 
to view caustic emission from both poles as the neutron star rotates past, in contrast to 
outer-gap models where an observer can view high-energy emission from only one pole.  The 
resulting profiles for viewing angles crossing caustics from both poles are astonishingly 
similar to the observed profile of the Crab (see Figure \ref{Crabpol}) and several 
other high-energy pulsars, and seems 
to do better than most of the other `bottom-up' models in reproducing the observed geometry.  
This seems to be telling us that both conventional polar-cap and outer-gap models are missing 
something critical in their description of the basic electrodynamics.  Recently, \citet{mh04} 
have shown that high-altitude radiation from the slot gap could form a physical basis for the
two-pole caustic model, producing very similar profiles.  

The geometry of emission at different wavelengths is also different in various models.  
In polar-cap models, the high-energy and radio emission are physically connected,
since the electron-positron pairs from the polar-cap cascades are thought to be a necessary 
ingredient for coherent radio radiation.  However, the relative geometry of the two
emission regions is not very constrained by observation.  Models for
radio emission morphology \citep{ran93,lm88} consist of core and conal 
components, emitted within a distance of about 10-100 stellar radii from the surface. 
If the polar-cap high-energy emission cone occurs at a lower altitude than the radio emission,
as would be expected, then the leading edge of the radio cone would lead the first 
high-energy peak.  The trailing edge of the radio conal emission would then have to be
undetected in the {\it CGRO} pulsars.  This type of radio geometry for young pulsars was
suggested by \citet{man96} and a class of pulsars with one-sided radio conal emission was
proposed by \citet{lm88}.  This picture has recently received some observational 
support by \citet{cmk01} and \citet{ck03}. 

High-energy emission in the outer gap is generally radiated
in a different direction from the radio emission, which allows these models to 
account for the observed phase offsets of the radio and high-energy pulses.  At the
same time, there will be fewer radio-high-energy coincidences and thus a larger
number of radio-quiet high-energy pulsars.  In the \citet{ry95}
geometrical outer-gap model, the observed radio emission originates 
from the magnetic pole
opposite to the one connected to the visible outer gap.  Many observer lines-of-sight
miss the radio beam but intersect the outer-gap high-energy beam, having a much
larger emission solid angle.  When the line-of-sight does intersect both, the radio pulse
leads the high-energy pulse, as is observed in most $\gamma$-ray and some X-ray pulsars.

In the two-pole caustic model of \citet{dr03}, trailing-edge radio conal emission 
would arrive in phase with the high-energy emission due to the fact that emission from a wide 
range of altitudes is compressed into a narrow phase range to form the caustic.  This may be a 
nice explanation for why the radio and high-energy emission tends to be phase-aligned in some 
of the fastest rotators, such as the Crab and the millisecond pulsars.

\subsection{Polarization Properties}

Phase-resolved polarimetry of rotation-powered pulsars has had enormous diagnostic capability 
at radio and optical wavelengths and could also be a powerful diagnostic in the X-ray range.  
Several X-ray polarimeters are planned for the near future (AXP, POGO and 
MEGA) which could measure X-ray polarization characteristics of the brightest sources, such as the Crab pulsar, Cyg X-1 and Her X-1.
The pulsed non-thermal radiation from relativistic particles in the magnetosphere is tightly 
beamed along the neutron-star magnetic field lines and thus the emitted radiation is believed 
to be highly polarized either parallel or perpendicular to the field lines. Since the field 
well inside the speed of light cylinder rotates as a solid body with the star, measurement of 
the polarization properties as a function of pulse phase can provide a 
multidimensional mapping of the pulsar emission.  The expected signature of emission near the poles 
of a dipole field, an `S'-shaped swing of the polarization position angle through the pulse 
profile \citep{rc69a}, has 
been seen from many radio pulsars and has generally been taken as proof that the radio emission 
originates from the open field lines of a magnetic dipole.  

\begin{figure} 
\epsfig{file=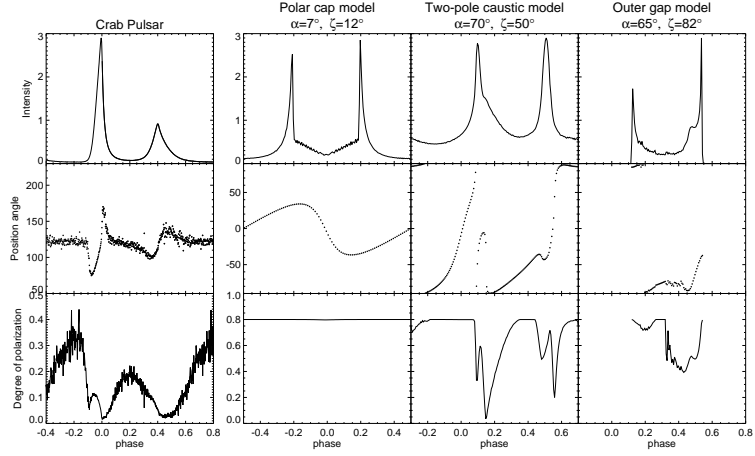, width=10cm}
\caption{Pulse profiles, polarization position angle and degree of polarization as a function 
of pulse phase for the Crab pulsar.  The left panel the observed optical pulse profile and 
polarization characteristics as measured by OPTIMA \citep{kel02}.  The right-hand panels are 
predicted profiles and corresponding polarization characteristics of the polar-cap, two-pole 
caustic and outer-gap model geometries \citep{dhr03}.}
\label{Crabpol}
\end{figure}

The different high-energy emission models share the common emission mechanisms of curvature, 
synchrotron and inverse Compton radiation from highly relativistic particles, albeit in very 
different locations.  These mechanisms all intrinsically produce highly polarized radiation (up 
to 70\%, depending on the particle spectrum).  Each of the models predicts a 
distinct variation of position angle and degree of polarization across the pulse profile.  
Figure \ref{Crabpol} shows simulations of expected position angle and percent polarization as a function of 
pulse phase for outer-gap and polar-cap models of the Crab pulsar.  Simulations are also shown 
for the two-pole caustic model \citep{dr03}.  The predicted position-angle swing in the polar 
cap model shows the classic S-shaped pattern of the rotating vector model since the observer is 
sweeping near the magnetic pole (phase 0).  The most rapid change in position angle occurs near 
the magnetic axis and there is no predicted decrease in degree of polarization.  The two-pole 
caustic model predicts rapid swings of position angle through the peaks, due to the large 
change in field orientation, as the viewing angle sweeps past the pole (phase 0) and approaches 
the trailing edge field line that is the origin of the caustic peaks \citep{dhr03}.  There is 
also a predicted decrease in percent polarization on the trailing edges of the peaks, as the 
observer is viewing emission in the caustics and from rotationally distorted field lines 
near the light cylinder that fold back and overlap.  The outer-gap model \citep{ry95} predicts 
position angle swings through the peaks, though they are smaller and occur only for carefully 
selected combinations of inclination 
and viewing angle \citep{dhr03}.  There would also be a predicted dip in degree of polarization, 
now primarily at the second peak, since the second peak in the outer-gap model originates from
the same caustic that is the origin of the dips at both peaks in the two-pole caustic model.   
The signature of caustic emission is thus a dip in the percent polarization and a rapid swing 
of the position angle at the pulse peak.  A recent measurement by OPTIMA \citep{kel02} of optical  
polarization 
position angle and degree of polarization as a function of pulse phase for the Crab pulsar  
exhibits sharp swings in position angle in the peaks and dips in degree in 
polarization just following each peak.  There appears to be a constant, unpulsed background 
component (at about $120^{\circ}$ position angle and 30\%) which if subtracted from the data, 
best matches the predictions of the two-pole caustic model.

\section{Thermal Emission from Cooling Neutron Stars}
\label{sec:thermal}

\subsection{Motivation and Background}
\label{sec:motivation}

The existence of magnetospheric emission from neutron stars, as
described in \S\ref{sec:magnetospheric}, came as an
observational surprise to neutron-star theorists.  By contrast, not
long after the existence of neutron stars was proposed \citep{bz34}, it
was also pointed out that such a star would form hot, with surface
temperatures over $10^6$~K, and that surface emission might be
detectable \citep{zwi38}.  The discovery of cosmic X-ray sources in the
1960's provided hope that neutron stars might one day be
observable, and detailed modeling of their potential observational
properties began \citep{mor64,tsu64}.  \cite{bw65} pointed out that the
equation of state of matter at densities comparable to those in atomic
nuclei, which cannot be studied in terrestrial laboratories, might be
constrained by observations of neutron-star cooling.  This is because
the magnitude of the stellar neutrino luminosity, which, in the early
history of the neutron star, should far exceed that of the photon
luminosity, should depend strongly on what species are present in the
core, and on how they interact.  Prodigious neutrino emission would
result in more rapid cooling than photon cooling alone.  Studying
neutron-star cooling began in quantatitive earnest with the computation
of the first cooling curves \citep{tc66}.

The first actual detections of thermal emission from the surfaces of
neutron stars came much later, following the launch of the {\it
Einstein} \citep{hcn80,hgsk85,mlt88} and {\it EXOSAT}
\cite[e.g.][]{bo87} X-ray observatories in the 1980's.  The situation
pre-1990's was reviewed by \cite{tsu86}.  The launch of {\it ROSAT} in
1990 heralded the next major progress
\cite[e.g.][]{fok92,ofz93,hr93}, with important
contributions made by {\it ASCA} \citep[e.g.][]{hw97,zpt98} as well.
By 1997, four {\it bona fide} detections of thermal emission from young
or middle-aged pulsars had been reported (from PSRs B0656+14,
B0833$-$45, B1055$-$52, and the Geminga pulsar) along with
interesting upper limits \citep[see][and references therein]{bt97}.  The recent
launches of the {\it Chandra} and {\it XMM-Newton} observatories have
already yielded important results
\citep[e.g.][]{pzs+01,ms02,kpsg02,spzt02,bcdm03}.  Recent reviews of
relevant observations include \cite{bp02,pzs02,pz03}.  Recent 
reviews of theory include \citet{tsu98}, \citet{pag98},
and \citet{yls99}.

As discussed in \S\ref{sec:magnetospheric}, thermal emission can also
be produced as a byproduct of emission processes in the magnetosphere.
Such emission is believed to be unrelated to that originating from
initial cooling.  Other heating mechanisms are predicted to matter only
in the photon-cooling era, i.e.  for $t>10^6$~yr
\citep[e.g.][]{aaps84a,usnt93,vle95,ll99}.  One notable
exception are magnetars, in which the decay of the field is thought to
provide a major energy source that dominates over all others (see
chapter 14).

\subsection{Theory of Neutron Star Cooling}
\label{sec:coolingtheory}

Neutron stars are born in the violent collapse of stars having ZAMS
mass $\sim$8--20~$M_{\odot}$, which have core temperatures prior to
collapse of $T_c \sim 10^9$~K.  Following collapse, it is expected that
the neutron star forms with central temperature $T_c \sim 10^{11}$~K,
although prodigious neutrino emission results in $T_c \sim 10^9 -
10^{10}$~K within a day or so.  Neutrino emission
dominates over photon emission until $T_c \sim 10^7-10^8$ K (depending
on leading neutrino emission process), which is reached at least $\sim
10^3$~yr, and more likely $\sim 10^5-10^6$ yr after birth.  The thermal
behavior before $10^6$~yr depends strongly on cooling model.  After $t
\simeq 10^{6}$~yr, various internal and external heating mechanisms
dominate thermal emission over the initial cooling.  Therefore, young,
i.e. having age $\lapp 10^6$~yr, neutron stars for which the surface
temperatures $T_s$ are expected to drop either slowly or quickly from
$T_s \sim 2 \times 10^6$ to $\sim 2 \times 10^5$~K, with core
temperatures dropping from $T_c \sim 10^9$ to $10^7-10^8$~K, are
interesting targets for studies of neutron-star interior physics.  Note
that in spite of these apparently high temperatures, from a statistical
mechanics point of view, neutron stars can be considered to be at
$T=0$, since for their average densities, the Fermi energy is $\sim
100-500$ MeV, while the internal temperature of a one-year old star is
certainly below $0.1$ MeV.  Next we consider in turn the main elements
that determine how a neutron star cools.

\subsubsection{The Stellar Core}
\label{sec:core}

During the first $\sim 10^5-10^6$~yr, a neutron star cools mainly via
neutrino emission from its innermost regions (which have densities $\rho
\gapp 10^{10}$~g~cm$^{-3}$).  For ages $t \gapp 10-50$ yrs, the
internal layers are isothermal.  They have an enormous
density gradient which results in different neutrino emissivities at
different radii.  As a rule, the most efficient neutrino emission is
produced in a stellar core, which extends from $\rho \approx \rho_0/2$
to the stellar center $\rho \sim (3-15)\, \rho_0$. Here, $\rho_0\approx
2.8 \times 10^{14}$~g~cm$^{-3}$ is the density of nuclear matter in
atomic nuclei. The composition and equation of state of supranuclear
matter ($\rho \gapp 2 \, \rho_0$) in neutron-star cores are largely
unknown but are of considerable interest to basic physics.

The main cooling mechanism in the core is neutrino emission via the
``Urca'' process\footnote{Urca is the name of a long-since-closed
casino in Rio de Janeiro, and was adopted as a name for these reactions
by Gamow \& Schoenberg (1941) \nocite{gs41} who saw a parallel between
how casinos extract money from players and how nature extracts energy
in these reactions.}.  In its simplest version adopted for neutron-star
cores, the Urca process is a sequence of a $\beta$-decay and an
electron capture, $n \longrightarrow p + e + \overline{\nu_e},$ and
$p+e \longrightarrow n + \nu_e$.  This is the so-called direct Urca
process \citep{lpph91}.  However, to proceed at interesting rates in
the neutron-star interior for a wide range of expected interior
properties, the Urca process requires a spectator reactant to
simultaneously conserve energy and momentum.  The process with the
spectator is called the ``modified'' Urca process \citep{cs64}.
Neutrons, protons, and other baryon spectators greatly suppress the
direct Urca rate (typically, by 5--7 orders of magnitude) because of
the lack of phase space due to degeneracy.  The most efficient direct
Urca cooling, without any spectator, is possible if the proton fraction
is sufficiently high.  This may happen in the cores of sufficiently
massive stars.  If hyperons are present in the core, they initiate
additional direct Urca processes \citep{pplp92}, as efficient (or
nearly as efficient) as the basic direct Urca process with nucleons.
If the direct Urca process is forbidden (e.g., if the proton
fraction is insufficiently high) but the dense matter contains pion or kaon
condensates, weaker processes similar to the direct Urca ones (though
stronger than the modified Urca ones) can open 
\citep[see, e.g. ][for a review]{pet92}.

%!cooling even more
%!efficient than even for exotic particles.  
%For example, for a core
%temperature of $\sim 10^9$~K, the neutrino emissivity in the presence
%of a kaon condensate is $\sim 10^{24}$~erg~cm$^{-3}$~s$^{-1}$, whereas
%for the direct Urca process, it is $\sim
%10^{27}$~erg~cm$^{-3}$~s$^{-1}$, compared with $\sim
%10^{20-21}$~erg~cm$^{-3}$~s$^{-1}$ for modified Urca cooling
%\citep[e.g.][]{pag98,ygk+03}.  Note that nucleon-nucleon bremsstrahlung
%($N + N \longrightarrow N + N + \nu + \overline{\nu}$) also occurs in
%the core, although at emissivities roughly a factor of $\sim$10 below
%those of the modified Urca process \citep{ygk+03}.

\subsubsection{Equation of State}
\label{sec:eos}

An equation of state (EOS) is the pressure-density-temperature
relationship of matter.  In a neutron star, most of the mass is at
densities two or three times $\rho_0$ (the nuclear matter density).  At
these densities, the EOS is unknown.  Nuclear physics can constrain
some aspects of the high-density EOS, however certain fundamental
parameters, such as the compression modulus, the bulk symmetry energy
and the effective nucleon mass remain weakly constrained \citep{lat92,lp01}
The EOS is crucial to understanding neutron-star structure; it
determines, among other things, the mass-to-radius relationship, the
stellar binding energy, the stellar moment of inertia and the relative
moments of different components, the minimum and maximum masses, and
the maximum angular velocity.  From a cooling point of view, the most
important parameters determined by the EOS are the density, pressure
and temperature profiles of the star, as well as its composition.  The
profiles are important because, for example, the location, amount and
type of superfluidities present depend on the density profile, via a
critical temperature/density relation (see \S\ref{sec:superfluidity}).
The composition is important as different species of particles will
react differently.  

Many EOSs have been proposed, and the range of possible
properties of neutron-star matter is large.  For example, the range of
predicted pressure of matter having density 
twice $\rho_0$ is approximately a factor of five, 
depending on the choice of plausible EOS, where plausible
means consistent with available laboratory data 
on atomic nuclei and nucleon scattering \citep{lp01}.
Phase transitions to kaon, pion or hyperon-rich matter are all possible
depending on the core density.  
%The effects of the large magnetic fields that are
%believed to exist in neutron stars are similarly unknown, but is
%thought to have a significant effect on Direct Urca processes
%\citep{by99}, as well as on the ratio of neutrons to protons and the
%neutron drip density \citep{sm01}.

\begin{figure}
\begin{center}
\epsfig{file=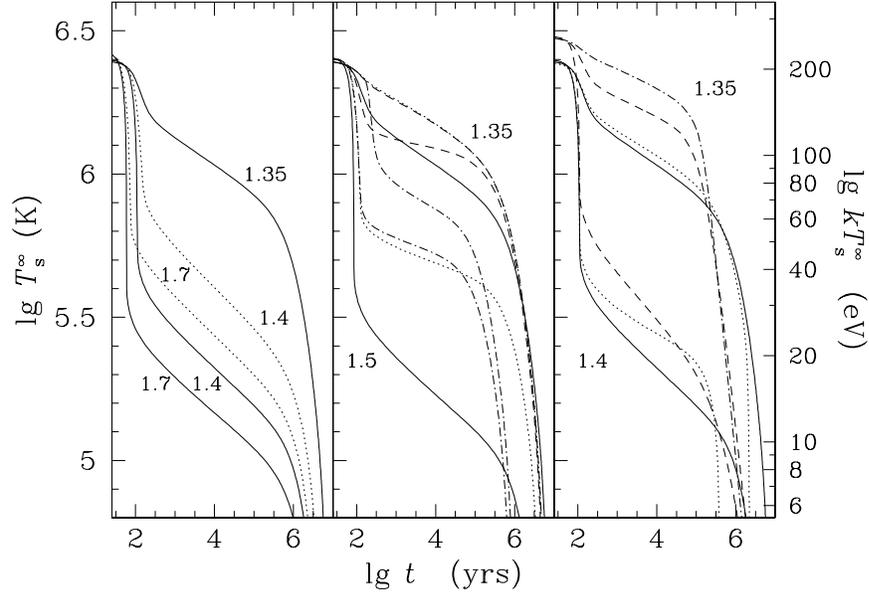, width=9cm, angle=270}
\caption{Neutron-star cooling curves for a variety of parameters and
models (courtesy D. Yakovlev).  All curves are for EOS ``A'' of
\cite{ykg01}, or ``model I'' of \cite{pal88} with compression modulus
of saturated nuclear matter 240~MeV.  Left: Curves for non-superfluid
stars for three masses (labelled in solar masses) are shown with solid
lines.  Dotted lines are for pion condensate in the core, for 1.4 and
1.7~$M_{\odot}$ stars (see \S\ref{sec:eos}).  Middle:  Curves for
superfluid neutron stars of masses 1.35 and 1.5~$M_{\odot}$.  Four
curves are shown for each mass:  solid curves are for non-superfluid
stars (as in the left plot); dotted curves are for proton superfluidity
model ``1p'' of \protect\citet{kyg02}; dashed curves are for proton
superfluidity model ``2p'' of the same reference; dot-dashed curves are
for the combination of proton superfluidity model ``p1'' and neutron
superfluidity model ``3nt,'' also from \protect\citet{kyg02}.  Note
that the 1p and 2p curves for the 1.35~$M_{\odot}$ star coincide as
both suppress modified Urca cooling.  Also, note that the 2p curve for
the 1.5~$M_{\odot}$ star sits among the 1.35~$M_{\odot}$ star curves.
Right:  Curves for non-superfluid stars for masses 1.35 and
1.4~$M_{\odot}$ for different accreted envelope masses and dipolar
magnetic fields:  solid curves are for zero accreted mass and zero
magnetic field $B$ (as in the left plot); dotted curves are for
$B=10^{14}$~G at the pole; dashed curves are for $B=0$ and accreted
envelope mass $10^{-7}$~$M_{\odot}$.  The dot-dashed curve is for a
1.35~$M_{\odot}$ solar mass star having $B=0$ and accreted envelope
mass $10^{-7}$~$M_{\odot}$, but with proton superfluidity included.  It
is shown to demonstrate the approximate upper bound for temperatures of
neutron stars having ages between $\sim$1 and $\sim$100~kyr.}
\label{fig:cooltheory}
\end{center}
\end{figure}

Figure~\ref{fig:cooltheory} (left) shows predicted cooling curves for a
commonly \citep[e.g.][]{ykg01} assumed core EOS \citep[originally
from][]{pal88} for three different masses.  Curves for
$M<1.35$~$M_{\odot}$ are very similar to that for
$M=1.35$~$M_{\odot}$.  Above this mass
(to be more exact, above 1.358\,$M_\odot$), for this EOS, direct Urca
cooling sets in; this is clear in the curves.  Also shown in the Figure
are two schematic curves assuming
a pion condensate (but no direct Urca if the pion condensate were
absent) in the core at $M > 1.35 \,M_\odot$.  The effect
is clearly enhanced cooling, but not as much as for direct Urca.

\subsubsection{Superfluidity and Superconductivity}
\label{sec:superfluidity}

Proton and neutron superfluidity in the neutron-star
interior can play a very significant role in cooling.
(Note that proton superfluidity means superconductivity.)  This is because
it suppresses traditional neutrino processes involving nucleons.
On the other hand, it produces a specific type of
neutrino emission \citep{frs76} associated with Cooper pairing of
nucleons.
Superfluidity also affects nucleon heat capacity.  However, its onset,
determined by a critical temperature, is very poorly known at
neutron-star densities.  
Overall, superfluidity in different forms can mimic enhanced cooling,
challenging the promise of
using cooling to determine the EOS.  See \citet{kyg02} and
\citet{ygkp02,ygk+04} for recent reviews of the subject.

The relevance of superfluidity can be summarized as follows.  In very
low-mass stars ($M \lapp 1.35$~$M_{\odot}$ for the EOS assumed here),
the effect of core singlet-state ($^1$S$_0$) proton superfluidity is
well determined and nearly independent of mass or EOS.  These stars
cool very slowly as even modified Urca processes are suppressed, so
that nucleon-nucleon bremsstrahlung, a slower process, dominates.  The
hottest young neutron stars are therefore plausibly explained as having
the lowest mass.  On the other hand, for very high-mass neutron stars
($M \gapp 1.5$~$M_{\odot}$), proton superfluidity becomes inefficient,
and one expects very rapid, model-independent cooling, because direct
Urca is too strong to be suppressed.  A neutron star that has cooled so
rapidly has not yet been observed (see \S\ref{sec:thermobs}).

For intermediate-mass neutron stars, the situation is highly model
dependent, and by assuming an EOS and proton superfluidity model, one
can, at least in principle, use observed surface temperatures to
determine neutron-star masses \citep[e.g.][]{ykg01}.  In these stars,
direct Urca processes would proceed unimpeded were it not for moderate
proton superfluidity suppression.  Figure~\ref{fig:cooltheory} (middle)
shows two different models for proton superfluidity \citep[see][and
references therein for definitions]{kyg02}.  
Mild $^3$P$_2$
pairing of neutrons in the stellar core results in very strongly
enhanced cooling  
of stars with $M \lapp 1.35\,M_{\odot}$
(Figure~\ref{fig:cooltheory}, middle) that is
inconsistent with the observations thus far (see
\S\ref{sec:coolingobservations}).

\subsubsection{The Stellar Envelope: Composition and Magnetic Field}
\label{sec:envelope}

Interpreting observed thermal emission from neutron stars means mapping
the observed surface temperature $T_s$ to that of the stellar core
$T_c$.  The temperature gradient from the surface inward is mainly in
the stellar envelope, the layer beneath the atmosphere down to the
isothermal internal region which effectively acts as a thermal
insulator for the bulk of the star.  The insulating envelope is
commonly defined as the region having density $\rho \lapp
10^{10}$~g~cm$^{-3}$.  This has thickness only a few tens of metres,
much smaller than the stellar radius.  The insulating envelope
thickness decreases as the star cools.

For envelopes made of iron, for $T_s \simeq 10^6$~K, $T_c \simeq
10^8$~K with $T_s \aproxpropto T_c^{1/2}$ for a typical surface gravity
\citep{gpe83}.  However, the mapping is highly sensitive to envelope
composition.  \cite{cpy97} show that even a tiny ($\sim
10^{-13}$~$M_{\odot}$) amount of accreted light-element matter, such as
from fallback after the supernova, from the interstellar medium or from
a binary companion, can have a substantial effect on $T_s$ for a given
$T_c$.  Specifically, accreted envelopes have lower thermal
conductivity. As a consequence, they lead to higher $T_s$ in the
neutrino-cooling era but lower $T_s$ in the photon-cooling era.
Figure~\ref{fig:cooltheory} (right) shows the effect of an accreted
envelope on predicted cooling curves of two different mass neutron
stars.

The magnetic field in the envelope can also play a major, though
complicated, role \citep{vl81,pag95,sy96,hh97a,py01,hh01}.  A
magnetic field affects all plasma components, especially the
electrons.  In particular, electron motion perpendicular to the field
lines becomes suppressed by classical Larmor rotation. In addition,
this motion is quantized in Landau levels.  Classical and quantum
effects greatly modify the electron thermal and electrical
conductivities.  Thus the effect of a magnetic field on the thermal
structure of the insulating envelope depends strongly on the field
geometry, with radial fields effectively reducing the insulation by the
envelope (quantum effect), and fields tangential to the surface
increasing it (classical effect).  Thus a dipolar field will have a
very different effect than an (often assumed) radial field
\citep[see][for a review]{py01}, with the tangential and radial regions
potentially cancelling local deviations when averaging over the surface
\citep{pag95,sy96,py01,hh01,pycg03}.  Figure~\ref{fig:cooltheory}
(right) shows the effect of a dipolar magnetic field of $10^{14}$~G (at
the pole) on cooling curves for two different masses.  Curves for
$10^{12}$~G are close to the zero field case.  The net effect is
clearly much more subtle than for light elements, mass, or EOS.
Note that the above consideration assumes to active magnetic
field decay of the type invoked in the ``magnetar'' model
(see Chapter 14).

\subsubsection{The Stellar Atmosphere}
\label{sec:atmosphere}

The neutron-star atmosphere, defined as the region having density $\rho
\lapp 10^2$~g~cm$^{-3}$, is typically only 0.1--10~cm thick, yet ultimately
determines many of the properties of the emerging thermal photon flux.  Although
in studying thermal radiation the assumption of a blackbody spectrum
is often made, every realistic atmosphere model predicts a significant
deviation from a Planckian curve.  The nature and degree of the
deviation, both from the continuum and in the form of lines (see
\S\ref{sec:speclines}) depends strongly on atmospheric
composition as well as on magnetic field.  A nice review of this
subject is given by \cite{zp02}.  We summarize the basic issues below.

\begin{figure}
\epsfig{file=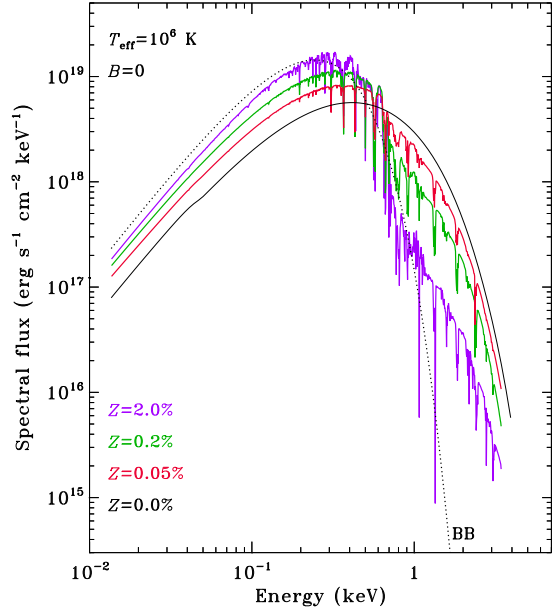, width=8cm}
\caption{
Spectra of emergent radiation in non-magnetic neutron star atmospheres
having $T_{\rm eff} = 10^6$~K for different metallicities $Z$ ($Z=2.0$\%
corresponds to solar metallicity.)  The corresponding blackbody is
shown with a dotted line.  From \protect\cite{zp02}.}
\label{fig:atmospheres}
\end{figure}

An unadulterated neutron star might be expected to have an iron
atmosphere, in keeping with the pre-collapse composition of the
progenitor core.  Heavy-element spectra, with even tiny (i.e.
$\sim$0.05\%) metallicities, are expected to contain many spectral
features and photoionization edges produced by ions at various stages
of ionization \citep{rom87,rr96a,zps96,rrm97}.  However, blurred to
lower resolution, a heavy-element atmosphere resembles a blackbody of
comparable effective temperature.  Figure~\ref{fig:atmospheres} shows
sample atmospheres containing different metal abundances, for
comparison with a blackbody spectrum.

On the other hand, a spectrum emerging from a light-element (H or He)
atmosphere is predicted to have few absorption lines at X-ray
energies \citep[e.g.][]{hlpc03} -- in fact none if the atoms are fully ionized, apart from
cyclotron lines.  Nevertheless, such an atmosphere has a clear signature: a
high-energy tail in the spectrum, resulting in a significant
overestimate, by as much as a factor of $\sim$3, of the surface
temperature if a blackbody model is fitted instead
\citep[see Fig.~\ref{fig:atmospheres};][]{rom87,rr96a,zps96}.  For example,
fitting a blackbody model to the X-ray spectrum of the Vela pulsar
yields $kT_s^{\infty} = 127$~eV, versus $kT_s^{\infty} = 59$~eV for a
hydrogen atmosphere \citep{pzs+01}.  Also, blackbody fits to hydrogen
atmosphere spectra yield emitting radii that are much smaller from
those inferred with hydrogen atmosphere models; for example,
\cite{ms02}, as revised by \cite{btgg03}, find $R^{\infty} = 156$~km and 8.5~km for hydrogen and
blackbody fits to the X-ray spectrum of PSR~B0656+14, respectively.
The high-energy tail is produced because light-element opacities
decrease rapidly with increasing energy, so that at higher energies one
is observing deeper layers of the star which are at higher
temperatures.  Although an unadulterated neutron star should have an
iron atmosphere, even tiny amounts (e.g. $10^{-20} \; M_{\odot}$) of
accreted interstellar material should result in a pure hydrogen
atmosphere, because of the short time scale for gravitational settling.
%\citep{ai80}.
%Accretion of material at the time of the supernova explosion, from a
%longer-lived fall-back disk of material, from the interstellar medium,
%or from a binary companion (relevant particularly for recycled
%pulsars) are all plausible sources of hydrogen.  
%This, together with the absence of any evidence for the numerous lines
%expected for heavy elements \citep[][and references therein]{pzs02,pz03}
%as well as the relative ease of calculability, make hydrogen atmosphere
%models particularly attractive.

As in the envelope, an appreciable magnetic field can also have a major
influence on the atmosphere.  The most important issue is the huge
difference that exists between the opacities for different
polarizations.  Specifically, the magnetic atmosphere is much more
transparent to photons with electric-field vector perpendicular to the
${\bf k \cdot B}$ plane than is a non-magnetic atmosphere.  Thus, the
emergent radiation is expected to be highly polarized (see
\S\ref{sec:pulsations}).  Furthermore, the emergent flux depends on the
direction of the magnetic field; hence for non-radial field geometries
pulsations are expected even for uniform temperature (which is, in any
case, implausible, since variations in surface magnetic field strength
result in temperature variations).

For magnetic field strengths greater than a quantum critical field $B_Q
\equiv 2 \pi m_e^2 c^3 /(e h) = 4.414 \times 10^{13}$~G, so-called
``vacuum polarization,'' that is the effective dielectric constant of
the vacuum, must be accounted for in the radiative transfer equations,
given the plasma densities in the envelope.  This is interesting
physically as, in spite of being predicted by QED theory, this effect
has not been confirmed experimentally.  The net effect of vacuum
polarization is to soften the high-energy tail of the hydrogen
spectrum, though the emission is still harder than the blackbody
spectrum of the same effective temperature
\citep{hl01a,oze01,ztst01,lh02,hl03,oze03,lh03a}.  
%Potentially observable
%cyclotron spectral features are also predicted in the presence of a
%substantial $B$ field \citep[e.g.][]{hl01a,ztst01,oze03}.

\subsection{Determinations of Neutron-Star Temperatures and Ages}
\label{sec:coolingobservations}

\subsubsection{Measuring Neutron-Star Temperatures}
\label{sec:temperatures}

There are several challenges to measuring neutron-star temperatures.  
First, the temperature measured by a distant observer $T_s^{\infty}$ is
not that at the neutron star surface $T_s$.  This is due to
gravitational redshifting; the two are related via $T_s^{\infty} =
T_s[1- 2GM/(R c^2)]^{1/2}$.  
Here, $R$ is a circumferential neutron-star radius
(which determines the proper length $2 \pi R$ of the stellar equator).
Similarly, the stellar luminosity as
measured by a distant observer is given by
$L_{\gamma}^{\infty} = 4 \pi \sigma T_s^4 R^2[1-2GM/(R c^2)]$,
where the $\gamma$ specifies that this is the bolometric photon (as
opposed to neutrino) luminosity.  
The apparent radius which would be measured
by a distant observer 
if the latter could resolve the star is
$R^{\infty} = R / [1-2GM/(Rc^2)]^{1/2}$.  Thus the
determination of $T_s$ from $T_s^{\infty}$ requires knowledge of
$M/R$, which is unknown, and which depends on the unknown EOS
in the stellar core
for any $M$ or $R$ that could be determined independently.

Particularly problematic in the pre-{\it Chandra} era is that young
neutron stars that are pulsars with high $\dot{E}$
are ubiquitously associated with pulsar
wind nebulae (see \S\ref{sec:pwn}).  These extended sources surround
the neutron star and contaminate measurements of the thermal emission
from the point source.  Also, sources
with high $\dot{E}$ produce copious non-thermal X-rays from
magnetospheric processes (see \S\ref{sec:magnetospheric}).  This
emission, for example, completely overwhelms any thermal emission from
the very young Crab pulsar, hence the availability of only an upper limit
on its thermal emission, obtained off-pulse
(Table~\ref{ta:thermal}).  Mitigation of both the above problems is
best done by avoiding high $\dot{E}$ sources, even though they are
often the youngest and hottest (see \S\ref{sec:ins}).

An additional important difficulty is the degeneracy between the
equivalent neutral hydrogen absorption toward the source, $N_h$,
and the measured temperature.  For an absorbed blackbody-like
spectrum, the low-energy exponential cutoff due to $N_h$ results
in higher inferred $T_s^{\infty}$ for smaller $N_h$, and vice
versa.  Generally it is necessary to constrain $T_s^{\infty}$ via
a contour plot in the $T_s^{\infty}$/$N_h$ plane \citep[e.g.][]{hw97,ztp99},
although frequently authors quote $T_s^{\infty}$ for only a single
assumed $N_h$ (see Table~\ref{ta:thermal}).

\subsubsection{Hot Spots and Cyclotron Resonance Scattering}
\label{sec:hotspots}

In many instances (see Table~\ref{ta:thermal} and \S\ref{sec:ccos}),
the effective radius of the star as determined by either a blackbody or
an atmosphere spectral fit is much smaller than the expected radius of
a neutron star, regardless of assumed EOS.  In at least one case,
RX~J1856.5$-$3754, this has led to claims that the thermal spectrum
therefore implies the target is not a neutron star at all, but a quark star
\citep{dmd+02}.

\begin{figure}
\epsfig{file=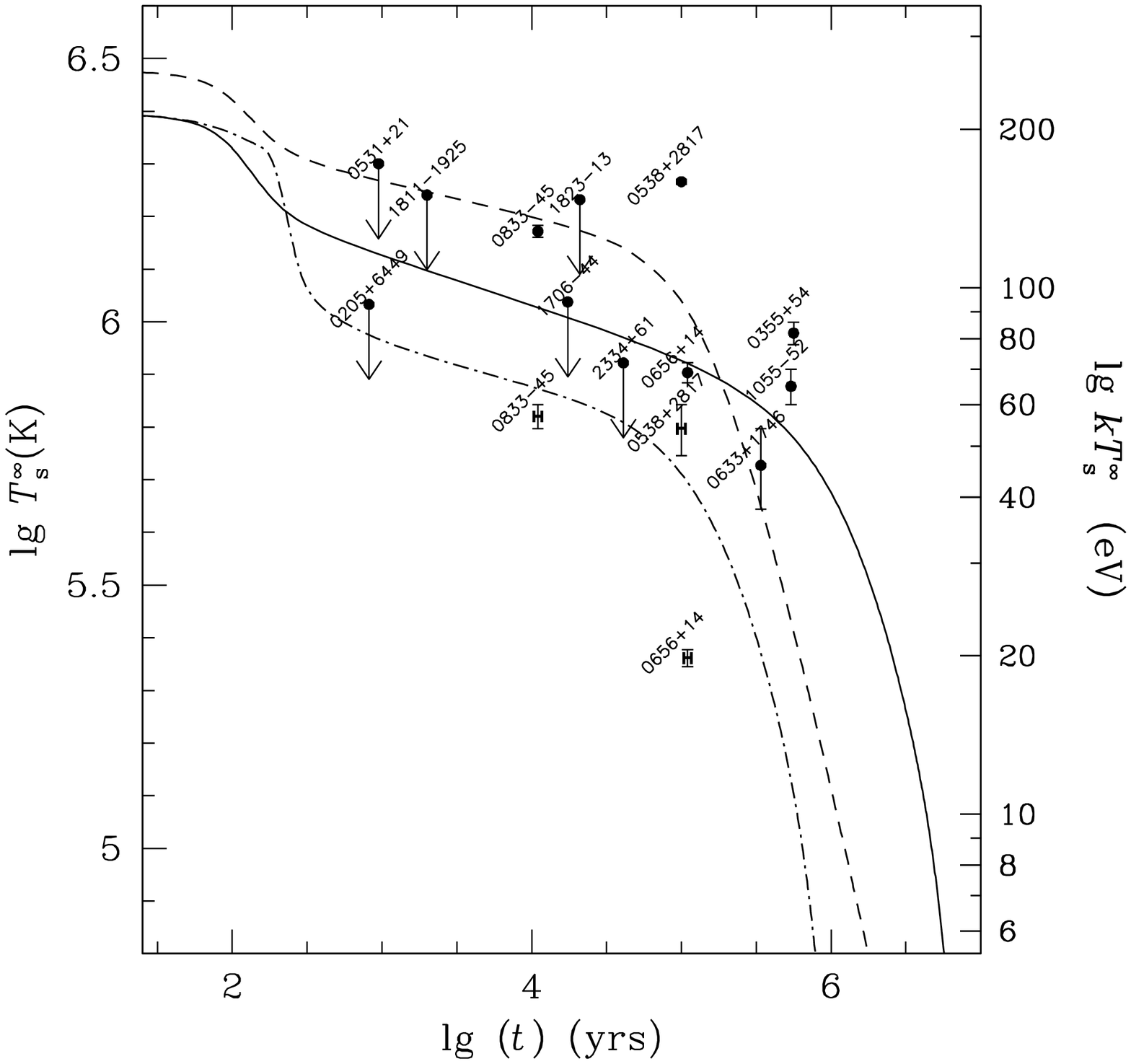, width=11.5cm}
\caption{Current best measured rotation-powered pulsar temperature
as a function of best age estimate (see Table~\ref{ta:thermal}).  
``H'' symbols denote that a $B=0$ hydrogen
atmosphere model was used; otherwise, a $B=0$ blackbody model was assumed.
The cooling curves shown are for 1.35~$M_{\odot}$ stars, as in Figure~\protect\ref{fig:cooltheory}: 
(solid) non-superfluid star with $B=0$ and no accreted mass 
(Fig.~\protect\ref{fig:cooltheory}, left);
(dot-dash) a combination 
proton and neutron superfluidity model 
(Fig.~\protect\ref{fig:cooltheory}, middle);
(dash) non-superfluid star with $B=0$ and accreted envelope mass $10^{-7}$~$M_{\odot}$ 
(Fig.~\protect\ref{fig:cooltheory}, right).  The curves are illustrative only and do
not represent fits to the data.}
\label{fig:coolobs}
\end{figure}

However, localized backheating of a neutron star surface by
relativistic \linebreak electron/positron pairs formed from conversion of
$\gamma$-rays produced in the magnetosphere in the polar-cap or
outer-gap accelerators (see \S\ref{sec:magnetospheric}) can produce ``hot
spots'' which emit thermal radiation.  Such hot spots have areas much
smaller than the stellar surface and could in principle outshine
surrounding thermal emission from initial cooling, thereby dominating the
X-ray spectrum.  This is the case for the thermal
emission detected from millisecond pulsars \citep[e.g. PSRs J0030+0451,
J0437$-$4715, J2124$-$3358; see e.g.][for a review]{bec01a}, which have
ages of $\sim 10^9$~yr hence should have cooled long ago.  When
hot spots dominate, studying thermal emission is mainly constraining
the physics of the magnetosphere, although upper limits on initial
cooling emission could still be interesting.  For example, the Vela
pulsar is a strong $\gamma$-ray source, yet a very faint thermal X-ray
source (see Tables~\ref{ta:nonthermal} and \ref{ta:thermal});
magnetospheric-acceleration-driven hot spots may well dominate the
thermal spectrum, but its low effective temperature is still
interesting (see Fig.~\ref{fig:coolobs}).  Thus, measured temperatures
that are associated with small emitting areas are most conservatively
taken to be upper limits on cooling temperatures.

Although one might expect stars with hot spots to have higher pulsed
fractions than those without, this is not necessarily the case due to
strong gravitational light bending expected in neutron stars (see
\S\ref{sec:pulsations}).  This and a less-than-favourable viewing angle
can significantly reduce the pulsed fraction seen from a star with hot
spots, as has been argued for RX~J1856.5$-$3754 \citep[e.g.][]{rgs02}.
The detection of a bow-shock H$\alpha$ nebula around this source
supports it being an off-beam rotation-powered pulsar \citep[see
\protect\ref{sec:ccos};][]{vk01}.  The same is true of other ``isolated
neutron stars'' (see \S\ref{sec:ins}) as well as of the ``central
compact objects'' (see \S\ref{sec:ccos}).

\citet{rud03} has argued that cyclotron-resonance scattering
of thermal photons by electrons and positrons within a few stellar
radii of an energetic rotation-powered pulsar results in a Planck-like
X-ray spectrum that is quite changed from the seed thermal spectrum.  A
pair plasma is thought to surround the star, maintained by conversion
of $\gamma$-rays from the star's polar-cap and/or outer-gap
accelerators.  This is because a thermal X-ray photon's energy, $\sim
1$~keV, is less than the cyclotron resonance energy at the stellar
surface, $eBh/2\pi mc = 11.6$~keV for $B=10^{12}$~G.  Thus, as the
photon passes through the magnetosphere, it goes through a resonance,
and the optical depth to cyclotron-resonance scattering becomes large.
\citet{rud03} describes the situation as if the X-rays are reflected
back from the surface of a ``lightly leaky {\it Hohlraum} container''
and only escape after multiple reabsorptions and re-emissions
from the stellar surface.  Thus, he argues, measurements of
thermal emission from neutron stars teach us only about the
stellar magnetosphere, not about initial cooling. 

\subsubsection{Pulsations and Polarization from Thermally Cooling Neutron Stars}
\label{sec:pulsations}

%The hallmark of most known neutron stars is their pulsations, as these
%determine the stellar rotation period $P$, and over time,
%the rate of spin-down $\dot{P}$.  The measurement of $P$ and
%$\dot{P}$ permits estimates of many interesting properties of
%the pulsars, including magnetic field and age (see \S\ref{sec:intro}).

The pulsations from the surface of a cooling neutron star have low
pulsed fraction and low harmonic content.  This is in strong contrast to
the high harmonic content and pulsed fractions seen from the
magnetospheric component of the emission, which is presumed to be
produced well above the stellar surface (see
\S\ref{sec:magnetospheric}).  Often young neutron stars exhibit
emission from both mechanisms, as is clear from the energy dependence
of their light curves (e.g. Fig.~\ref{fig:velalc}).

The low harmonic content results primarily from the presumably
low variation of temperature over the neutron-star surface.
However, pulsations are further reduced
by the strong gravitational bending
of light near the surface of the neutron star
\citep[e.g.][]{pag95,pod00}.  This becomes important as the general
relativistic compactness parameter $p~\equiv Rc^2/2GM \rightarrow 1$ 
(a Newtonian star is described by $p \rightarrow \infty$).  Due
to bending, although the observer may be facing one side of the neutron
star, she/he 
may still detect significant emission from the opposite
side.  Simulations of well defined infinite contrast spots on neutron
stars show that the effect is so strong that even for $p=4$, the
maximum pulsed fraction is less than 0.4 for {\it any} size emitting
spot \citep{pod00}.
 
Measured temperature is expected to be dependent on pulse phase, as the
thermal conductivity of the neutron-star surface depends on the
magnetic field strength, which varies over the surface
\citep{pag95,pod00}.  However, generally, only phase-averaged
temperatures are measured, both because of faint signals, and also
because gravitational light bending makes it difficult to do
otherwise.  Clearly this complicates the interpretation since the
averaged value depends on the distribution of temperature over the
surface.  However, this effect is likely to be small compared to other
systematic effects.

X-ray emission from the surface of a cooling neutron star endowed with
a magnetic field at the surface of magnitude $\gapp 10^{11}$~G is
expected to be significantly polarized.  This is because atmospheric
opacities depend strongly on polarization when the photon energies are
much smaller than the electron cyclotron energy.  The opacity to light with
its electric field vector oriented perpendicular to the magnetic field
is smaller by a factor of approximately the squared ratio of the photon
energy to the cyclotron energy, if the ratio is small, relative to that
for light with electric-field parallel to $B$ \citep{lcrt74}.  Thus,
radiation with electric-field vector perpendicular to $B$ escapes from
greater atmospheric depths where the temperature is higher, hence
results in a higher flux.  Polarizations from localized regions on the surface
should be very high, $\gapp 50\%$ \citep{kan75,ps78}.  However, an
observer sees radiation averaged over a hemisphere, and even beyond due
to gravitational light bending, which is strongest for large $M/R$.
This will reduce the observed polarization fraction
\citep[e.g.][]{pz00}.  Similarly, the different magnitudes and
directions of the lines of a dipolar $B$ field at the neutron-star
surface might be thought to reduce polarization \citep{pz00}.  However,
the QED effect of vacuum polarization, in which the region surrounding
the star is effectively birefringent, has recently been shown to
counteract the GR and varying $B$-field effects, because the observed
polarization direction of rays from the surface is correlated with the
$B$-field direction far from the surface; a ray bundle that we observe
passes through only a small $B$-field solid angle as it leaves the
magnetosphere \citep{hs02}.  This results in very large net
polarizations as a function of pulse phase, and even phase-averaged
polarization fractions of $\sim 10-20$\%
\citep{hsl03,lh03b}.  Measurement of polarization light curves should thus
allow sensitive constraints on the $B$ field, as well as on $M/R$ and the
magnetic geometry of
the neutron star.  Hence, thermally cooling neutron stars and magnetars
(Chapter 14) are expected to be very interesting targets for future
X-ray polarimetry experiments.

\subsubsection{Features in Neutron Star Thermal Spectra}
\label{sec:speclines}

Features in the X-ray spectrum of neutron stars have been an
astrophysical ``Holy Grail'' since their X-ray emission was
discovered.  This is because the presence of lines of identifiable
origin, with known inertial reference frame energies, would allow the
determination of the stellar redshift, $z$, and hence, at least in
principle, the mass-to-radius relationship via $M/R =
(c^2/2G)[1-(1+z)^{-2}]$, which would constrain EOS.  \citep[The above
is true only under the assumption that General Relativity correctly
describes the spacetime around the neutron star;][]{dp03}.
Atmosphere models with even as little as 0.05\% metal abundance predict
prolific line production due to the many transitions available for the
various atomic ionization stages \citep{rrm97}.  Hydrogen atmosphere
models, by contrast, predict few lines.  The relevance of spectral
features to constraining the neutron-star age is thus that (i) they
can help determine the atmosphere composition which can help
establish
what continuum model to fit; and (ii) they could, in principle, distort
the continuum given poor spectral coverage or resolution.

\begin{figure}
\epsfig{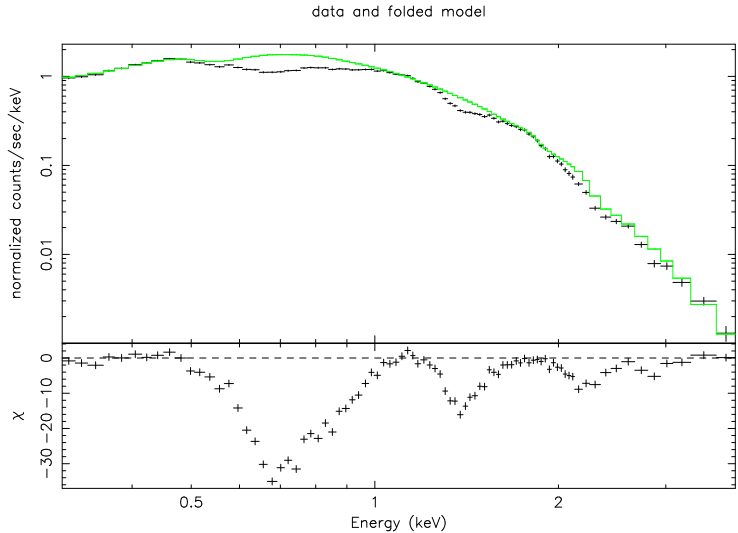}
\caption{
{\it XMM-Newton} spectrum of 1E~1207.4$-$5209.
The best-fit continuum curve
is represented by the sum of two
blackbody functions. 
Four absorption features appear at the harmonically
spaced energies of $\sim$0.7 keV, $\sim$1.4 keV, $\sim$2.1 keV
and $\sim$2.8 keV 
\protect\citep{dmc+04}.}
\label{fig:lines}
\end{figure}

\cite{spzt02}, using {\it Chandra}'s ACIS CCD detectors, observed two
absorption features, at 0.7 and 1.4~keV, in the X-ray spectrum of
1E~1207.4$-$5209, in the supernova remnant PKS 1209$-$51/52
(G296.5+10.0).  The nature of this object is not understood; see
\S\ref{sec:ccos}.  \cite{mdc+02} confirmed the existence of the lines
using {\it XMM-Newton}.  A subsequent long {\it XMM-Newton} exposure
revealed an additional absorption feature at 2.1~keV, as well as a
possible feature at 2.8~keV \citep[see
Fig.~\ref{fig:lines};][]{bcdm03,dmc+04}.  These suggest harmonics of a
cyclotron line at 0.7~keV.  A cyclotron fundamental at this energy
implies a magnetic field of $8 \times 10^{10}$~G for an electron line,
or $2 \times 10^{14}$~G for a proton line.  However, \citet{spzt02} and
\citet{hm02a} argue against the cyclotron interpretation, and suggest
possible spectral line identification for the most secure 0.7 and
1.4~keV features.  In any case, why such lines are not seen, in spite
of many comparably sensitive observations of other neutron stars,
remains a mystery.  This source is discussed further in
\S\ref{sec:ccos}.

Very recently, features in spectra of three ``isolated neutron stars'' have been
reported.  These are discussed in \S\ref{sec:ins}.

\subsubsection{Neutron Star Ages}
\label{sec:ages}

The standard age estimator for rotation-powered pulsars assumes frequency evolution
of the form $ \dot\Omega = k \Omega^n,$
where $n$ is the ``braking index,'' and $k$ is a constant that depends on
the magnetic moment of the neutron star.  
Assuming $k$ and $n$ to be constant, 
the braking index can be determined
from a measurement of the second time derivative of the frequency, 
$\ddot{\Omega}$, 
via
$n= \nu \ddot{\Omega} / \dot{\Omega}^2$.
The age derived by integrating the above
differential equation is:
\begin{equation}
\tau = \frac{P}{(n-1)\dot{P}} \left[1 - \left(\frac{P_0}{P}\right)^{n-1}\right],
\label{eq:spin}
\end{equation}
where $P_0$ is the spin period of the pulsar at the time it became a dipole
rotator, generally presumed to coincide with the supernova event.
For a simple vacuum dipole spin-down
model, $n=3$.  For $P_0 \ll P$ and $n=3$, Equation~\ref{eq:spin} reduces to
$\tau_c = P/2\dot{P}$, the often-used pulsar characteristic age.
For the five pulsars for which a constant value of $n$ has been
measured \citep{lps88,kms+94,lpgc96,dnb99,ckl+00}, the observed
values are in the range 1.4--2.91.  Though pulsars clearly do not rotate like
perfect vacuum dipoles, the range of braking 
indexes is limited and observationally well constrained.

The situation is less clear for the initial spin period, $P_0$.  This
can be determined from Equation~\ref{eq:spin} if the age is known and $n$ measured.
This is only the case for the Crab pulsar, whose estimated
$P_0\sim19$~ms has led to the generally made assumption that $P_0 \ll P$
for all but the very fastest pulsars.  However, the initial spin period
distribution of neutron stars is not well predicted by theory, since
the rotation rates of the cores of the massive progenitors are largely
unknown \citep{es78}.  Also, circumstances at core collapse could
significantly affect the neutron-star spin independent of the angular
momentum properties of the progenitor \citep[e.g.][]{sp98}.
As an example, consider PSR~J1811$-$1925 in the supernova remnant
G11.2$-$0.3 \citep[\protect{Fig.~\ref{g11_fig}};][]{ttd+99,krv+01}.  
%Figure~\ref{fig:g11per} shows the
%true age of PSR~J1811$-$1925 in units of $\tau_c$ as a function of
%$P_0$, in units of $P$ for different values of $n$.  
The supernova remnant age is well determined to be $\sim$2~kyr, while $\tau_c =
24$~kyr, implying $P_0 \simeq P = 65$~ms for reasonable $n$.  In this case, $\tau_c$ is
certainly over an order of magnitude different from the true age.
Similar arguments hold for PSR~J0205+6449 in the historic supernova
remnant 3C~58 \citep{mss+02}.

\begin{table*}
\caption{Observations of Thermally Cooling Young Rotation-Powered Pulsars}
\label{ta:thermal}
\vspace*{50pc}
\end{table*}
\nocite{shm02,sla94,wop+04,mkz+03,hw97,ms02,pzs+01,pz03,ghd02,gsk+03,bbt96}

\subsection{Current Status of Observations Vs Neutron Star Cooling Theory}
\label{sec:thermobs}

{\it Chandra} and {\it XMM-Newton} have made by far the best
measurements of neutron-star spectra thus far, and so offer the best
opportunity for constraining models of neutron-star cooling.  The
current status of observations of thermally cooling young neutron stars
has recently been reviewed by \citet{kyg02}, \citet{ykhg02} and
\cite{ttt+02}, and is summarized in Figure~\ref{fig:cooltheory}.
There, the data points are actual temperature measurements or upper
limits; those marked with an ``H'' are for hydrogen atmosphere models,
while the rest are for simple blackbody models.  
%Unfortunately, the
%literature is not always consistent regarding assumed models.
%Generally, the model yielding the more ``realistic'' radius, that is,
%the one closest to the expected neutron-star radius, is reported.  
The uncertainty from possible nebular contamination is not indicated; for
sources with nebulae, the observations made with {\it Chandra}, given
its superb spatial resolution, are superior to those with other
telescopes.  Age uncertainties are not indicated, but as discussed in
\S\ref{sec:ages}, can be an order of magnitude.  
For PSRs J0538+2817, B0656+14, and B0833$-$45 (Vela), we show both the
blackbody and hydrogen atmosphere temperatures, to illustrate the substantial
difference.  The plotted data are summarized in Table~\ref{ta:thermal}.

A striking property of the data is its scatter.  Although the above
caveats must be applied, the plot remains log-log, and it is hard to
explain all the data with identical neutron stars sharing a common mass
and hence EOS and interior structure.  
\cite{kyg02} argue that on the basis of the data together with models
that account for most of the phenomena affecting cooling (they do not
consider ``exotic'' particles, but do consider neutron and proton
superfluidity in detail), there are effectively three types of cooling
neutron stars \citep[see also][]{ttt+02}:  (i) low-mass, slowly cooling
stars where there is no direct Urca process occurring; (ii)
medium-mass, moderately cooling stars in which some nucleon direct Urca
processes occur, but for which proton superfluidity suppresses the bulk
of the possible Urca processes; and (iii) high-mass, rapidly cooling
stars for which suppression of direct nucleon Urca processes is weak.
Of course the mass cuts corresponding to these different categories
depend strongly on the EOS and on the possible types of superfluidity.
As an example, for the EOS ``A'' of \cite{ykg01}, for neutron stars
having age 4~kyr, for a 2p proton superfluid, the low-mass/medium-mass
cut is 1.36~M$_{\odot}$, while the medium/high mass cut is
1.64~M$_{\odot}$.  

What is clear is that we have not yet seen any very rapidly cooling
objects.  However this may be less of a statement about the interior
physics as it is about the mass function of neutron stars, i.e. there
may be very few high-mass neutron stars produced for purely
astrophysical reasons.  According to \cite{kyg02}, the Vela pulsar (PSR
B0833$-$45, fit with a hydrogen atmosphere) and Geminga (PSR
J0633+1746, fit with a blackbody), as well as presumably PSR~J0538+2817
(with a hydrogen atmosphere) the coolest objects, are prime examples of
moderately cooling stars for which some direct Urca reactions are
occurring, and their masses could be determined assuming some EOS, a
model of proton superfluidity \citep[but see][]{ttt+02} and the
dependence of the critical temperature for proton superfluidity on
density.  Note that the hydrogen-atmosphere model fit to PSR~B0656+14
is generally ignored, as it yields a radius much greater than is
possible for a neutron star.  The apparently higher temperature
objects, like PSR B1055$-$52, must, in this case, be low-mass neutron
stars, possibly affected by some internal heating mechanism or by
polar-cap heating.  The pulsar in 3C~58 recently received
fanfare for its relatively low temperature in spite of having likely
been born in 1181 AD, hence being very young \citep{shm02}.  Though the
Vela and Geminga pulsars nominally have lower reported temperatures
hence might be argued as being closest to rapidly cooling objects
\citep{ykhg02}, in fact the 3C~58 upper limit was for a blackbody
spectrum.  A hydrogen-atmosphere fit to the data (as was done for Vela)
could possibly bring the upper limit down to a very constraining level.
Indeed, the Vela temperature when fit with a blackbody is well above
that of the 3C~58 pulsar.

\section{``Isolated Neutron Stars''}
\label{sec:ins}

A relatively recently discovered class of X-ray sources are believed to
be thermally cooling neutron stars.  Their nature is still not well
understood.  These are currently being referred to in the literature as
``isolated neutron stars'' (INSs). This at first glance may seem like a
misnomer since this title does not distinguish them from isolated
rotation-powered pulsars.  Actually, as we describe below, it is
possible that at least a subset of the INSs are conventional
rotation-powered pulsars.  However, the sources known as INSs are
currently distinguished as showing no evidence for radio pulsations,
nebulosity, or accretion, and X-ray emission characterized by very soft
spectra that are well described by a blackbody, with no apparent
magnetospheric contributions.  These very properties, as well as their
proximity, make them enticing targets, since many of the observational
difficulties inherent in using thermal emission to constrain the
nuclear EOS that are described above in \S\ref{sec:thermal} are not
present.  The name INS therefore does apply to rotation-powered pulsars
as well, however the latter are a very special class of the INS
population:  long after rotation-powered pulsars have spun down past
the ``death line,'' beyond which they no longer produce observable
magnetospheric emission (this typically takes $\sim$30~Myr), ``dead''
neutron stars will still populate the Galaxy.  INSs, if identified as
this ``dead'' component, may well number as high as $10^9$ in the
Galaxy, much more than the $\sim 10^5$ active radio pulsars
\citep{lml+98}.

Prior to the launch of
the {\it ROSAT} satellite, theoretical predictions following an idea
first put forth by \citet{ors70} were that several thousand otherwise
dead neutron stars would be detectable in soft X-rays by {\it ROSAT},
due to their accretion of material from the interstellar medium
\citep{tc91,bm93}.  In fact, well after the rise and fall of {\it
ROSAT}, only 7 objects \citep[Table~\ref{ta:ins}; see][for a
review]{ttzc00} having the predicted properties (such as high X-ray to
optical flux ratio) have been found in spite of very careful searches
\citep[e.g.][]{dan98a,dan98b,rfbm03}.  This is likely due mainly to the
much higher than expected average neutron star space velocity
\citep[e.g.][]{ll94,hp97,acc02}, which hampers standard Bondi-Hoyle
accretion, although other explanations have been suggested
\citep[e.g.][]{pnr+03}.

In fact, of the objects in Table~\ref{ta:ins}, it now appears likely that {\it
none} is accreting from the ISM; rather, they are visible
because of thermal emission from the neutron-star surface.  If so,
these are unlikely to be extremely old neutron stars
as they are too bright; rather they would have to be either
off-beam conventional rotation-powered pulsars having ages $\sim 10^6$~yr and
magnetic fields $10^{11} - 10^{12}$~G, or even younger quiescent
magnetars having magnetic fields 2--3 orders of magnitude higher.

\begin{table}
\caption{Reported Isolated Neutron Stars$^a$}
\begin{center}
\begin{tabular}{ccccc}\hline\hline
Name$^b$ &  P  & ROSAT PSPC rate & $m_V$  & log $f_x/f_{opt} $ \\
     & (s) & (cps) & & \\ \hline
%MS 0317.7$-$6647$^*$ & ... & 0.03 & ... & $>1.8$ \\
RX J0420.0$-$5022 & 3.45 & 0.11 & ... & $>3.3$ \\
RX J0720.4$-$3125 & 8.39 & 1.69 & 26.8 & 4.6 \\
RX J0806.4$-$4132 & 11.4 & 0.38 & ... & $>3.4$ \\
RX J1308.6+2127 & 10.31 & 0.29 & 28.7 & 5.0\\
RX J1605.3+3249 & ... & 0.88 & 27.1 & 4.4 \\
RX J1856.5$-$3754 & ... & 3.64 & 25.8 & 4.4 \\
1RXS J214303.7+065419 & ... & 0.18 & ... & $>3$ \\\hline\hline
\end{tabular}
\end{center}
$^a$Data from \citet{ttzc00}, \citet{kkv03}, and \citet{hab03,hab04}; see text for more
information.\\
$^b$Source names ending with ``*'' are uncertain members of the class.
\label{ta:ins}
\end{table}

Although 4/7 of the INSs have now been shown to exhibit pulsations, none has had
its spin-down rate measured.  Their ages are thus
not known, which hinders efforts to use them in cooling studies.
However, in these sources, one hope is that spectral modeling, coupled
with sensitive, high-resolution X-ray observations, could determine
their radii.  If spectral lines were detectable as well, the
stellar masses could in principle be known independently from the lines'
gravitational redshift (\S\ref{sec:speclines}), thus providing a direct
constraint on the neutron-star mass/radius relation and testing EOSs.
Spectral modeling in these sources is greatly aided by their low equivalent
neutral hydrogen column densities, typically $1-2 \times 10^{20}$~cm$^{-2}$.
Although these ambitious goals have not yet been realized, the recent
discoveries of absorption features in the spectra of three INSs is
quite encouraging \citep{hsh+03,hztb04,vkd+04}.

The current situation is nicely illustrated by the example of
RX~J1856.5$-$3754, which is the best studied of the INSs.  This object
was discovered in {\it ROSAT} data \citep{wwn96} and optical
observations with the {\it Hubble Space Telescope} revealed a faint
blue counterpart \citep{wm97}.  This suggested
emission that is the Raleigh-Jeans tail of a $kT \simeq 50-60$~eV
blackbody.  The {\it HST} detection enabled an astrometric study of the
source that detected both parallax and proper motion \citep{wal01}.
The source is evidently extremely close to the Sun, having parallax
$8.5 \pm 0.9$~mas \citep{kva02,wl02} or distance $117 \pm 12$~pc.  The
proper motion of the source, if traced backward, appears to intersect
the Upper Sco OB association, if it has a radial velocity of
$\sim$30~km~s$^{-1}$ \citep{wal01,kva02,wl02}.  Assuming the neutron
star was born in the association implies an age of 0.4~Myr.

Because of the proximity of the source and the chance for the discovery
of spectral features, a 500~ks Director's Discretionary Time
observation with the {\it Chandra X-ray Observatory} was performed in
2002.  The X-ray emission was found to be unpulsed, with stringent
upper limits set on the pulsed fraction \citep[$<$1.3\% for frequencies
less than 50~Hz;][]{bhn+03}.  The spectral results of the DDT
observation were a surprise.  Not only were no features detected, but
the resulting X-ray spectrum is extremely well modeled by a simple
blackbody, with no evidence for the atmospheric distortion so clearly
predicted by theory \citep[see
\S\ref{sec:atmosphere};][]{dmd+02,br02,bhn+03}.  In particular, the
emitting radius for a single blackbody fit is $4.3(d/117 \; {\rm
pc})$~km, smaller than expected for any plausible neutron star.  This
led to speculation that the object is not a neutron star but rather a
self-bound quark or strange star \citep{dmd+02}.  A very recently
revised parallax distance of 175~pc would imply a radius of 7.2~km,
more in line with expectations for a neutron star (D. Kaplan, personal
comm.).  In any case, it is possible that the emitting region is simply
a hot spot on a larger stellar surface; this is not completely
precluded by the low pulsed fraction upper limits because of the strong
effects of gravitational light bending \citep{bhn+03}.  Interestingly,
the optical emission is too bright to be the low-energy extrapolation
of the X-ray spectrum.   A two-component blackbody model is necessary:
the lower temperature, perhaps from the bulk of the surface and
accounting for the optical emission, has $kT^{\infty} < 33$~eV, and the
higher temperature, perhaps from a hot spot and accounting for the
X-ray emission, has $kT^{\infty} \simeq 63.5$~eV.  In this case, a
strange star is not required \citep{dmd+02,br02,bhn+03}.  Of course,
the standard caveats regarding the use of blackbody models apply (see
\S\ref{sec:atmosphere}).  \citet{vk01} discovered, using H$\alpha$
imaging, a cometary shaped nebula having RX~J1856.5$-$3754 at its apex,
similar to those detected around some conventional radio pulsars (see
\S\ref{sec:pwn}).  This suggests that RX~J1856.5$-$3754 is a
conventional neutron star whose emission beam does not intersect our
line of sight.

Other sources in Table~\ref{ta:ins} are not yet as well characterized
as RX~J1856$-$3754.  We discuss each briefly, in order of the degree to
which it has been studied.

RX~J0720.4$-$3125 is a 8.39-s soft-X-ray pulsar \citep{hmb+97}
plausibly identified as an off-beam conventional rotation-powered
pulsar and not an accretor \citep[see][and references
therein]{kvm+03}.  No spin-down rate has yet been determined, although
interesting upper limits on $\dot{P}$ have been established, and imply
that $B \lapp 3.5 \times 10^{13}$~G \citep{chzz04}.  Its optical
through X-ray spectrum is well modelled by two blackbodies \citep[as in
RX~J1856$-$3754;][]{mzh03} plus a possible power-law component, similar
to what is observed in young and middle-aged radio pulsars (e.g.
PSR~B0656+14).  A large inferred velocity seems to preclude it being an
accretor \citep{mzh03}.  Intriguingly, \citet{hztb04} recently report
the discovery of a phase-dependent absorption line it this pulsar's
spectrum.  The line has energy $\sim$271~eV and an equivalent width
that varies from $-31$~eV to $-58$~eV depending on pulse phase.  The
line appears best interpreted as cyclotron resonance of charged
particles, presumably either electrons or protons.  For the latter, a
$\sim 5 \times 10^{13}$~G magnetic field is implied.  For electrons,
the field must be 2000 times smaller.

RX~J1308+2127 (RBS 1223) was identified by \citet{shs+99} in {\it
ROSAT} data as a relatively bright INS candidate, based on it soft
spectrum and the high X-ray to optical flux ratio made possible by the
good positional localization.  \citet{hsh+03} showed the source is
pulsed with $P=10.3$~s.  \citet{kkv02} reported a possible optical
counterpart having flux in excess of the extrapolation of the X-ray
blackbody spectrum, as in RX~J1856$-$3754 and RX~J0720.4$-$3125.  A
broad absorption line feature has been reported in this source
\citep{hsh+03}, and interpreted as a proton cyclotron line in a
$>10^{13}$~G field.  \citet{hsh+03} reported the discovery of a broad
absorption feature in this object's X-ray spectrum.  The line energy 
is at $\sim$300~eV and has equivalent width $\sim -$150~eV.  This
line is interpreted by the authors as being due to cyclotron
resonance absorption, in the $10^{13} - 10^{14}$~G range if protons,
and in the $10^{10} - 10^{11}$~G range if electrons.

RX~J1605.3+3249 (RBS~1556) was identified by \citet{mhz+99} as being a
possible INS, because of the soft source spectrum ($kT \simeq 92$~eV),
the lack of X-ray variability, and the absence of any optical
counterpart.  \citet{kkv03} reported the detection of a faint, blue
optical counterpart, which they argue confirms the INS interpretation.
The optical counterpart has flux that is over an order of magnitude
above the Rayleigh-Jeans extrapolation of the X-ray blackbody.  This is
like, though somewhat larger than, the excesses seen in the other
sources, and argues either for the X-rays being from a hot polar cap, or,
for the optical emission having a non-thermal component.  Either is
consistent with an off-beam rotation-powered pulsar.
Very recently, \citet{vkd+04} report the detection of a broad absorption
feature at 450~eV.  As for RX~J0720.4$-$3125 and RX~J1308+2127, the
origin of the line is as yet unclear, although \citet{hl04} show on the
basis of approximate calculations that all can be plausibly explained
by proton cyclotron absorption.

RX~J0806.4$-$4132 was reported by \citet{hmp98} as an interesting INS
candidate.  It was seen in two {\it ROSAT} All Sky Survey observations
at the same flux, and having a soft spectrum, with $kT \simeq 78$~eV.
\citet{hmp98} detected no optical counterpart, indicating a high X-ray
to optical flux ratio.  \citet{hz02} using {\it XMM-Newton}
observations of this source, report a possible 11.4~s periodicity at
the 3.5$\sigma$ level having a small $\sim$6\% pulsed fraction, and a
blackbody temperature of $kT = 94$~eV.  The improved {\it XMM-Newton}
position should allow deeper searches for a possible optical
counterpart.

RX~J0420.0$-$5022 is a {\it ROSAT} soft-spectrum source showing
evidence for a 22.7~s periodicity, a high X-ray to optical flux ratio,
no apparent variability, and a soft spectrum that is well described by
a blackbody of temperature $\sim$57~eV \citep{hpm99} All argue for it
being an INS,  although the period is much longer than has been seen
for any rotation-powered pulsar.

%MS~0317.7$-$6647 is an uncertain member of the INS class.  Discovered
%by \citet{swp+95}, the source appears to be a compact object of some
%type, as it has X-ray to optical flux ratio $\gapp 60$, and no radio
%emission.  \citet{swp+95} suggest that this source could be like the
%other INSs or it could be a very distant high-mass X-ray binary in the
%galaxy NGC~1313.  Evidence for variability in the {\it ROSAT} data
%argue for the HMXB interpretation.

\section{``Central Compact Objects''}
\label{sec:ccos}

There are several putative compact objects that have been studied
extensively and which have potential to constrain neutron-star cooling
theory (\S\ref{sec:thermal}), yet which are poorly understood.  By
definition, these sources, known as ``Central Compact Objects'' (CCOs),
are found near the centers of supernova remnants (SNRs), although this
could well be a selection effect resulting from SNR X-ray studies.  The
CCOs share the following distinguishing properties: unusual X-ray
spectra that, fit with blackbodies, and, in some cases, hydrogen
atmospheres, imply very small radii, and very high effective
temperatures (see Table~\ref{ta:cco}).  The small radii are
inconsistent with these objects being neutron stars for any EOS, while
the temperatures are inconsistent with any neutron-star cooling
models.  Other distinguishing characteristics are high X-ray to optical
luminosity ratios, no evidence for pulsations (except in one case --
see below), no evidence for a wind as seen in conventional young
rotation-powered pulsars (\S\ref{sec:pwn}), and no evidence for any
companion star that could be powering the X-ray emission via
accretion.  We discuss the five best studied CCOs briefly below.  For a
more detailed review, see \citet{psgz02}.

\begin{table}
\caption{Central Compact Objects$^{a,b}$}
\begin{center}
\begin{tabular}{ccccccc}\hline\hline
Name                   & SNR &  P  &  $kT^{\infty}_{bb}$   &  $R^{\infty}_{bb}$ & $kT^{\infty}$ & $R^{\infty}$ \\
                       &     &     & (keV)   & (km)  & (keV) & (km)  \\\hline
CXO J082157.5$-$430017 & Pup A & ... & 0.4 & 1.4 &  0.2 &  10\\
CXO J085201.4$-$461753 & G266.1$-$1.2 & ... & 0.4 & 0.3 & 0.3 & 1.5  \\
RX  J121000.8$-$522625 & G296.5+10.0 & 0.424 s & 0.25 & 1.6 & 0.1 & 1  \\
CXO J161736.3$-$510225 & RCW 103 & 6 hr ? & 0.4--0.6 & 0.2--1.6 & 0.3 & 1--8 \\
CXO J232327.9+584843   & Cas A & ... &  0.5 & 0.5 & 0.3 & 1 \\\hline\hline
\end{tabular}
\end{center}
$^a$See \protect\citet{psgz02}, \protect\citet{pzs02} and references therein.\\
$^b$Note: $kT^{\infty}_{bb}$ and $R^{\infty}_{bb}$ refer to blackbody models, while
$kT^{\infty}$ and $R^{\infty}$ refer to magnetic hydrogen atmosphere models.
\label{ta:cco}
\end{table}

\begin{figure}
\epsfig{file=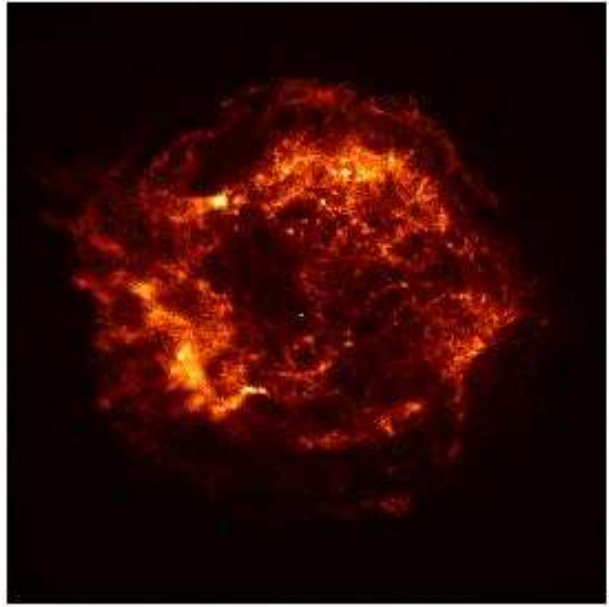, width=9cm}
\vspace{-0.5in}
\caption{``First light'' {\it Chandra} X-ray image of the supernova
remnant Cas~A \protect\citep[after][]{tan99}.  The image is 6$'$ on a side.
The previously unknown point source at the center of the remnant is
obvious.}
\label{fig:casA}
\end{figure}

The ``first light'' of {\it Chandra} targetted
the 300-yr old supernova remnant Cassaeopia A, and revealed, for the
first time, the presence of an X-ray point source near the center
\citep[Fig.~\ref{fig:casA};][]{tan99}.  This object, CXO
J232327.9+584843, seems likely to be the long-sought-after compact
object formed in this oxygen-rich remnant, clearly a result of the
supernova explosion of a massive star.
The spectral properties of this source, as well as the absence of any
detectable pulsations, make it very different from canonical young
neutron stars like the Crab pulsar.  Specifically, \citet{pza+00} and
\citet{cph+01} showed that for either a simple blackbody or hydrogen
atmosphere model, the temperature of the emission ($kT =
0.25-0.35$~keV) is very high, and the emitting area ($R = 0.2-0.5$~km)
too small to be consistent with surface thermal emission from initial
cooling.  A power-law fit yields a photon index ($\Gamma = 2.6-4.1$),
significantly higher than those seen in pulsars.  The difference is
further evidenced by the absence of any synchrotron nebula around the
object (Fig.~\ref{fig:casA}), as well as the absence of any radio
emission from the source \citep{mcd+01}.  The very high X-ray to
optical flux ratio, however, does implicate a compact object and render
an accreting binary scenario problematic \citep{kkm01,rws01}.  The
source spectrum is consistent with it being an ``anomalous X-ray
pulsar'' (see Chapter 14), that is, endowed with an ultrahigh magnetic
field ($10^{14}-10^{15}$~G), or with it being a conventional $B\sim 10^{12}$-G neutron
star having a cooling iron surface and hot hydrogen or helium polar
caps \citep{pza+00,cph+01}.  More recent {\it XMM-Newton} observations
of the source support these conclusions \citep{mti02}.  However, if
it is an AXP, why it should be so faint and show no pulsations, and, if
it has hot polar caps, why they should be so hot in the absence of
any evidence for the pulsar mechanism, is a mystery.

CXO~J082157.5$-$430017 is located in the Puppis~A SNR, though
significantly off-center \citep{pkwc82}.  From {\it ROSAT} and {\it
ASCA} observations \citep{pbw96,ztp99}, the object's X-ray spectrum is
well modeled by a blackbody, however the best-fit radius, 1.4~km, is
much smaller than that of neutron star.  A magnetized ($B\gapp
6\times10^{12}$~G) hydrogen atmosphere model yields a more
neutron-star-like radius of 10~km, and a temperature more in line with
that expected from cooling emission \citep{zpt99}.  However, neither
pulsations nor evidence for a surrounding nebula has been seen
\citep{gbs00,psgz02}, arguing that if this is a rotation-powered
pulsar, it must have very low $\dot{E}$.  This argues against the
heated-polar-cap interpretation, as in Cas A.

SNR G266.1$-$1.2, also known affectionately as ``Vela Jr,'' is in the
direction of the southeast corner of the Vela SNR \citep{asc98}.  The
possible detection of a radioactive $^{44}$Ti line suggested an age of
$\sim$700~yr, and a distance of $\sim$200~pc.  However, subsequent {\it
ASCA} observations yielded a $N_h$ more consistent with a distance of
$\gapp 1$~kpc, and an age of several thousand years \citep{she+01}.
The point source CXO J085201.4$-$461753 is near the center of the SNR,
and has an X-ray to optical flux ratio only consistent with a compact
object identification \citep{pskg01}.  This source is not seen to
pulsate and shows no evidence of a pulsar wind nebula in {\it Chandra}
data \citep{kpsg02}.  The source spectrum, fit with a blackbody model,
yields a radius of only $\sim 0.3$~km.  Atmosphere models do not yield
more reasonable radii \citep{kpsg02}.

RX~J121000.8$-$522625 is a compact source near the center of the large
SNR G296.5+10.0 (also known as PKS~1209$-$51/52).  The source was first
noticed in {\it Einstein} IPC observations made by \citet{hb84}.
\citet{mbc96} used {\it ROSAT} to show that a blackbody model fit the
spectral data well.  They further demonstrated the absence of any
extended emission, and set an optical limit of $V \gapp 25$.  They did
not detect any radio emission from the source at 4.8~GHz, and
\citet{kmj+96} saw no pulsed radio emission at 436~MHz.  \citet{vka+97}
and \citet{zpt98} presented {\it ASCA} observations of the source which
again supported the neutron-star interpretation.  {\it Chandra}
observations suggested a periodicity in the source, with $P=424$~ms
\citep{zpst00}.  \citet{pzst02}, using a second {\it Chandra}
observation, reported a spin-down age of 200--900~kyr, much larger than
the 3--20~kyr age inferred for the SNR.  However, subsequent
observations have not found simple spin down as in rotation-powered
pulsars, unless the source is an active glitcher \citep{zps04}.  This
is currently not understood.  The source spectrum is thermal, and the
best-fit continuum blackbody model yields a small 1.6-km emitting
radius \citep{pzst02}.  This source also shows absorption features in
its X-ray spectrum (Fig.~\ref{fig:lines}; \S\ref{sec:speclines}).

1E 161348$-$5055, the central source in the supernova remnant RCW~103
was the first proposed cooling neutron star \citep{tg80}.
\citet{gpv99} showed that the source is X-ray variable by as much as an
order of magnitude on a time scale of years.  A $\sim$6~hr periodicity
was tentatively reported in data from {\it Chandra} and {\it ASCA}
observations \citep{gpgz00}, and a 2002 {\it Chandra} observation
showed a clear 6.4-hr periodicity and apparently phase correlated X-ray
flux variations \citep{sgg+02}.  This strongly suggests that this
object is, in fact, an accreting neutron star having a very low-mass
companion.

\section{Pulsar Wind Nebulae}
\label{sec:pwn}
\subsection{Physical Overview}
\label{sec:pwnov}

Only a small fraction ($\lsim 10 \%$) of the spin-down energy of a young
pulsar is 
converted into observable pulsed emission (\S\ref{sec:magnetospheric}).
It is generally 
accepted that most of the energy leaves the pulsar's
magnetosphere in the form of a magnetized wind \citep{mic69}. In the 
ideal (${\bf E \cdot B}=0 $) MHD approximation to the aligned
rotating magnetic dipole \citep{gj69,ckf99}, 
it appears that the poloidal field ($B_p$, 
the field perpendicular to the toroidal component $B_{\phi}$)
goes from a nearly dipolar structure within the light cylinder 
to a split monopole structure outside the light cylinder (i.e. $B_p=B_r
\propto r^{-2}$, a purely radial field whose sign abruptly changes
at the equator). The toroidal
component $B_{\phi}$, which is small out to the light cylinder
% (sweepback 
%of the field lines due to rotational inertia of particles in the magnetosphere
%is not a significant factor until the velocity required for co-rotation
%approaches $c$, i.e. near the light cylinder, 
%\citet[eg.][]{ry95} and sect. 1.2.3), 
grows rapidly
outside the light cylinder so that for radii 
greater than a few times $R_{LC}$, $B_{\phi} >> B_p$.
``Cold" (i.e. non-radiating) charged particles flow outward
with this magnetic field forming a magnetized wind which is ultimately 
accelerated to very high energies 
($\gamma >> 1$). This highly relativistic magnetized wind eventually 
interacts with 
the surrounding medium, and emits synchrotron radiation from 
radio to $\gamma$-ray wavelengths. This synchrotron emission caused
by the pulsar wind is what is generally meant by the term
pulsar wind nebula (PWN). However, the winds from rapidly moving pulsars
may produce a bow shock in the ISM causing $H\alpha$ emission in what is 
essentially a thermal process. 
Optical filaments in the Crab nebula also appear to 
be thermal emitters \citep[eg.][]{fk82}, and in principal there could also be thermal X-ray 
emission at the outer edges of young PWNe, although this is difficult to 
distinguish from the surrounding supernova remnant (SNR) 
and has yet to be observed unambiguously . 

The details of the structure and luminosity of the PWN should
depend on the pulsar's spin-down energy history and space velocity
as well as the density profile of the surrounding medium. It may also
depend upon the magnetic inclination angle of the pulsar, although 
how the wind properties depend on this angle is poorly understood and it is possible
that the deviation from the aligned rotator model is negligible in terms
of overall growth and energetics of the PWN.
What we observe is dependent on the observer's viewing angle,
with Doppler boosting being a non-negligible factor in the surface brightness
distribution of the PWN. 

Since a pulsar is born in a supernova, the initial environment that the
pulsar wind encounters is the (nearly) freely expanding supernova ejecta. 
The pulsar may be born with a large space velocity and overtake
the expanding ejecta shell within a few tens of thousands of years, after
which the wind environment is probably the ISM 
%(or, in principle, if the pulsar is 
%rapid enough, the cavity caused by wind loss from the pre-supernova star).  
There appears to be a minimum spin-down energy necessary to create
bright PWNe with a significant drop in PWN
emission efficiency from pulsars with log $\dot E \lsim 36$
\citep{fs97,gsf+00,got03}, 
and generally only the youngest isolated pulsars are
observed to have PWNe. However, millisecond pulsars can also have
a substantial $\dot E$, and both $H\alpha$ \citep{bbm+95} and X-ray \citep{sgk+03}
nebulae have been observed around them.
 
\begin{figure}
\centering
\epsfig{file=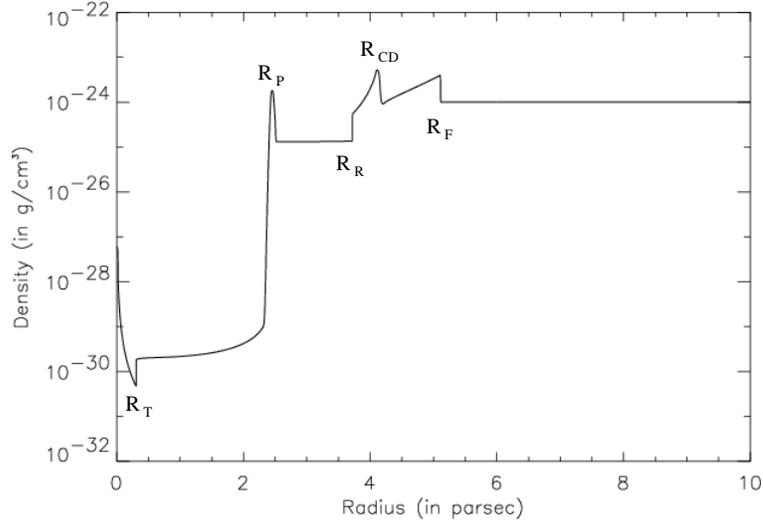, width=10cm}
\caption{Density profile of a spherically symmetric PWN expanding within
a pre-Sedov SNR showing the PWN termination shock $R_T$, the PWN forward shock $R_P$, 
the SNR shell reverse shock $R_R$, the contact discontinuity between the
SNR ejecta and the ISM $R_{CD}$, and the SNR shell forward shock $R_F$. Adapted
from \citep{vagt01}}
\label{density_profile}
\end{figure}

The standard theoretical picture for PWNe from young pulsars 
\citep{ps73,rg74,rc84,kc84a} is of a synchrotron bubble being
blown at the center of an expanding supernova remnant. Initially, the outer
edge expands supersonically into the supernova ejecta, so is presumably bounded
by a forward shock ($R_P$ of Fig.~\ref{density_profile}). The corresponding 
reverse shock ($R_T$) is near the center
of the nebula, where the cold relativistic wind from 
the pulsar is terminated.  (Note that the wind is cold only in the sense that
$kT << (\gamma-1) mc^2$ and hence in the co-moving frame of the bulk flow 
the particles have very little time to radiate before 
encountering the shock.)  The bulk flow energy is then 
converted into random particle motion with a power-law distribution
of particle energies. The wind continuously injects high-energy electrons
and positrons as well as magnetic field into the bubble. The particles gyrate
in the magnetic field, emitting synchrotron radiation from the radio through
soft $\gamma$-ray regions of the spectrum. 

%There is presumably a high-energy cut-off corresponding to the maximum energy
%to which the electrons are accelerated, and possibly a low energy cut-off as well.

\subsection{Observational Properties of PWNe}
\label{sec:pwnobs}

Except for a few cases (notably the Crab and Vela nebulae), we have
data on PWN emission only at centimeter wavelengths and/or in
X-rays below $\sim 10$ keV.  The radio properties of PWNe have been
outlined by \citet{ws78} among others.  A source can be identified as a
radio PWN if it has an amorphous (i.e. non-shell) morphology, flat,
non-thermal spectrum (energy spectral index $\alpha \simeq 0.0-0.3$,
where flux $S=\nu^{-\alpha}$\footnote{In radio work, the spectral index
is often defined as $S=\nu^{\alpha}$.  However we use the minus sign
here to be consistent with the X-ray definition.}), and high ($\gsim
10\%$) fractional linear polarization. The X-ray emission also has a
spectrum described by a power law, except the photon index $\Gamma
\equiv 1+\alpha$ is generally between $\sim 1.5 - 2.2$. Hence there is a need
for one or more spectral breaks such that the net change in
spectral index is $\Delta \alpha = 0.5-0.9$. The X-ray nebula
tends to be smaller than the radio nebula, and generally has a more
defined morphology, sometimes taking the form of a thick torus with
perpendicular jets.

The efficiency with which the spin-down energy is converted into synchrotron
radiation can be quite high. However, it is often difficult to 
determine since the observations tend to be restricted to two narrow bands.
In general, one can find four observed quantities, with associated errors, 
quoted in the literature (or minor variants): radio 
flux density at $\sim 1$~GHz, $S_{\rm 1GHz}$,
radio energy spectral index $\alpha_r$, 
1-10 keV X-ray flux $F_x$ (corrected for interstellar absorption), and 
X-ray photon spectral index $\Gamma = 1+\alpha_x$. From these, one 
can infer a break frequency where the spectral slope changes:
\begin{equation}
\nu_b \equiv \left( {S_x \nu_x^{\alpha_x}\over 
S_r \nu_r^{\alpha_r}}\right)^{1/(\alpha_x - \alpha_r)},
\end{equation}
%%= \left(4.1\times 10^5 F_X(2-\Gamma)(2.42\times 10^{19)$$ 
where $S_x$ and $S_r$ are the flux densities at $\nu_x$ and $\nu_r$, respectively.
Possible physical meanings of this break frequency will be discussed in 
\S\ref{sect:pwnspec}. 
At centimeter wavelengths, the spectral index $\alpha_r < 1$, hence the bulk of
the energy emitted in this band is at higher frequencies.
The radio luminosity $L_r$ therefore depends on what is chosen as the upper 
frequency limit $\nu_{ul}$ of the radio band:
\begin{equation}
L_r = 1.1\times 10^{30} {S_{\rm 1GHz} d^2\nu_{ul}^{1-\alpha_r}\over1-\alpha_r}
{\rm erg}\,{\rm s}^{-1},
\end{equation}
where $d$ is in kpc, $S_{\rm 1GHz}$ is in Jansky, and $\nu_{ul}$ is in GHz.
A natural upper limit would be the break frequency, but that
is usually not well known due to uncertainties in the radio and X-ray 
spectral indices and the large extrapolation.  Often, an upper limit of 100 GHz 
is randomly chosen in order to compare different nebulae \citep{gsf+00}.

The X-ray luminosity $L_x$ can be defined as the total power emitted from the break 
energy up to a high-energy cutoff  $E_C$:
\begin{equation}
L_x = 1.1\times 10^{32}{F_{x12}d^2(E_C^{2-\Gamma}-E_B^{2-\Gamma})\over
10^{2-\Gamma} - 1} {\rm erg}\,{\rm s}^{-1}, 
\end{equation}
where $F_{x12}$ is the 1--10 keV flux in units of $10^{-12} {\rm erg}\, 
{\rm cm}^{-2}\,{\rm s}^{-1}$ and $E_C$ and $E_B=h\nu_b$ are in keV. 
If the photon index $\Gamma < 2$, the 
X-ray luminosity depends critically on the high-energy cutoff, which
is only known in a very few cases. For the Crab and Vela PWNe, it is
in the few to tens of MeV range. If $\Gamma > 2$, then the bulk 
of the energy is emitted below the X-ray band, peaking near the 
break frequency. Note that the X-ray luminosity is usually much higher
than the radio luminosity. Since the synchrotron lifetime of the X-ray
emitting particles is often much shorter than the age of the
nebula, the properties of the X-ray emission can be highly 
dependent on the current spin-down
energy of the pulsar. The radio emitting particles generally have synchrotron
lifetimes longer than the age of the nebula, and are therefore an indication of
the total number of lower-energy relativistic particles injected into the nebula
throughout its lifetime. 

All known pulsars with detected X-ray or radio PWNe 
are listed in Table~\ref{pwn_table}. 
Table~\ref{pwn_table2} lists other X-ray or radio PWNe from which pulsations
have not been detected, while Table~\ref{pwn_table3} lists pulsars 
with $H\alpha$ PWNe. Note that the isolated neutron star RX J1856.5$-$3754
also has a $H\alpha$ nebula associated with it (see \S\ref{sec:ccos}).
A source is considered a PWN if it has one or more of the following observational 
signatures:
a known pulsar embedded in a radio and/or X-ray nebula 
which appears to be morphologically related to the pulsar; an isolated, 
filled-center radio nebula with the properties mentioned above;
a composite supernova remnant defined as a SNR shell with either a central 
radio component whose spectrum is flatter than the shell, a compact 
but extended central X-ray component with a power-law spectrum, or both;
an X-ray point source with a very high ratio of X-ray to optical flux and
associated extended X-ray emission with a power-law spectrum.
The X-ray morphologies are quite varied, but can be loosely placed 
within the following categories: torus + weak jet (T), 
jet/trail dominated (J), and mixed or uncertain (M). 
In the mixed case, sometimes there appears to be a bow-shock or torus as well 
as a bright trail or jet-like component, but sometimes there is just an 
indistinct blob. Although a dominant toroid is generally
assumed to be the typical structure, only a few of the brightest sources 
have been clearly determined to have toroidal morphologies.
A reference for X-ray morphology is given where available, otherwise
for the radio morphology.

There seems to be a change in PWN morphology from toroidal
to more of a jet or trail when the pulsar
spin-down energy drops below $\log \dot E \sim 36.5$ (in erg~s$^{-1}$). 
This probably simply reflects the
on-average greater age of these pulsars so they have moved substantially 
from their birth sites. The radio and X-ray
PWNe largely disappear once the spin-down energy drops below $\log \dot E \sim 35.5$. 
Those few sources listed as being PWNe whose pulsars' spin-down energies are below this 
tend to have barely detectable X-ray PWN. The fact that they are observable at all, 
along with the existence of $H\alpha$ nebula seen around pulsars whose $\log \dot E < 34.0$ 
attest to the persistance of the relativistic wind generation mechanism 
long after the effects on the surrounding medium are readily apparent.

\begin{table}[t]
\caption{Known Pulsars With Synchrotron Wind Nebulae}
\label{pwn_table}
%\vspace{0.4cm}
\begin{center}
\begin{tabular}{|l|l|c|c|c|c|c|l|}
\hline\hline
Name&Pulsar &  log($\dot E$) & $d$ & M$^a$
& S$^b$ & $\gamma^c$ & Reference\\
 & & (${\rm erg}\,{\rm s}^{-1}$) &(kpc) & & & &  \\
\hline
N157B & J0537$-$6910 & 38.7 &50&M&?&Y& Wang et al. 01\\
Crab & B0531+21 & 38.7 &2.0&T &N &Y & Weisskopf et al. 00\\
SNR 0540-69.3 & B0540$-$69 & 38.2 &50&T&Y &Y & Gotthelf \& Wang 00 \\
3C 58 &J0205+6449 & 37.4 &4.5&M &N &N & Murray et al. 02\\
G106.6+3.1&J2229+6114&37.4& 4.0&T &N? &Y & Halpern et al. 01\\
G320.4$-$1.2 &B1509$-$58 & 37.2& 5.0&M &Y&Y & Gaensler et al. 02\\
G292.0+1.8&J1124$-$5916&37.1& 6.5 &T&Y&N & Hughes et al. 01\\
G54.1+0.3 & J1930+1852&37.1& 9.1&T &N &N & Lu et al. 02\\
Kookaburra&J1420$-$6048&37.0& 5.6&M &? &Y & Roberts et al. 01\\
Kes 75 & J1846$-$0258&36.9& 21&M &Y &N & Helfand et al. 03\\
Vela X & B0833$-$45&36.8&0.29&T&Y &Y & Helfand et al. 01\\
G11.2$-$0.3 & J1811$-$1925&36.8& 5&M & Y &N & Roberts et al. 03\\
CTB 80 &B1951+32 & 36.6& 3.2&M&Y &Y & Safi-Harb et al. 95\\
G343.1$-$2.3 &B1706$-$44 & 36.5& 2.3&M &N &Y & Gotthelf et al. 02\\
GeV J2020+3658&J2021+3651&36.5& 12&T &? &Y& Hessels et al. 03\\
G18.0$-$0.7&B1823$-$13 &36.5&3.9&J&N &? & Gaensler et al. 03\\
Duck & B1757$-$24 & 36.4& 5.2&J &O&N & Kaspi et al. 01\\
3EG J1027$-$5817& J1016$-$5857&36.4& 8.0&M&?&Y & Camilo et al. 01\\
Mouse&J1747$-$2958&36.4&5.0&J &N &? & Gaensler et al. 03\\
G292.2$-$0.5&J1119$-$6127&36.4&6&M&Y &N & Gonzalez \& Safi-Harb 03\\
G308.8$-$0.1&J1341$-$6220& 36.1& 11&?&Y &N& Kaspi et al. 92\\
G270.3$-$1.0 &B0906$-$49 &35.7&2.6&?(J)&N &N& Gaensler et al. 98\\
W44&B1853+01 &35.6&3.1&J &Y &Y& Petre et al. 02\\
G341.2+0.9& B1643$-$43 &35.6&5.8&? &Y &N & Giacani et al. 01\\
Black Widow&B1957+20 &  35.0& 2.5&J&N&N& Stappers et al. 03\\
%%G114.3+0.3&B2334+61 & 34.8& 2.5&?&N&N& Kulkarni et al. 93 \\
Geminga& J0633+1746 & 34.5 & 0.16& J & N & Y & Caraveo et al. 03\\
%%S147& J0538+2817 & 34.3 & 1.2 & M &Y& N & Romani \& Ng 03 \\
%%Guitar&B2224+65&33.1&2.0&?(J)&N&N & Romani et al. 97 \\
\hline\hline 
\end{tabular}
\end{center}
$^a$X-ray Morphology. T = torus + weak jet, J = jet or trail dominated, 
M = mixed or uncertain.  A ? implies that no X-ray nebula has been clearly 
detected yet,
but if there is a clear radio or H$\alpha$ morphology, that is indicated in 
parentheses.\\
$^b$Indicates if there is an associated SNR shell. O means it is outside
the shell, ? means there is a nearby structure that could be interpreted as a related SNR shell.\\
$^c$Indicates coincident $\gamma$-ray source. ? implies
outside nominal $\gamma$-ray error box but possibly still related.\\ 
\end{table}
\nocite{wgc+01,wht+00,gw00,mss+02,hcg+01,gak+02,hsb+01,lwa+02,rrj01,hcg03,hgh01}
\nocite{rtk+03,sof95,ghd02,hrr+03,gsk+03,kggl01,cbm+01,gvc+03,gs03,kmj+92,gsfj98}
\nocite{pks02,gfgv01,sgk+03,cbd+03}
%%\nocite{kpha93,rn03,rcy97}
\begin{table}[t]
\caption{PWNe With No Detected Pulsar}
\label{pwn_table2}
\vspace{0.4cm}
\begin{center}
\begin{tabular}{|l|l|c|c|c|c|l|}
\hline\hline
Nebula & other name & $d$ & M & $S$ & $\gamma$ & Reference \\
 & &(kpc) & & & & \\
\hline
G0.13$-$0.11 & & ? & M & N & ? & Wang et al. 02\\
G0.9+0.1 && 8 &T&Y &N & Porquet 03\\
G7.4$-$2.0&GeV J1809$-$2327 &1.9&J & ?&Y & Braje et al. 02\\
G16.7+0.1 & & 2.2 &M& Y&N & Helfand et al. 03\\
G18.5$-$0.4&GeV J1825$-$1310 & 4.1&J&? & Y & Roberts et al. 01\\
G20.0$-$0.2 & & 5.4 &?&N&N & Becker \& Helfand 85a\\
G21.5$-$0.9 & & 5.5 &M&N&N & Slane et al. 00\\
G24.7+0.6 & & & ? & N & N & Reich et al. 84\\
G27.8+0.6 & & & ? & N & N & Reich et al. 84\\
G39.2$-$0.3 & 3C 396  & 7.7&M&Y&Y & Olbert et al. 03\\
G63.7+1.1 & & 3.8 & ? & N & N & Wallace et al. 97 \\
G74.9+1.2&CTB 87  & 12&M &N &Y & Mukherjee et al. 00\\
G119.5+10.2&CTA 1 & 2.1 &M&Y &Y & Slane et al. 03\\
G189.1+3.0&IC 443 & 1.5 & J &Y &? & Olbert et al. 01\\
G279.8$-$35.8&B0453$-$685&50&M&Y&N& Gaensler et al. 03\\
G291.0$-$0.1&MSH 11$-$6{\it 2} & ? &J&Y&Y& Harrus et al. 03\\
G293.8+0.6 & & & ? & Y & N & Whiteoak \& Green 96 \\
G313.3+0.1&Rabbit &?& J&N&Y & Roberts et al. 99\\
G318.9+0.4 & & & ? & Y & N & Whiteoak \& Green 96 \\
G322.5$-$0.1 & & & ? & Y & N & Whiteoak \& Green 96 \\
G326.3$-$1.8&MSH 15$-$5{\it 6} &4.1&?&Y &N & Dickel et al. 00\\
G327.1$-$1.1&  &8.8 &M&Y&N & Bocchino \& Bandiera 03\\
G328.4+0.2& MSH 15$-$5{\it 7}& $>17$&M&N &N& Hughes et al. 00\\
G359.89$-$0.08 & & 8 & J & N & Y & Lu et al. 03 \\
\hline\hline 
\end{tabular}
\end{center}
See footnotes to Table~\ref{pwn_table}.
\end{table}
\nocite{pdw03,brrk02,hag03,rrk01,bh85a,scs+00,oka+03,mght00,szh+03,ocw+01,wll02}
\nocite{ghrb03,hsg+03,rrjg99,dms00,bb03,hsp00,lwl03,wg96,rfs84,wlt97}

\begin{table}[t]
\caption{Pulsars with H$\alpha$ Bow-Shock Nebulae }
\label{pwn_table3}
\vspace{0.4cm}
\begin{center}
\begin{tabular}{|l|l|c|c|c||l|}
\hline\hline
Name & Pulsar & log($\dot E$)& $d$ & v & Reference \\
  & & (${\rm erg}\,{\rm s}^{-1}$) & (kpc) & (${\rm km}\,{\rm s}^{-1}$)& \\
\hline
&J0740$-$28&35.1&1.9&260 & Jones et al. 02\\
Black Widow&B1957+20 &35.0& 2.5&220 & Stappers et al. 03\\
&J0437$-$4715&33.6&0.14&94 & Bell et al. 95\\
&J2124$-$3358&33.6&0.27&61 & Gaensler et al. 02\\
Guitar&B2224+65 & 33.1& 3.2&1725 & Romani et al. 97 \\
%%RX J1856.5$-$3754&&&0.14&220 & van Kerkwijk \& Kulkarni 01\\
\hline\hline 
\end{tabular}
\end{center}
\end{table}
\nocite{jsg02,sgk+03,bbm+95,gjs02,rcy97}
%%\nocite{vk01}

\subsection{Evolution of a PWN  in a SNR}
\label{sect:pwnev}

The standard evolutionary model of PWNe is a spherical bubble
being blown at the center of a spherical SNR shell 
(Fig.~\ref{density_profile}). Observationally, 
this picture does not accurately represent the true state of
affairs. Many PWNe, including the ``canonical" example of the Crab,
do not have observable SNR shells, and so there is no observational
evidence for the expanding blast wave caused by the supernova. 
However, presumably there is an external blast wave 
although the surface brightness of the
shell might be very low. PWNe appear to be highly aspherical, 
sometimes with thick torii,
sometimes dominated by narrow, jet-like features. How this 
asphericity may affect the evolution of the PWN is not well understood,
although it is plausible that the mere addition of a filling factor to
the standard equations could adequately account for the morphological
variations \citep{rtk+03}. Pulsars are generally high-velocity
objects, so they rapidly leave their birth site, eventually overtaking the
supernova blast wave. The interaction of the pulsar wind with the swept-up
shell material can reenergize the shell, which seems to be occuring in the
case of CTB80 \citep{sfs89,vag+03}.
To reach the shell usually takes a few tens of thousand years, and so for the
first few thousand years assuming the pulsar is near the center of the 
SNR is reasonable. Well before the pulsar reaches the SNR shell, it
will overtake the outer edge of its original bubble, severely distorting the
shape, changing from a spherical/toroidal structure to more of a bow-shock.  
Since the forward shock of the PWN is generally accelerating within the 
expanding ejecta \citep{rc84}, this will usually not happen until after the PWN forward
shock encounters the SNR reverse shock. Analytical and 
numerical models of this transition suggest it
occurs at roughly half the pulsar crossing time of the nebula \citep{vdk03}. 

Therefore, for the first few thousand years of a pulsar's life, the assumptions
(other than sphericity) of the standard models are probably valid. The 
evolutionary picture has been described analytically \citep[e.g.][]{ps73,rc84}
and numerically \citep[e.g.][]{vagt01,bcf01}. 
Since the content and structure of the pulsar wind are poorly
understood, it is generally assumed that a spherical outflow of energy
equal to the spin-down energy of the pulsar (the observed pulsed electromagnetic
radiation being a negligibly small fraction of the energy) 
continuously flows outwards from the 
pulsar magnetosphere. This energy is split between the kinetic energy of the
bulk particle flow and the magnetic-field energy. At some radius which is small
compared to the total radius of the nebula, the electron/positron component
of the bulk flow is randomized and takes on a power-law distribution of 
energies. If this transition is caused by a standing shock, then for efficient
shock acceleration of the particles to occur, the flow energetics must
be dominated by the particle component at the shock. 
\citet{kc84a}, expanding on the work of \citet{rg74}, constructed a 
steady state MHD model of the Crab nebula, assuming a positronic 
wind terminated by an MHD shock whose downstream properties are
determined by boundary conditions imposed by the size, X-ray luminosity 
and expansion velocity of the nebula. The upstream flow properties
can then be determined from the Rankine-Hugoniot relations. The
ratio of magnetic to particle energy flux $\sigma$ in the unshocked wind of the
Crab is inferred to be $\sigma \sim 0.003$ \citep{kc84}. This leads to the well-known 
``sigma problem" of pulsar wind theory, since mechanisms for accelerating
the particles within the magnetosphere
%%ic gaps 
generally require $\sigma > 1$
% for the particle energy density to be low enough to allow the gaps to operate 
(see \S\ref{sec:magnetospheric}). 
This suggests
that the bulk of the acceleration occurs somewhere in the wind zone
outside of the pulsar magnetosphere, for which there is no generally
accepted model \citep[e.g.][]{mel98,ks03}. 
	
Although shock theory requires the ratio of magnetic to particle energy 
just downstream of the wind termination shock to be small, it is expected that
the magnetic-field will grow rapidly towards equipartition in the 
downstream flow. This allows the assumption that the magnetic and particle
energy densities are roughly equal throughout the bulk of the PWN. For the
first few days after the supernova explosion, the magnetic field density 
rapidly increases within the nebula, but then the expansion of the nebula
causes the magnetic-field strength to decrease with time despite
continuous injection of magnetic energy. 

The subsequent evolution of the
PWNe can then be separated into two eras of the pulsar spin-down $\dot E$ and 
two eras of the SNR expansion \citep[e.g.][]{rc84}. 
For times $t < \tau \equiv P_0/2 \dot P_0$, 
the initial spin-down time of the pulsar (note that this is 
probably the case for systems where the true age 
$t << \tau_c \equiv P/2 \dot P$, the 
{\it observed} characteristic age of the pulsar), 
energy will be injected at a nearly
constant rate ($\dot E \simeq \dot E_0$), 
and the radius of the PWN expands as $R\propto t^{6/5}$ while
the average magnetic field strength will decrease as $B\propto t^{1.3}$.
For large initial spin periods ($P_0 \gsim 50$~ms), the time of nearly 
constant energy input can last for several thousand years. During this
time, the outer edge of the PWN bubble is moving supersonically 
compared to the surrounding medium, and so a forward shock should form,
heating the swept-up ejecta which is in a thin shell around the 
PWN. The total mass of the swept-up 
material is initially fairly small, and so the thermal luminosity 
of this material may not be very large.
At later times ($t > \tau$ or $t \sim \tau_c$) when $\dot E << \dot E_0$, the expansion slows to 
$R\propto t$ if the SNR is still in its free expansion phase. 

The blast wave of the surrounding SNR shell initially expands almost
freely into the ISM \citep[or into a surrounding bubble blown by the 
wind of the pre-supernova progenitor star][]{che82}, 
shock heating the swept-up material.  As the
swept-up mass builds, a reverse shock forms and begins to propagate 
towards the center of the SNR, heating the ejecta. This reverse 
shock will reach the center of the SNR when the swept-up ISM mass
is much greater than the ejecta mass, marking the beginning of the
so-called Sedov-Taylor phase of the SNR \citep{sed59}.
Until this time, there are four shocks in the system: the forward shock of the
SNR blast wave propagating into the ISM, the reverse shock of the 
SNR blast wave propagating back into the freely expanding ejecta interior to 
the shock, the forward shock of the PWN bubble, and the PWN reverse shock/wind 
termination shock (Figure~\ref{density_profile}). 
When the SNR reverse shock encounters the 
PWN forward shock, numerical simulations show the PWN is initially compressed, 
and then its radius
oscillates a few times over the next several thousand years before 
settling down into a more compact and 
probably distorted shape \citep{vagt01}. 
During the compression, the magnetic field is     
enhanced causing an increase in the synchrotron luminosity and 
filaments can develop due to Rayleigh-Taylor instabilities \citep{bcf01}.
If the pulsar wind luminosity is still near its initial luminosity 
($\dot E \simeq \dot E_0$) after this
compression, the PWN begins to grow again, but now the expansion
is subsonic with $R\propto t^{11/15}$. If, on the other hand, $\dot E <<
\dot E_0$ then the expansion of the PWN goes as $R\propto t^{1/3}$ \citep{rc84}. 

\subsection{Emission Spectra of PWNe}
\label{sect:pwnspec}

The relativistic particle gas that fills the PWN bubble loses energy 
through synchrotron radiation and adiabatic expansion. Since the observed 
radiation spectrum has the form of a broken power law, 
a power-law electron energy spectrum is suggested, 
with the particle spectral index
$s$ (defined as $N(E)\propto E^{-s}$) related to the radiation spectral 
index $\alpha$ by $s=2\alpha +1$. 
The simplest model is of a single electron population injected into
a uniformly magnetized nebula with a single spectral index $S$ and a 
high energy cut-off $E_{ec}$ \citep{che00}. 
Synchrotron radiation is usually the most important
cooling process for the X-ray emitting electrons, 
with the typical timescale $t_c$ for
cooling dependent on the electron energy and the magnetic field.
At high energies,  $t_c << t$, the age of the PWN, 
and an approximate steady state can be
assumed between the injection and cooling of the electrons. In this regime,
the average particle spectrum in the nebula is $S+1$ which corresponds
to a steepening of the observed spectrum by $\Delta \alpha = 0.5$. A break 
in the particle spectrum occurs at an energy where $t_c \sim t$, 
resulting in a break in the observed photon spectrum at:
$$\nu_b = {1.68 \over B^3 t^2} {\rm GHz}, $$
where $B$ is in Gauss and $t$ is in years. Note that since $B$ decreases
with time, this break frequency actually increases as the PWN ages. 
At low energies ($t_c >> t$), cooling losses are negligible, and the 
particle spectral index is simply that of the injection spectrum. 
At late times when $\dot E << \dot E_0$, the initial cooling of the large 
number of particles injected at early times can leave an imprint on the 
particle spectrum in the form of a separate break below $\nu_b$ whose 
frequency decreases with time due to losses from the adiabatic expansion of 
the nebula. This results in an intermediate particle spectral index 
$S > s > S+1$ between the two breaks \citep{rc84}. The passage of the reverse
shock causes a brightening of the nebula and an enhancement of the magnetic 
field. The high-energy electrons all rapidly lose their energy, resulting in a 
spectral cutoff of the electron population. After the reverse shock passes, 
the low-energy spectrum is dominated by the old electrons with enhanced
emission, while the high-energy spectrum is from newly injected particles,
with a middle range below the post-reverse-shock cooling break which could have 
a fairly steep spectrum, i.e. $s_m > S+1$.  

It is tempting to identify the radio emission with the uncooled $s=S$ spectrum
and the X-ray emission with the cooled $s=S+1$ spectrum. However, this 
predicts $\Delta \alpha = 0.5$ and in many cases it is significantly greater \citep{wspb97}. 
If the X-ray or radio emitting particles were in the adiabatically cooled regime, $\Delta \alpha 
< 0.5 $. The only time a single injection spectrum would fit this
situation is after the reverse shock passage if the cooling break is 
above the soft X-ray band. In this case, there should be a spectral 
hardening somewhere in the hard X-ray or soft $\gamma$-ray band.
Another possibility is that the injection spectrum has changed over time. 
However, a correlation between PWNe X-ray spectral indices
and $\dot E$ showing a softening with increasing $\dot E$ has been claimed \citep{got03}, 
implying the spectral index is decreasing with age. This would again lead
to $\Delta \alpha < 0.5$.  A third possibility is that more than one
electron population is being injected into the nebula, one dominant at 
low energies and a second at high energies \citep{bnc02}. It is now clear that many, if not
most, PWNe consist of both equatorial outflows and polar outflows, the latter 
in the form of collimated jets. It might be expected that the two flows would 
inject particles into the PWN with different spectral slopes. Such a situation
would often require the polar flows to be relatively underluminous in the X-ray
region. 
A more fundamental problem is the flatness of the PWN spectra in radio. 
Both theoretical models \citep{kgga00} and observations of other astrophysical 
shocks which produce radio emitting electrons (eg. SNR shells and AGN jets)
tend to produce emission spectral indices of $\alpha \sim 0.6$ and then 
cool to $\alpha\sim 1.1$. The only workable mechanism for producing 
the radio spectra of PWN currently in the literature is one developed
for the Crab where a 
low-energy cut-off which is rapidly cooled during the early spin-down 
era \citep{ato99}. However, this requires some fine tuning of the 
magnetic field evolution of the PWN as well as a non-standard
spin-down history for the pulsar, and it is not clear whether this 
can be consistently applied to other PWN. Alternative models for the 
acceleration of the radio emitting particles at sites in the
nebula other than the termination shock should be explored.

In the above discussion, it has been implicitly  assumed that the 
electrons have been injected evenly throughout the nebula, the so-called 
one-zone model \citep{che00}. In a real PWN, the electrons are injected at
the termination shock within a few tenths of a parsec 
from the pulsar and then propagate outwards, initially at the typical 
post-shock velocity $v\simeq c/3$ but then with the bulk velocity 
typically decreasing as $1/r^2$ (this is true whatever the emission geometry,
since the nebula will have an $r^2$ expansion as long as the external 
confinement is spherically symmetric; this is probably not the case for some
rPWN -- see \S\ref{sect:pwnduck}).
If the PWN bubble is large enough, the X-ray emitting particles will
have a chance to cool significantly before reaching the outer edge, and it 
is expected that the X-ray spectral index will increase as a function of
distance from the pulsar. For all energies where the travel time to the outer
edge of the PWN is more than
the synchrotron cooling time, then the PWN should decrease in size 
with increasing energy. Therefore, we should expect the radio PWN to be larger 
than the X-ray PWN. \citet{kc84} modelled this in 
detail, and found the integrated X-ray spectral index should be 
$\Delta \alpha = (4 + \alpha / 9)$
larger than that of the uncooled particles emitting 
just downstream of the termination shock. For reasonable values of 
$\alpha= 0.0 - 1.0$, $\Delta \alpha = 0.44 - 0.56$, very close to 
the simple, one-zone model estimate of $\Delta \alpha = 0.5$.  
The spatial resolution of 
{\it Chandra} and {\it XMM-Newton} has allowed 
this increase in spectral index with radius to be observed in several PWNe
\citep[e.g.][]{wht+00,scs+00,pdw03}. 

\subsection{The Crab and Other Classic ``Plerions"}
\label{sect:pwncrab}

Until recently, observational and theoretical studies of PWNe have
been dominated by the Crab nebula. It contains the most 
energetic pulsar known in the Galaxy, it is bright at all wavelengths, 
has a precisely known age due to historical records of the associated 
supernova, and has been intensely studied for over 100 years. It had
been the defining member of the class of ``supernova remnants" with
no observable shells, sometimes called
{\it plerions} meaning filled center \citep{wei78}, 
and provided the conclusive 
proof that neutron stars are born in Type II supernovae. Ironically, 
although it has been known as the remnant of SN 1054 A.D. for many years, 
it is something of a misnomer to call it a SNR since the nebula is
a result of the pulsar wind and not the blast wave of the supernova
itself.  Why there is no observable radio shell is still uncertain, 
with explanations including that it is in a low density 
region of the ISM and that SN 1054 was an anomalously low-energy 
event \citep{fkcg95}. There are several other 
PWNe in the Galaxy with no observable shell, typified by 3C~58 \citep{ra85}. 
They tend to have much lower 
X-ray luminosites and harder spectra than the Crab. The inferred break energies 
between the radio and X-ray spectra of these other PWNe also tend to be    
much lower than that of the Crab's breaks \citep{wspb97}. 
Whether these differences
can be ascribed completely to the difference in spin-down energies
is debatable, but since observations of the Crab tend
to be of much higher quality than of all other PWNe, attempting
to interpret what is viewed by scaling from the Crab is somewhat
inevitable. The PWN whose properties are most similar to the
Crab's is that around PSR B0540$-$69, often referred to as the
Crab's twin, in the Large Magellenic Cloud \citep{chr84,msk93,gw00}.  

\begin{figure}
\centering
\epsfig{file=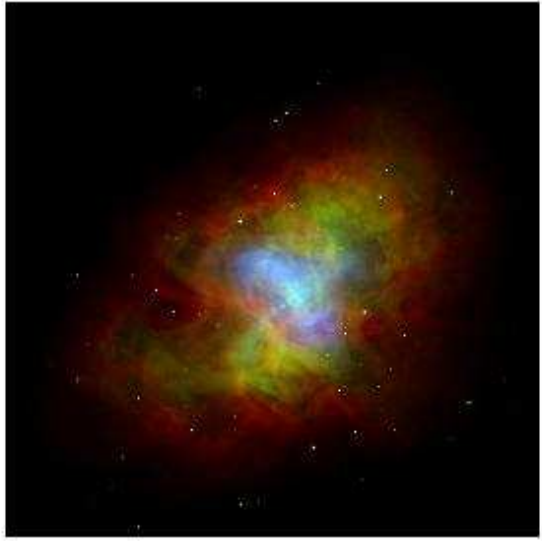, width=8cm}
\vspace{-0.5in}
\caption{A composite image of the Crab Nebula showing X-ray in blue, optical in green, and radio in red.
(Credits: X-ray: NASA/CXC/ASU/J. Hester et al.; Optical: NASA/HST/ASU/J. Hester et al.;
Radio: VLA/NRAO) (from Chandra website http://chandra.harvard.edu)} 
\label{crab_fig}
\end{figure}

At radio energies, the Crab appears as an ovoid blob with a
superimposed filamentary structure.  The radio spectral index is
remarkably spatially uniform \citep{bkf+97}, and the emission is mostly
constant except for a slow expansion \citep{vrv92} and a long term
overall decline in its flux \citep{ar85}. Recently, variations in the
form of small ripples in the nebula have been detected whose general
morphology correspond to the regions of activity seen in the optical
and X-ray nebulae \citep{bfh01}. The radio emission is highly linearly
polarized and well organized, showing that the emission mechanism is
almost certainly synchrotron \citep{bk90a}.

At optical energies the Crab 
is somewhat smaller than in the radio, and consists of thermal
filaments superimposed on a polarized non-thermal continuum.  Near the
pulsar, there is an underluminous region probably associated with the
unshocked wind.  Outside of this zone, there is a series of bright,
non-thermal enhancements, referred to as wisps,  that have been known
for a long time to be variable on timescales of months
\citep{lam21,ow56,sca69}. Monitoring observations with the Hubble Space
Telescope have  shown the enhancements, the innermost one starting at
the edge of the underluminous zone which is presumably at the wind
termination shock, move outwards over several weeks to months
\citep{hss96b}. One model of the radial placement of the wisps posits
that if the wind contains a significant fraction of ions, the
post-shock ions would cause enhancements in the magnetic field at
multiples of their cyclotron radius, and hence be sites of enhanced
synchrotron emission \citep{ga94}.

In X-rays, the Crab has a very definite toroidal morphology with 
polar jets. Again, surrounding the pulsar is an underluminous zone ending
in a bright ring with knotty enhancements which have been observed with 
{\it Chandra} to form and move outwards at $v \sim 0.5 c$ and then diffuse, similar
to the optical wisps \citep{hmb+02}. Outside of this region is a thick torus, 
which ends rather
abruptly well within the boundaries of the optical and radio nebula
(Fig.~\ref{crab_fig}). There is 
clear evidence of spectral steepening with increasing radius in the torus.
The jets emit a much smaller fraction of the X-ray luminosity than does
the torus, and overall have a harder spectrum. Outward motion of features within
the jets has also been observed. The proper motion of the pulsar is along the 
line of the jets, suggesting the natal kick was along the spin axis
of the pulsar \citep{cm99,wht+00}.

The unpulsed, and hence presumably nebular, emission spectrum is seen to extend
smoothly from the X-rays all the way to $\sim 25$~MeV. Above this is 
a sharp, possibly variable cut-off \citep{dhm+96} indicating some 
limiting factor in the 
particle acceleration mechanism. Above $\sim 100$ MeV, a rising unpulsed
component is seen which can be observed all the way into the TeV energy
range \citep{wee91}. This is most likely due to inverse Compton scattering 
off the synchrotron emitting electrons in the nebula.   

Since the Crab is energetically unique among Galactic pulsars, it is 
valid to wonder if its properties can be legitimately scaled down over 
two orders of magnitude to other pulsars with more typical spin-down energies. 
3C~58 was the second plerion to be identified and is thought to be a result
of SN 1181. 
Its pulsar, which was only recently discovered through X-ray 
and very deep radio observations \citep{mss+02,csl+02}, has a spin-down energy of
$2.7  \times 10^{37} {\rm erg}\, {\rm s}^{-1}$, second
only to the Crab among Galactic PWNe.  Although its spin-down energy is 
only a factor of $\sim 15$ less than the Crab's and the nebular radio 
luminosity is a factor of $\sim 10$ less than the Crab nebula, the nebular 
X-ray luminosity is a factor of $\sim 1000$ less \citep{bhs82,mss+02}. There
appears to be an unusually sharp spectral break \citep{gs92} at $\sim 50$~GHz.
The radio luminosity appears to be increasing, suggesting the unseen
SNR reverse shock may already be affecting the PWN \citep{ar85a}. 

The Vela pulsar is generally considered to be the prototype of pulsars
with $\tau_c\sim 10,000$ yr and spin-down energies in the $10^{36} -
10^{37} {\rm erg}\, {\rm s}^{-1}$ range. Its PWN appears very bright and can
be well-resolved at X-ray energies because of its proximity to Earth,
being only $\sim 290$~pc away \citep{dlrm03}. There is an associated
radio SNR shell, but it is patchy, uneven,  and diffuse, as is typical
of older remnants \citep{cc76}, having three sections called Vela X,Y,
and Z. \citet{wp80} noted that the Vela X region had radio properties
more similar to the Crab and other plerions than to ordinary SNR
shells.  Due to the age of the Vela SNR, it is likely that the 
reverse shock encountered the PWN long ago, although there may still be
some transient effects. The morphology of the Vela PWN in X-rays is a
double arc with narrow, jet-like enhancements, again near the line of
proper motion. While this has sometimes been interpreted as a bow-shock
morphology, it is now usually interpreted as a double torus
\citep{hgh01}. An apparently variable, extended jet feature is
sometimes prominent \citep{ptks03} outside of the arcs.  There is no
clear indication of an underluminous region near the pulsar, although
that may be due to an unfortunate viewing angle, and the observed
double-arc system may be the location of the termination shock.

\begin{figure}
\centering
\epsfig{file=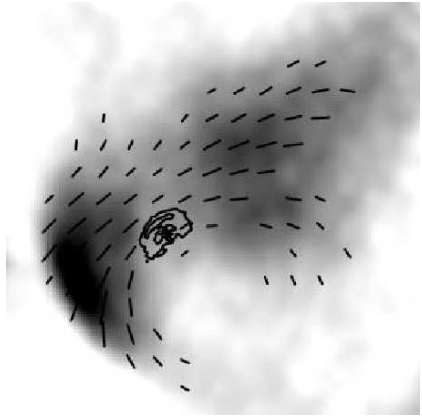, width=5.5cm}
\caption{5 GHz {\it ATCA} image of the Vela radio lobes near
the pulsar with polarization vectors 
showing how highly structured the magnetic field is. The contours
near the center are of the X-ray nebula as seen with {\it Chandra}. Note this is 
only a small part of the Vela X radio region. Adapted from Fig. 10 of
\citet{dlmd03}}
\label{vela_fig}
\end{figure}

The physical size of the bright part of the Vela X-ray PWN 
is about an order of magnitude
smaller than the Crab, and its X-ray efficiency in the 1--10 keV band
is more than two orders of magnitude less. The X-ray spectrum is
flatter and the high-energy cut-off is probably less than 10 MeV
\citep{dhss96}.  Whether these are indicative of a general trend of
lower efficiency, spectral hardening, and lower maximum particle
energies with lower $\dot E$ is difficult to say, although there is
some evidence for a dependency of spectral index on $\dot E$
\citep{got03} and the maximum potential drop across the open field
lines $E_{max} \propto \dot E^{1/2}$ so there is a theoretical
expectation for a lower high-energy cutoff \citep{dhm+96}.  The X-ray
PWN is centered between two radio lobes which are much larger than the
X-ray PWN \citep[Fig.~\ref{vela_fig}][]{dlmd03}, and are themselves just a small part near the
edge of the filamentary, flat spectrum Vela X radio complex. 
It is likely the reverse shock has displaced
the larger radio nebula which is why the proper motion vector of the
pulsar is not directed away from the center of Vela X \citep{bcf01}.

\subsection{Young Composite SNR}
\label{sect:pwng11}

The standard theoretical picture of a PWN growing within the expanding
shell of the supernova blast wave has been around since the early
eighties. Although quite a few young composite SNR were known at that
time \citep{hb87}, in embarrassingly few cases had the exciting pulsar
been observed.  While the unknown radio beaming could be invoked to
account for the lack of observable radio pulsations, this situation
made it difficult to confront theoretical expectations with
observational realities. Since the mid-nineties, several pulsars have
been discovered within young composites by deep radio searches
\citep{ckl+00,cmg+02} or with sensitive X-ray searches
\citep{ttd+97,gvbt00}, representing different stages and scenarios in the 
early evolution of PWN systems.

Perhaps the system whose observational
properties best match the assumptions made in the models is 
G11.2$-$0.3. This bright, remarkably 
circular remnant is often associated with
a ``Guest Star" observed by the Chinese in 386 A.D. \citep{cs77} and 
the age inferred from expansion measurements of shell support this 
association \citep{tr03}. HI absorption measurements
place the remnant at a distance of $\sim 5$~kpc \citep{ggts88}.
A 65-ms X-ray pulsar is within a few arcseconds of the geometrical
center of the shell \citep{krv+01} which has a 
characteristic age of 24,000 yr;  over ten times that of the
SNR \citep{ttd+97} strongly implying $P_0 \simeq P$ and $\dot E \simeq 
\dot E_0$. 
High-resolution radio and X-ray observations 
have separated the PWN from the shell and measured its spectrum \citep{trk02,rtk+03}. 

\begin{figure}
\centering
\epsfig{file=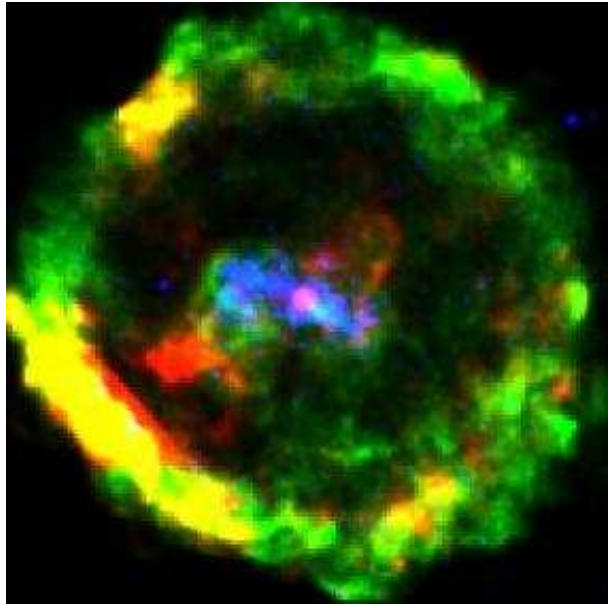, width=8cm}
\caption{{\it Chandra} and VLA image of G11.2$-$0.3 and its central PWN showing the
relationship between the soft thermal X-ray emission (red), the radio 
synchrotron emission (green) and the hard, non-thermal X-ray emission (blue).
The point source at the center is the pulsar, only seen in X-rays. 
Adapted from Fig. 1 of \citet{rtk+03}.}
\label{g11_fig}
\end{figure}

We therefore have in G11.2$-$0.3 an example of a composite remnant
where the assumptions of near spherical symmetry of the shell and negligible 
displacement of the pulsar from its birthsite are demonstrably true, the
energy output history is well constrained to be nearly
constant, the age is known, and the distance is fairly well determined. 
Given its age and estimates of the swept-up mass, 
it should be nearing the Sedov phase, so the reverse shock 
should be nearing the PWN. However, the radio PWN is one of the largest
relative to the shell and there is marginal evidence it is 
expanding \citep{tr03}, suggesting the reverse
shock has not yet begun to crush the PWN bubble. It therefore 
should still be expanding supersonically, and may be shock-heating the 
surrounding ejecta. There is thermal X-ray emission which seems to be
morphologically related to the radio PWN, but may also be a region 
of enhanced shell emission seen in projection (Fig.~\ref{g11_fig}). The 
velocity of the shell
inferred from the thermal emission is consistent with that inferred from
the expansion rate and HI distance estimates \citep{tr03}. 

Current estimates of the X-ray and radio spectral indices are consistent
with a single cooling break in between. However, given the age of the nebula, 
the implied magnetic field is unusually high. If it uniformly 
fills a region of the apparent bubble size, then the required magnetic
energy is much larger than could have been supplied by the pulsar. 
This may indicate that the pulsar wind is highly non-spherical, predominantly
coming from within $\sim 10^{\circ}$ of either the poles or the equator. 
The narrowness of the X-ray emission may support this interpretation 
\citep{rtk+03}. Varying X-ray spots are seen in this nebula as well, with
apparent motions of $\sim c$, much faster than the expected $c/3$ of 
a standard post-shock flow. So even with this most ``simple" of 
PWN/SNR systems where our knowledge is most complete, there are anomalies 
which are not well understood.

Kes 75 is a young ($\tau \lsim 2000$~yr) distant \citep[$d \sim
19$~kpc][]{bh84} composite SNR in many ways similar to G11.2$-$0.3. It
contains a 324~ms X-ray pulsar with the youngest known characteristic
age, $\sim 700$~yr \citep{gvbt00}.  The {\it
Chandra} image of Kes 75 \citep{hcg03} of both the shell and PWN in
both X-rays and radio of Kes 75 look very similar to G11.2$-$0.3, and
the pulsars powering both PWN are very X-ray efficient and have
unusually broad profiles. However, on closer examination, there are
remarkable differences. The X-ray efficiency in the 0.5--10~keV band of
the PWN in Kes 75 is $\sim 6.5\%$, which is the highest known,  while
that of G11.2$-$0.3 is only $\sim 0.2\%$, a much more typical value.
The pulsar's spin parameters imply both an unusually
large magnetic field ($B\sim 5\times 10^{13}$~G) and that $P_0 << P$,
so that the energy input into the PWN has changed dramatically over its short
history.  The implied size of the shell of Kes 75 requires an enormous
expansion velocity ($v\sim 13,000\,{\rm km\,s}^{-1}$) and therefore
implies a remarkably energetic explosion or else the neutron star was
born spinning very rapidly and a significant fraction of its rotational
kinetic energy was somehow transferred to the expanding ejecta
\citep{hcg03}.

SNR G292.0+1.8 is an example of a slightly older ($\tau\sim 3000$~yr)
composite system.  {\it Chandra} imaging shows a remnant filled with ejecta
material, but there is also a clear X-ray PWN slightly offset from the 
geometrical center \citep{hsb+01}. 
In this case there is no clear morphological separation between the radio 
PWN and shell and it seems likely that the SNR reverse shock has recently begun
to interact with the PWN \citep{gw03}.

\subsection{The Duck and Other PWNe with Trail Morphologies} 
\label{sect:pwnduck}

\begin{figure}
\centering
\epsfig{file=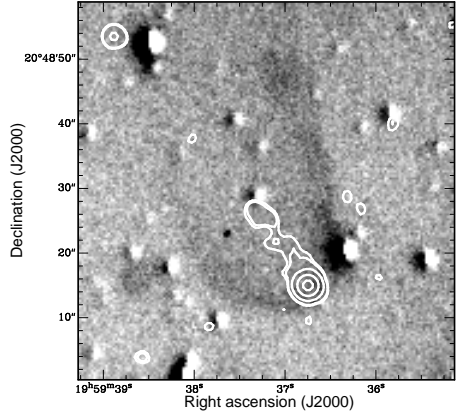, width=6cm}
\caption{$H\alpha$ image of Black Widow pulsar PSR~B1957+20
with {\it Chandra} X-ray contours.
Adapted from Fig. 2 of \citet{sgk+03}.}
\label{duck_fig}
\end{figure}

Several PWNe show a predominantly trail morphology at X-ray, and
sometimes radio, wavelengths extending from an X-ray point source,
presumably the pulsar, back towards the apparent birth-site. This has
been observed in pulsars both within the SNR shell, such as PSR
B1853+01 in W44 \citep{pks02}, and outside such as coming from PSR
B1757$-$24, the head of the Duck system \citep{kggl01}.
X-ray trails have also been seen from sources
with H$\alpha$ bow-shock nebulae, such as the 
Black Widow binary millisecond pulsar system PSR B1957+20
\citep[][see Fig.~\ref{duck_fig}]{sgk+03}.  The apparent radio bow-shock PWNe  source associated
with the possibly variable $\gamma$-ray source GeV J1809$-$2328 also
shows a clear point source and trail morphology in X-rays
\citep{brrk02}.  These type of PWNe are the most common around pulsars
with spin-down energies $\log \dot E \lsim 36.5$ (in erg~s$^{-1}$).
These nebulae are sometimes referred to as bow-shock
nebulae or ram-pressure confined nebulae, however it is not clear if either
of these terms accurately reflect the physical situation in respect
to the X-ray and radio nebulae. Therefore, we
will simply refer to these nebulae as rapidly-moving PWNe (rPWN), where the
pulsar's motion is rapid relative to the radial expansion of the nebula and
probably relative to the local sound speed.  

The Duck is an illustrative case. Upper limits on the proper motion of
the pulsar \citep{gf00,tbg02} rule out the X-ray emission being an
actual trail, since the synchrotron lifetime of the X-ray emitting
particles is too short for the pulsar to have moved the length of the
X-ray emitting region in the required time \citep{kggl01}. The X-ray
emission is therefore due to a collimated outflow, somehow constrained
to move within a cavity created by the pulsar's passage.  A spectacular
{\it Chandra} X-ray image of the Mouse shows the structure of the X-ray
nebula to consist of a narrow tail region in a broader, fainter,
bow-shaped nebula which may be too faint to see clearly in the Duck
\citep{gvc+03}.  The extremely long radio tail of the Mouse
\citep{pk95} is another example where it seems something must continue
to confine the flow well beyond where ram-pressure from the pulsar's
passage should no longer be relevant.  The extraordinary jet/trail
coming from PSR B1509$-$58 \citep{gak+02} also demonstrates that in
many cases pulsar outflows can be narrowly confined over large
distances even while still within the parent SNR.

The Black Widow pulsar PSR 1957+20 \citep{fst88},
so called because its wind is believed to be
destroying its white dwarf companion,
is interesting for several
reasons. Unlike the other sources with X-ray trails, it is a 
millisecond pulsar, demonstrating that old, recycled pulsars
also have relativistic winds.
It also has an $H\alpha$ nebula \citep{kh88} which clearly delineates
the forward bow-shock as well as a narrow trail of non-thermal 
X-ray emission \citep{sgk+03} which can serve as a test bed for
simulations of rPWNe \citep{buc02}.

\subsection{The $\gamma$-Ray Connection}
\label{sect:pwngam}

The magnetosphere of a young pulsar is one of the few places in the 
universe where particles can be accelerated such that
they will emit high-energy $\gamma$-ray emission through either the 
synchrotron process, curvature radiation, or 
inverse Compton scattering. Approximately 
1/3 of the known PWNe are coincident with sources of emission at $E > 100$~MeV
observed by the EGRET instrument on the {\it Compton Gamma Ray Observatory}
\citep{hbb+99}. 
Some of these are the known $\gamma$-ray pulsars \citep{tbb+99},
and this may be an
indication that the potential needed to produce pulsed $\gamma$-ray 
emission observable at typical Galactic distances is similar to that
needed to produce observable PWNe. However, several of the PWNe 
are associated with unidentified EGRET sources which appear to
be variable \citep{rrk01,ntgm03} on timescales of a few months, similar
to the synchrotron cooling timescale at these energies, 
indicating that the PWN may be the source of emission.  Practically
speaking, $\gamma$-ray sources can serve as tracers of PWNe. Six of the
PWNe listed in Tables~\ref{pwn_table} and \ref{pwn_table2} were initially 
discovered through
X-ray imaging of a $\gamma$-ray error box. 
In addition, the shape of PWNe can be used to infer the viewing angle, and
possibly the magnetic inclination angle, of the pulsar \citep[e.g.][]{nr04}
which is critical information for testing models of pulsed $\gamma-$ray emission
(\S\ref{sec:magnetospheric}).
Observations from the upcoming
GeV energy telescopes {\it AGILE} and {\it GLAST} will be able to 
distinguish between pulsed and unpulsed emission from many of these
sources, verify any variability, and locate more 
PWNe buried deep in the Galactic plane. 

Steady, unpulsed emission up to several TeV has been observed towards
three pulsars \citep[Crab, Vela, and PSR B1706$-$44][]{wee91,ykd+97,kht+93}) 
with PWNe, presumably caused by inverse Compton scattering
by electrons in the inner nebula. This is direct evidence of very high 
energy acceleration and puts constraints on the magnetic field in the inner
nebula. As the new generation of ground based Cerenkov air shower 
telescopes comes on line, many more PWNe will undoubtably be observed 
at energies of $\sim 100$ GeV to a few TeV. 

\subsection{Current Trends and Future Directions}
\label{sect:pwnfut}

The high-resolution X-ray imaging capabilities of the {\it Chandra} and
{\it XMM-Newton} telescopes are allowing detailed study of the
structure of many PWNe. The presence of collimated jet-like outflows as
a common feature of PWNe is a surprise which most models of PWNe,
dominated by a toroidal equatorial flow, did not allow for. New
analytical and  numerical models are now being developed with varying
amount of power being lost through polar outflows
\citep{kb03,kl03,sts+03}.  PWNe have also been shown to be
dynamical X-ray systems with observable variations on timescales as
short as a few weeks \citep{hmb+02}. As quantitative measurements of
these variations are made in more PWNe the need for the further
development of models of acceleration in the inner nebula becomes more
acute.  

The remarkable diversity of PWNe has also been a surprising result of
the new X-ray images. Many of the observations currently archived are
too short to distinguish different spectral characteristics of the
various features within PWNe. There is a suggestive correlation between
the X-ray spectral index of PWNe and the magnetospheric potential
\citep{got03}, but whether this will be born out by further
observations and whether there is a different correlation for jet-like
and toroidal structures will require deep observations of many more
PWNe. The structure of the PWNe also seems to depend on the magnitude and
direction of the pulsar's velocity, but so far only a few young pulsars
have had their proper motions measured.  Radio interferometry campaigns
now underway \citep[e.g.][]{bfg+03} promise to greatly increase the
number of measured pulsar transverse velocities, allowing 
the determination of how well jets are aligned with velocity vectors
\citep[e.g.][]{dlrm03}, if at all.
New 2-D and 3-D numerical simulations of PWN evolution are beginning to
yield intriguing results. For example: relaxing the spherical symmetry
outflow constraint but keeping axial symmetry leads to a
multiplicity of shock regions that can mimic the Crab morphology
\citep{kl03}.  Toroidal magnetic outflows can cause an elongation of
PWNe with spherical particle outflows, reminiscent of the shape of 3C58
\citep{van03}, and interactions with the reverse shock can lead to
complicated, asymmetric morphologies \citep{bcf01,bbda03}.

\section{X-rays from Rotation-Powered Pulsars in Binary Systems}

For completeness, we mention briefly another related source of X-rays
from rotation-powered pulsars -- shock emission in binary systems.
Although these systems do not obviously satisfy the mandate of this
chapter on ``isolated neutron stars,'' the nature of their X-ray
emission is very similar to that in a PWN.  For high-$\dot{E}$ pulsars
in binary systems, the confining medium can be the wind of the
companion star.  In this case, the confinement, particularly for
eccentric binaries, is strongly orbital-phase dependent, leading to a
PWN of luminosity (and, presumably, size) that varies systematically
with orbital phase.  Even circular orbits are of interest since the
orientation of the PWN, which, in this case, is not expected to be
isotropic, changes with orbital phase.  In principle, such systems are
an excellent diagnostic of shock acceleration and the pulsar wind as
they have regular, repeating and predictable dynamical properties.
They also represent a new form of ``X-ray binary,'' one not powered by
accretion.  However, only two such systems have been detected in
X-rays:  PSRs 1259$-$63 and B1957+20.  Their rarity is because of the
need to have a high $\dot{E}$ pulsar in a binary system in which the
two components come sufficiently close to each other for the shock to
be strong, but not too close so that accretion occurs.

The 48-ms radio pulsar PSR~B1259$-$63 is in a 3.4~yr binary orbit
having eccentricity 0.87 \citep{jml+92}.  From radio timing, the pulsar
is known to have $B=3 \times 10^{11}$~G and $\tau_c = 3 \times
10^5$~yr.  The pulsar's companion, identified by its location within
the $< 1^{\prime \prime}$ pulsar timing error box, is the 10th mag B2Ve
star SS~2883.  It has mass $\sim 10$~M$_{\odot}$ and radius $\sim
6$~R$_{\odot}$, deduced from its spectral type.  The system provides
an evolutionary link between the rotation-powered pulsars and the
high-mass X-ray binaries.  For PSR~B1259$-$63, near periastron, the
pulsar approaches its companion to within $\sim$25 Be star radii.
\citet{crj94} detected variable X-rays from the system near apastron.
This ruled out standard wind accretion scenarios, magnetospheric 
emission, and emission from the companion.  \citet{crj94} suggested either
some form of non-standard accretion or the pulsar wind shocked 
by the companion wind as possible mechanisms for the X-ray emission.
\citet{tak94} considered pulsar/Be star wind interactions in detail, in
particular, shock emission at the location of pressure balance between
the pulsar and Be star winds.  They suggested this mechanism produced
the apastron X-rays, and predicted PSR~1259$-$63 would be a moderately
strong, unpulsed X-ray source near periastron, unable to accrete due to
the shock distance from the pulsar being much larger than the accretion
radius.  \citet{ktn+95} and \citet{hnt+96} reported on {\it ASCA}
observations around the periastron of 1994; \citet{hck+99} summarize
all the X-ray observations including two more made at the subsequent
apastron.  The source was clearly seen to increase in intensity by a
factor of $>10$ near periastron relative to apastron, with the peak
luminosity $\sim 10^{34}$~erg~s$^{-1}$.  The emission was well
described at each epoch by a power law, however the photon index
clearly varied from 1.6 to 2.0, with the emission softest at
periastron.  \citet{ta97} considered the X-ray radiation mechanisms and
interaction geometry in detail, showing that a
synchrotron/inverse-Compton scattering model of emission of
electron/positron pairs accelerated at the inner shock front of the
pulsar cavity and adiabatically expanding in the MHD flow explains well
the observed time-variable X-ray flux and spectrum.  They conclude that
most likely the Be-star spin axis is misaligned with the orbital
angular momentum, and its mass outflow rate was constant over the
$\sim$2-yr period in question.  \citet{csmc95} argue that an unusual
Galactic X-ray source, LSI+61$^{\circ}$303, is a similar system to
PSR~B1259$-$63, but with the radio pulsations permanently eclipsed.

PSR~B1957+20 is a 1.6-ms recycled radio pulsar in a 9-hr binary system
with a low-mass companion \citep{fst88}.  The pulsar has $\dot{E}=
10^{35}$~erg~s$^{-1}$.  For $\sim$10\% of every orbit, the radio
pulsations are eclipsed by the wind from the companion, which is being
ablated and, eventually, evaporated, by the pulsar wind.  For this
reason, the pulsar is sometimes known as the ``Black Widow,'' as it
seems to be destroying the star that gave it new life as a recycled
pulsar.  The system is surrounded by an H$\alpha$ bow shock nebula,
oriented with apex in the direction of the pulsar's known proper
motion, a result of the interaction of the pulsar wind with the ambient
interstellar medium \citep{kh88}.  \citet{at94} argued that intra-binary
shock emission ought to be observable, because of the interaction of
the pulsar wind with material being ablated off the companion.  They
also argued that X-rays might originate from the bow shock.  Of course,
magnetospheric X-rays might be detectable from this pulsar as well.
The first X-ray detection of PSR~B1957+20 came from {\it ROSAT},
although only handful of photons were detected \citep{fbgb92,kpeh92},
and the conclusions that could be drawn were limited.  More recently,
{\it Chandra} observed the PSR~B1957+20 system (see Fig.~\ref{duck_fig}), 
and detected both a
point source as well as an extended X-ray tail \citep{sgk+03}.  The
tail emission is likely related to the bow shock nebula, and represents
the first proof that millisecond pulsars have relativistic winds like
their much younger counterparts.  The origin of the point-source
emission is less clear as the observation's time resolution was
insufficient to detect pulsations.  Planned {\it XMM-Newton}
observations should be able to decide with certainty whether the point
source is due to magnetospheric emission, or if it is from the
intra-binary shock, as in PSR~B1259$-$63.

\bigskip
The authors thank B. Gaensler and D. Yakovlev for discussions, for
careful reading of the manuscript and for many helpful comments.  We
also thank P. Jaikumar, D. Lai, M. Lyutikov, G. Pavlov, S. Ransom, R. Turolla, M. van Kerkwijk, and S. Zane for helpful
conversations, and M. Strickman and J. Dyks for help with figures.

%\bibliographystyle{apj}
%
%\bibliography{journals_apj,alice/alicerefs,mallory/malloryrefs,modrefs,psrrefs,crossrefs}

\begin{thebibliography}{382}
\expandafter\ifx\csname natexlab\endcsname\relax\def\natexlab#1{#1}\fi

\bibitem[{{Aller} \& {Reynolds}(1985{\natexlab{a}})}]{ar85a}
{Aller}, H.~D. \& {Reynolds}, S.~P. 1985{\natexlab{a}}, in The Crab Nebula and
  Related Supernova Remnants (Cambridge University Press), 75--78

\bibitem[{{Aller} \& {Reynolds}(1985{\natexlab{b}})}]{ar85}
{Aller}, H.~D. \& {Reynolds}, S.~P. 1985{\natexlab{b}}, ApJ, 293, L73

\bibitem[{Alpar {et~al.}(1984)Alpar, Anderson, Pines, \& Shaham}]{aaps84a}
Alpar, M.~A., Anderson, P.~W., Pines, D., \& Shaham, J. 1984, ApJ, 276, 325

\bibitem[{Arons(1981)}]{aro81}
Arons, J. 1981, ApJ, 248, 1099

\bibitem[{Arons(1983)}]{aro83b}
---. 1983, ApJ, 266, 215

\bibitem[{Arons \& Scharlemann(1979)}]{as79}
Arons, J. \& Scharlemann, E.~T. 1979, ApJ, 231, 854

\bibitem[{Arons \& Tavani(1994)}]{at94}
Arons, J. \& Tavani, M. 1994, ApJS, 90, 797

\bibitem[{Arzoumanian {et~al.}(2002)Arzoumanian, Chernoff, \& Cordes}]{acc02}
Arzoumanian, Z., Chernoff, D.~F., \& Cordes, J.~M. 2002, ApJ, 568, 289

\bibitem[{Aschenbach(1998)}]{asc98}
Aschenbach, B. 1998, Nature, 141

\bibitem[{{Atoyan}(1999)}]{ato99}
{Atoyan}, A.~M. 1999, A\&A, 346, L49

\bibitem[{Baade \& Zwicky(1934)}]{bz34}
Baade, W. \& Zwicky, F. 1934, Proc. Nat. Acad. Sci., 20, 254

\bibitem[{Bahcall \& Wolf(1965)}]{bw65}
Bahcall, J.~N. \& Wolf, R.~A. 1965, ApJ, 142, 1254

\bibitem[{{Bandiera} {et~al.}(2002){Bandiera}, {Neri}, \& {Cesaroni}}]{bnc02}
{Bandiera}, R., {Neri}, R., \& {Cesaroni}, R. 2002, A\&A, 386, 1044

\bibitem[{{Baring} \& {Harding}(2001)}]{bh01}
{Baring}, M.~G. \& {Harding}, A.~K. 2001, ApJ, 547, 929

\bibitem[{Becker \& Helfand(1984)}]{bh84}
Becker, R.~H. \& Helfand, D.~J. 1984, ApJ, 283, 154

\bibitem[{{Becker} \& {Helfand}(1985)}]{bh85a}
{Becker}, R.~H. \& {Helfand}, D.~J. 1985, ApJ, 297, L25

\bibitem[{Becker {et~al.}(1982)Becker, Helfand, \& Szymkowiak}]{bhs82}
Becker, R.~H., Helfand, D.~J., \& Szymkowiak, A.~E. 1982, ApJ, 255, 557

\bibitem[{Becker(2001)}]{bec01a}
Becker, W. 2001, in AIP Conf. Proc. 599: X-ray Astronomy: Stellar Endpoints,
  AGN, and the Diffuse X-ray Background, ed. N. White, G.~Malaguti, \& G.~G.
  Palumbo, 13--24

\bibitem[{Becker \& Aschenbach(2002)}]{ba02}
Becker, W. \& Aschenbach, B. 2002, in Neutron Stars, Pulsars, and Supernova
  Remnants, ed. W.~Becker, H.~Lesch, \& J.~Tr\"umper (Garching bei M\"unchen:
  Max-Plank-Institut f\"ur extraterrestrische Physik), 64

\bibitem[{Becker {et~al.}(1996)Becker, Brazier, \& Tr\"umper}]{bbt96}
Becker, W., Brazier, K. T.~S., \& Tr\"umper, J. 1996, A\&A, 306, 464

\bibitem[{Becker \& Pavlov(2002)}]{bp02}
Becker, W. \& Pavlov, G.~G. 2002, in The Century of Space Science, ed.
  J.~Bleeker, J.~Geiss, \& M.~Huber (Dordrecht: Kluwer)

\bibitem[{Becker \& Tr\"{umper}(1997)}]{bt97}
Becker, W. \& Tr\"{umper}, J. 1997, A\&A, 326, 682

\bibitem[{{Becker} \& {Tr\"{u}mper}(1999)}]{bt99}
{Becker}, W. \& {Tr\"{u}mper}, J. 1999, A\&A, 341, 803

\bibitem[{Becker {et~al.}(2004)Becker, Weisskopf, Tennant, Jessner, Dyks,
  Harding, \& Zhang}]{bwt+04}
Becker, W., Weisskopf, M.~C., Tennant, A.~F., Jessner, A., Dyks, J., Harding,
  A.~K., \& Zhang, S.~N. 2004, ApJ, submitted

\bibitem[{Bell {et~al.}(1995)Bell, Bailes, Manchester, Weisberg, \&
  Lyne}]{bbm+95}
Bell, J.~F., Bailes, M., Manchester, R.~N., Weisberg, J.~M., \& Lyne, A.~G.
  1995, ApJ, 440, L81

\bibitem[{Bietenholz {et~al.}(2001)Bietenholz, Frail, \& Hester}]{bfh01}
Bietenholz, M.~F., Frail, D.~A., \& Hester, J.~J. 2001, ApJ, 560, 254

\bibitem[{Bietenholz {et~al.}(1997)Bietenholz, Kassim, Frail, Perley, Erickson,
  \& Hajian}]{bkf+97}
Bietenholz, M.~F., Kassim, N., Frail, D.~A., Perley, R.~A., Erickson, W.~C., \&
  Hajian, A.~R. 1997, ApJ, 490, 291

\bibitem[{Bietenholz \& Kronberg(1990)}]{bk90a}
Bietenholz, M.~F. \& Kronberg, P.~P. 1990, ApJ, 357, L13

\bibitem[{Bignami {et~al.}(2003)Bignami, Caraveo, De~Luca, \&
  Mereghetti}]{bcdm03}
Bignami, G.~F., Caraveo, P.~A., De~Luca, A., \& Mereghetti, S. 2003, Nature,
  423, 725

\bibitem[{Blaes \& Madau(1993)}]{bm93}
Blaes, O. \& Madau, P. 1993, ApJ, 403, 690

\bibitem[{Blondin {et~al.}(2001)Blondin, Chevalier, \& Frierson}]{bcf01}
Blondin, J.~M., Chevalier, R.~A., \& Frierson, D.~M. 2001, ApJ, 563, 806

\bibitem[{{Bocchino} \& {Bandiera}(2003)}]{bb03}
{Bocchino}, F. \& {Bandiera}, R. 2003, A\&A, 398, 195

%\bibitem[{{Bogovalov} \& {Khangoulian}(2002)}]{bk02}
%{Bogovalov}, S.~V. \& {Khangoulian}, D.~V. 2002, MNRAS, 336, L53

\bibitem[{{Braje} \& {Romani}(2002)}]{br02}
{Braje}, T.~M. \& {Romani}, R.~W. 2002, ApJ, 580, 1043

\bibitem[{{Braje} {et~al.}(2002){Braje}, {Romani}, {Roberts}, \&
  {Kawai}}]{brrk02}
{Braje}, T.~M., {Romani}, R.~W., {Roberts}, M.~S.~E., \& {Kawai}, N. 2002, ApJ,
  565, L91

\bibitem[{Brinkmann \& \"Ogelman(1987)}]{bo87}
Brinkmann, W. \& \"Ogelman, H. 1987, A\&A, 182, 71

\bibitem[{{Brisken} {et~al.}(2003{\natexlab{a}}){Brisken}, {Fruchter}, {Goss},
  {Herrnstein}, \& {Thorsett}}]{bfg+03}
{Brisken}, W.~F., {Fruchter}, A., {Goss}, W., {Herrnstein}, R., \&
  {Thorsett}, S. 2003{\natexlab{a}}, AJ, 126, 3090

\bibitem[{{Brisken} {et~al.}(2003{\natexlab{b}}){Brisken}, {Thorsett},
  {Golden}, \& {Goss}}]{btgg03}
{Brisken}, W.~F., {Thorsett}, S.~E., {Golden}, A., \& {Goss}, W.~M.
  2003{\natexlab{b}}, ApJ, 593, L89

\bibitem[{{Bucciantini}(2002)}]{buc02}
{Bucciantini}, N. 2002, A\&A, 387, 1066

\bibitem[{{Bucciantini} {et~al.}(2003){Bucciantini}, {Blondin}, {Del Zanna}, \&
  {Amato}}]{bbda03}
{Bucciantini}, N., {Blondin}, J.~M., {Del Zanna}, L., \& {Amato}, E. 2003,
  A\&A, 405, 617

\bibitem[{{Burwitz} {et~al.}(2003){Burwitz}, {Haberl}, {Neuh{\" a}user},
  {Predehl}, {Tr{\" u}mper}, \& {Zavlin}}]{bhn+03}
{Burwitz}, V.,{Haberl},F.,{Neuh{\" a}user},R.,{Predehl},P.,{Tr{\"u}mper}, J., \& {Zavlin}, V. 2003, A\&A, 399, 1109

%\bibitem[{Camilo {et~al.}(2001)Camilo, Bell, Manchester, Lyne, Possenti,
%  Kramer, Kaspi, Stairs, D'Amico, Hobbs, Gotthelf, \& Gaensler}]{cbm+01}
%Camilo, F., Bell, J.~F., Manchester, R.~N., Lyne, A.~G., Possenti, A., Kramer,
%  M., Kaspi, V.~M., Stairs, I.~H., D'Amico, N., Hobbs, G., Gotthelf, E.~V., \&
%  Gaensler, B.~M. 2001, ApJ, 557, L51

\bibitem[{Camilo {et~al.}(2001)}]{cbm+01}
Camilo, F. et al. 2001, ApJ, 557, L51

\bibitem[{Camilo {et~al.}(2000)Camilo, Kaspi, Lyne, Manchester, Bell, D'Amico,
  McKay, \& Crawford}]{ckl+00}
Camilo, F., Kaspi, V.~M., Lyne, A.~G., Manchester, R.~N., Bell, J.~F., D'Amico,
  N., McKay, N. P.~F., \& Crawford, F. 2000, ApJ, 541, 367

\bibitem[{Camilo {et~al.}(2002{\natexlab{a}})Camilo, Lorimer, Bhat, Gotthelf,
  Halpern, Wang, Lu, \& Mirabal}]{clb+02}
Camilo, F., Lorimer, D.~R., Bhat, N. D.~R., Gotthelf, E.~V., Halpern, J.~P.,
  Wang, Q.~D., Lu, F.~J., \& Mirabal, N. 2002{\natexlab{a}}, ApJ, 574, L71

\bibitem[{Camilo {et~al.}(2002{\natexlab{b}})Camilo, Manchester, Gaensler,
  Lorimer, \& Sarkissian}]{cmg+02}
Camilo, F., Manchester, R., Gaensler, B., Lorimer, D., \& Sarkissian, J. 
2002{\natexlab{b}}, ApJ, 567, L71

%\bibitem[{Camilo {et~al.}(2002{\natexlab{c}})Camilo, {Stairs}, {Lorimer},
%  {Backer}, {Ransom}, {Klein}, {Wielebinski}, {Kramer}, {McLaughlin},
%  {Arzoumanian}, \& {M{\" u}ller}}]{csl+02}
%Camilo, F., {Stairs}, I.~H., {Lorimer}, D.~R., {Backer}, D.~C., {Ransom},
%  S.~M., {Klein}, B., {Wielebinski}, R., {Kramer}, M., {McLaughlin}, M.~A.,
%  {Arzoumanian}, Z., \& {M{\" u}ller}, P. 2002{\natexlab{c}}, ApJ, 571, L41

\bibitem[{Camilo {et~al.}(2002{\natexlab{c}})}]{csl+02}
Camilo, F. et al. 2002{\natexlab{c}}, ApJ, 571, L41

\bibitem[{Campana {et~al.}(1995)Campana, Stella, Mereghetti, \& Colpi}]{csmc95}
Campana, S., Stella, L., Mereghetti, S., \& Colpi, M. 1995, A\&A, 297, 385

\bibitem[{{Caraveo} {et~al.}(2003){Caraveo}, {Bignami}, {DeLuca}, {Mereghetti},
  {Pellizzoni}, {Mignani}, {Tur}, \& {Becker}}]{cbd+03}
{Caraveo}, P.~A., {Bignami}, G.~F., {DeLuca}, A., {Mereghetti}, S.,
  {Pellizzoni}, A., {Mignani}, R., {Tur}, A., \& {Becker}, W. 2003, Science,
  301, 1345

\bibitem[{Caraveo \& Mignani(1999)}]{cm99}
Caraveo, P.~A. \& Mignani, R.~P. 1999, A\&A, 344, 366

\bibitem[{{Chabrier} {et~al.}(1997){Chabrier}, {Potekhin}, \&
  {Yakovlev}}]{cpy97}
{Chabrier}, G., {Potekhin}, A.~Y., \& {Yakovlev}, D.~G. 1997, ApJ, 477, L99

\bibitem[{{Chakrabarty} {et~al.}(2001){Chakrabarty}, {Pivovaroff}, {Hernquist},
  {Heyl}, \& {Narayan}}]{cph+01}
{Chakrabarty}, D., {Pivovaroff}, M., {Hernquist}, L., {Heyl}, J., \&
  {Narayan}, R. 2001, ApJ, 548, 800

\bibitem[{Chanan {et~al.}(1984)Chanan, Helfand, \& Reynolds}]{chr84}
Chanan, G.~A., Helfand, D.~J., \& Reynolds, S.~P. 1984, ApJ, 287, L23

\bibitem[{Chen \& Ruderman(1993)}]{cr93}
Chen, K. \& Ruderman, M. 1993, ApJ, 408, 179

\bibitem[{Cheng {et~al.}(1986)Cheng, Ho, \& Ruderman}]{chr86a}
Cheng, K.~S., Ho, C., \& Ruderman, M. 1986, ApJ, 300, 500

\bibitem[{Cheng {et~al.}(2000)Cheng, Ruderman, \& Zhang}]{crz00}
Cheng, K.~S., Ruderman, M.~A., \& Zhang, L. 2000, ApJ, 537, 964

\bibitem[{Cheng \& Zhang(1999)}]{cz99}
Cheng, K.~S. \& Zhang, L. 1999, ApJ, 515, 337

\bibitem[{Chevalier(1982)}]{che82}
Chevalier, R.~A. 1982, ApJ, 259, 302

\bibitem[{{Chevalier}(2000)}]{che00}
{Chevalier}, R.~A. 2000, ApJ, 539, L45

\bibitem[{Chiu \& Salpeter(1964)}]{cs64}
Chiu, H.~Y. \& Salpeter, E.~E. 1964, Phys. Rev. Lett., 12, 413

\bibitem[{Clark \& Stephenson(1977)}]{cs77}
Clark, D. \& Stephenson, F. 1977, MNRAS, 179, 87P

\bibitem[{Clark \& Caswell(1976)}]{cc76}
Clark, D.~H. \& Caswell, J.~L. 1976, MNRAS, 174, 267

\bibitem[{Comella {et~al.}(1969)Comella, Craft, Lovelace, Sutton, \&
  Tyler}]{ccl+69}
Comella, J.~M., Craft, H., Lovelace, R., Sutton, J., \& Tyler,
  G. 1969, Nature, 221, 453

\bibitem[{Cominsky {et~al.}(1994)Cominsky, Roberts, \& Johnston}]{crj94}
Cominsky, L., Roberts, M., \& Johnston, S. 1994, ApJ, 427, 978

\bibitem[{Contopoulos {et~al.}(1999)Contopoulos, Kazanas, \& Fendt}]{ckf99}
Contopoulos, J., Kazanas, D., \& Fendt, C. 1999, ApJ, 511, 351

\bibitem[{Crawford {et~al.}(1998)Crawford, Kaspi, Manchester, Lyne, Camilo, \&
  D'Amico}]{ckm+98}
Crawford, F., Kaspi, V.~M., Manchester, R.~N., Lyne, A.~G., Camilo, F., \&
  D'Amico, N. 1998, in Proc. of the Elba Workshop: Neutron Stars and
  Supernova Remnants, Vol.~69 (Memorie della Societa' Astronomica Italiana),
  951--954

\bibitem[{Crawford \& Keim(2003)}]{ck03}
Crawford, F. \& Keim, N.~C. 2003, ApJ, 590, 1020

\bibitem[{{Crawford} {et~al.}(2001){Crawford}, {Manchester}, \&
  {Kaspi}}]{cmk01}
{Crawford}, F., {Manchester}, R.~N., \& {Kaspi}, V.~M. 2001, AJ, 122, 2001

\bibitem[{Cropper {et~al.}(2004)Cropper, Haberl, Zane, \& Zavlin}]{chzz04}
Cropper, M., Haberl, F., Zane, S., \& Zavlin, V. 2004, MNRAS, 351, 1099

\bibitem[{{Danner}(1998{\natexlab{a}})}]{dan98a}
{Danner}, R. 1998{\natexlab{a}}, A\&AS, 128, 331

\bibitem[{{Danner}(1998{\natexlab{b}})}]{dan98b}
---. 1998{\natexlab{b}}, A\&AS, 128, 349

\bibitem[{Daugherty \& Harding(1982)}]{dh82}
Daugherty, J.~K. \& Harding, A.~K. 1982, ApJ, 252, 337

\bibitem[{Daugherty \& Harding(1996)}]{dh96}
---. 1996, ApJ, 458, 278

\bibitem[{{de Jager} {et~al.}(1996{\natexlab{a}}){de Jager}, {Harding},
  {Michelson}, {Nel}, {Nolan}, {Sreekumar}, \& {Thompson}}]{dhm+96}
{de Jager}, O.~C., {Harding}, A.~K., {Michelson}, P.~F., {Nel}, H.~I., {Nolan},
  P.~L., {Sreekumar}, P., \& {Thompson}, D.~J. 1996{\natexlab{a}}, ApJ, 457,
  253

\bibitem[{{de Jager} {et~al.}(1996{\natexlab{b}}){de Jager}, {Harding},
  {Sreekumar}, \& {Strickman}}]{dhss96}
{de Jager}, O.~C., {Harding}, A.~K., {Sreekumar}, P., \& {Strickman}, M.
  1996{\natexlab{b}}, Astronomy \& Astrophysics Supplements, 120, C441+

\bibitem[{De~Luca {et~al.}(2004)De~Luca, Mereghetti, Caraveo, Moroni, Mignani,
  \& Bignami}]{dmc+04}
De~Luca, A., Mereghetti, S., Caraveo, P.~A., Moroni, M., Mignani, R.~P., \&
  Bignami, G.~F. 2004, A\&A, 418, 625

\bibitem[{{Dedeo} \& {Psaltis}(2003)}]{dp03}
{Dedeo}, S. \& {Psaltis}, D. 2003, Phys. Rev. Lett., 90, 141101

\bibitem[{Deeter {et~al.}(1999)Deeter, Nagase, \& Boynton}]{dnb99}
Deeter, J.~E., Nagase, F., \& Boynton, P.~E. 1999, ApJ, 512, 300

\bibitem[{Dickel {et~al.}(2000)Dickel, Milne, \& Strom}]{dms00}
Dickel, J.~R., Milne, D.~K., \& Strom, R.~G. 2000, ApJ, 543, 840

\bibitem[{{Dodson} {et~al.}(2003{\natexlab{a}}){Dodson}, {Legge}, {Reynolds},
  \& {McCulloch}}]{dlrm03}
{Dodson}, R., {Legge}, D., {Reynolds}, J.~E., \& {McCulloch}, P.~M.
  2003{\natexlab{a}}, ApJ, 596, 1137

\bibitem[{{Dodson} {et~al.}(2003{\natexlab{b}}){Dodson}, {Lewis}, {McConnell},
  \& {Deshpande}}]{dlmd03}
{Dodson}, R., {Lewis}, D., {McConnell}, D., \& {Deshpande}, A.~A.
  2003{\natexlab{b}}, MNRAS, 343, 116

%\bibitem[{{Drake} {et~al.}(2002){Drake}, {Marshall}, {Dreizler}, {Freeman},
%  {Fruscione}, {Juda}, {Kashyap}, {Nicastro}, {Pease}, {Wargelin}, \&
%  {Werner}}]{dmd+02}
%{Drake}, J.~J., {Marshall}, H.~L., {Dreizler}, S., {Freeman}, P.~E.,
%  {Fruscione}, A., {Juda}, M., {Kashyap}, V., {Nicastro}, F., {Pease}, D.~O.,
%  {Wargelin}, B.~J., \& {Werner}, K. 2002, ApJ, 572, 996

\bibitem[{Drake {et al.}(2002)}]{dmd+02}
Drake, J. J. et al. 2002, ApJ, 572, 996

\bibitem[{Dyks {et~al.}(2004)Dyks, Harding, \& Rudak}]{dhr03}
Dyks, J., Harding, A.~K., \& Rudak, B. 2004, ApJ, 606, 1125

\bibitem[{Dyks \& Rudak(2003)}]{dr03}
Dyks, J. \& Rudak, B. 2003, ApJ, 598, 1201

\bibitem[{{Endal} \& {Sofia}(1978)}]{es78}
{Endal}, A.~S. \& {Sofia}, S. 1978, ApJ, 220, 279

\bibitem[{{Fesen} \& {Kirshner}(1982)}]{fk82}
{Fesen}, R.~A. \& {Kirshner}, R.~P. 1982, ApJ, 258, 1

\bibitem[{Fierro {et~al.}(1998)Fierro, Michelson, Nolan, \& Thompson}]{fmnt98}
Fierro, j.~M., Michelson, P.~F., Nolan, P.~L., \& Thompson, D.~J. 1998, ApJ,
  494, 734

\bibitem[{{Finley} {et~al.}(1992){Finley}, {Ogelman}, \& {Kiziloglu}}]{fok92}
{Finley}, J.~P., {Ogelman}, H., \& {Kiziloglu}, U. 1992, ApJ, 394, L21

\bibitem[{Finzi \& Wolf(1969)}]{fw69}
Finzi, A. \& Wolf, R.~A. 1969, ApJ, 155, 107

\bibitem[{Flowers {et~al.}(1976)Flowers, Ruderman, \& Sutherland}]{frs76}
Flowers, E.~G., Ruderman, M., \& Sutherland, P.~G. 1976, ApJ, 205, 241

\bibitem[{Frail {et~al.}(1995)Frail, Kassim, Cornwell, \& Goss}]{fkcg95}
Frail, D.~A., Kassim, N.~E., Cornwell, T.~J., \& Goss, W.~M. 1995, ApJ, 454,
  L129

\bibitem[{Frail \& Scharringhausen(1997)}]{fs97}
Frail, D.~A. \& Scharringhausen, B.~R. 1997, ApJ, 480, 364

\bibitem[{Freire {et~al.}(2001)Freire, Kramer, Lyne, Camilo, Manchester, \&
  D'Amico}]{fkl+01}
Freire, P.~C., Kramer,M.,Lyne,A.,Camilo,F.,Manchester,R., \&
  D'Amico, N. 2001, ApJ, 557, L105

\bibitem[{Fruchter {et~al.}(1992)Fruchter, Bookbinder, Garcia, \&
  Bailyn}]{fbgb92}
Fruchter, A.~S., Bookbinder, J., Garcia, M.~R., \& Bailyn, C.~D. 1992, Nature,
  359, 303

\bibitem[{Fruchter {et~al.}(1988)Fruchter, Stinebring, \& Taylor}]{fst88}
Fruchter, A.~S., Stinebring, D.~R., \& Taylor, J.~H. 1988, Nature, 333, 237

\bibitem[{Gamow \& Scheonberg (1941)}]{gs41}
Gamow, G. \& Schoenberg, S. 1941, Phys. Rev., 59, 539

\bibitem[{Gaensler {et~al.}(2002)Gaensler, Arons, Kaspi, Pivovaroff, Kawai, \&
  Tamura}]{gak+02}
Gaensler, B.~M., Arons,J.,Kaspi,V.,Pivovaroff,M.,Kawai,N., \&
  Tamura, K. 2002, ApJ, 569, 878

\bibitem[{Gaensler {et~al.}(2000{\natexlab{a}})Gaensler, Bock, \&
  Stappers}]{gbs00}
Gaensler, B.~M., Bock, D. C.-J., \& Stappers, B.~W. 2000{\natexlab{a}}, ApJ,
  537, L35

\bibitem[{{Gaensler} \& {Frail}(2000)}]{gf00}
{Gaensler}, B.~M. \& {Frail}, D.~A. 2000, Nature, 406, 158

\bibitem[{{Gaensler} {et~al.}(2003{\natexlab{a}}){Gaensler}, {Hendrick},
  {Reynolds}, \& {Borkowski}}]{ghrb03}
{Gaensler}, B.~M., {Hendrick}, S.~P., {Reynolds}, S.~P., \& {Borkowski}, K.~J.
  2003{\natexlab{a}}, ApJ, 594, L111

\bibitem[{{Gaensler} {et~al.}(2002){Gaensler}, {Jones}, \& {Stappers}}]{gjs02}
{Gaensler}, B.~M., {Jones}, D.~H., \& {Stappers}, B.~W. 2002, ApJ, 580, L137

\bibitem[{{Gaensler} {et~al.}(2003{\natexlab{b}}){Gaensler}, {Schulz}, {Kaspi},
  {Pivovaroff}, \& {Becker}}]{gsk+03}
{Gaensler}, B.~M., {Schulz}, N., {Kaspi}, V., {Pivovaroff}, M., \&
  {Becker}, W. 2003{\natexlab{b}}, ApJ, 588, 441

\bibitem[{Gaensler {et~al.}(1998)Gaensler, Stappers, Frail, \&
  Johnston}]{gsfj98}
Gaensler, B.~M., Stappers, B.~W., Frail, D.~A., \& Johnston, S. 1998, ApJ, 499,
  L69

\bibitem[{Gaensler {et~al.}(2000{\natexlab{b}})Gaensler, Stappers, Frail,
  Moffett, Johnston, \& Chatterjee}]{gsf+00}
Gaensler, B.~M., Stappers, B.~W., Frail, D.~A., Moffett, D.~A., Johnston, S.,
  \& Chatterjee, S. 2000{\natexlab{b}}, MNRAS, 318, 58

\bibitem[{{Gaensler} {et~al.}(2003{\natexlab{c}}){Gaensler}, {van der Swaluw},
  {Camilo}, {Kaspi}, {Baganoff}, {Yusef-Zadeh}, \& {Manchester}}]{gvc+03}
{Gaensler}, B.~M., {van der Swaluw}, E., {Camilo}, F., {Kaspi}, V.~M.,
  {Baganoff}, F.~K., {Yusef-Zadeh}, F., \& {Manchester}, R.~N.
  2003{\natexlab{c}}, ApJ, in press (astro-ph/0312362)

\bibitem[{{Gaensler} \& {Wallace}(2003)}]{gw03}
{Gaensler}, B.~M. \& {Wallace}, B.~J. 2003, ApJ, 594, 326

\bibitem[{Gallant \& Arons(1994)}]{ga94}
Gallant, Y.~A. \& Arons, J. 1994, ApJ, 435, 230

\bibitem[{{Garmire} {et~al.}(2000){Garmire}, {Pavlov}, {Garmire}, \&
  {Zavlin}}]{gpgz00}
{Garmire}, G.~P., {Pavlov}, G.~G., {Garmire}, A.~B., \& {Zavlin}, V.~E.
\newblock IAU circular 7350

\bibitem[{{Giacani} {et~al.}(2001){Giacani}, {Frail}, {Goss}, \&
  {Vieytes}}]{gfgv01}
{Giacani}, E.~B., {Frail}, D.~A., {Goss}, W.~M., \& {Vieytes}, M. 2001, AJ,
  121, 3133

\bibitem[{Goldreich \& Julian(1969)}]{gj69}
Goldreich, P. \& Julian, W.~H. 1969, ApJ, 157, 869

\bibitem[{{Gonzalez} \& {Safi-Harb}(2003)}]{gs03}
{Gonzalez}, M. \& {Safi-Harb}, S. 2003, ApJ, 591, L143

\bibitem[{{Gotthelf}(2003)}]{got03}
{Gotthelf}, E.~V. 2003, ApJ, 591, 361

\bibitem[{{Gotthelf} {et~al.}(2002){Gotthelf}, {Halpern}, \& {Dodson}}]{ghd02}
{Gotthelf}, E.~V., {Halpern}, J.~P., \& {Dodson}, R. 2002, ApJ, 567, L125

\bibitem[{Gotthelf {et~al.}(1999)Gotthelf, Petre, \& Vasisht}]{gpv99}
Gotthelf, E.~V., Petre, R., \& Vasisht, G. 1999, ApJ, 514, L107

\bibitem[{Gotthelf {et~al.}(2000)Gotthelf, Vasisht, Boylan-Kolchin, \&
  Torii}]{gvbt00}
Gotthelf, E.~V., Vasisht, G., Boylan-Kolchin, M., \& Torii, K. 2000, ApJ, 542,
  L37

\bibitem[{Gotthelf \& Wang(2000)}]{gw00}
Gotthelf, E.~V. \& Wang, Q.~D. 2000, ApJ, 532, L117

\bibitem[{{Green} {et~al.}(1988){Green}, {Gull}, {Tan}, \& {Simon}}]{ggts88}
{Green}, D.~A., {Gull}, S.~F., {Tan}, S.~M., \& {Simon}, A. J.~B. 1988, MNRAS,
  231, 735

\bibitem[{Green \& Scheuer(1992)}]{gs92}
Green, D.~A. \& Scheuer, P. A.~G. 1992, MNRAS, 258, 833

%\bibitem[{Greiveldinger {et~al.}(1996)Greiveldinger, Camerini, W., Markwardt,
%  \"{O}gelman, Safi-Harb, Finley, Tsuruta, Shibata, Sugawara, Sano, \&
%  Tukahara}]{gcf+96}
%Greiveldinger, C., Camerini, U., W., F., Markwardt, C.~B., \"{O}gelman, H.,
%  Safi-Harb, S., Finley, J.~P., Tsuruta, S., Shibata, S., Sugawara, T., Sano,
%  S., \& Tukahara, M. 1996, ApJ, 465, L35

\bibitem[{Greiveldinger {et al.}(1996)}]{gcf+96}
Greiveldinger, C. et al. 1996, ApJ, 465, L35

\bibitem[{Grindlay {et~al.}(2002)Grindlay, Camilo, Heinke, Edmonds, Cohn, \&
  Lugger}]{gch+02}
Grindlay, J.~E., Camilo,F.,Heinke,C.,Edmonds,P.,Cohn, H., \&
  Lugger, P. 2002, ApJ, 581, 470

\bibitem[{{Gudmundsson} {et~al.}(1983){Gudmundsson}, {Pethick}, \&
  {Epstein}}]{gpe83}
{Gudmundsson}, E.~H., {Pethick}, C.~J., \& {Epstein}, R.~I. 1983, ApJ, 272, 286

\bibitem[{Haberl(2003)}]{hab03}
Haberl, F. 2003, in High Energy Studies of Supernova Remnants and Neutron
  Stars, ed. W.~Becker \& W.~Hermsen, in press (astro-ph/0302540)

\bibitem[{Haberl(2004)}]{hab04}
Haberl, F. 2004, in Memorie della Societa Astronomica Italiana

\bibitem[{{Haberl} {et~al.}(1997){Haberl}, {Motch}, {Buckley}, {Zickgraf}, \&
  {Pietsch}}]{hmb+97}
{Haberl}, F., {Motch}, C., {Buckley}, D.~A.~H., {Zickgraf}, F.-J., \&
  {Pietsch}, W. 1997, A\&A, 326, 662

\bibitem[{{Haberl} {et~al.}(1998){Haberl}, {Motch}, \& {Pietsch}}]{hmp98}
{Haberl}, F., {Motch}, C., \& {Pietsch}, W. 1998, Astronomische Nachrichten,
  319, 97

\bibitem[{{Haberl} {et~al.}(1999){Haberl}, {Pietsch}, \& {Motch}}]{hpm99}
{Haberl}, F., {Pietsch}, W., \& {Motch}, C. 1999, A\&A, 351, L53

\bibitem[{Haberl {et~al.}(2003)Haberl, Schwope, Hambaryan, Hasinger, \&
  Motch}]{hsh+03}
Haberl, F., Schwope, A.~D., Hambaryan, V., Hasinger, G., \& Motch, C. 2003,
  A\&A, 403, L19

\bibitem[{{Haberl} \& {Zavlin}(2002)}]{hz02}
{Haberl}, F. \& {Zavlin}, V.~E. 2002, A\&A, 391, 571

\bibitem[{Haberl {et~al.}(2004)Haberl, Zavlin, Tr\"uper, \& Burwitz}]{hztb04}
Haberl, F., Zavlin, V.~E., Tr\"uper, J., \& Burwitz, V. 2004, A\&A, 419, 1077

\bibitem[{{Hailey} \& {Mori}(2002)}]{hm02a}
{Hailey}, C.~J. \& {Mori}, K. 2002, ApJ, 578, L133

\bibitem[{{Halpern} {et~al.}(2001){Halpern}, {Camilo}, {Gotthelf}, {Helfand},
  {Kramer}, {Lyne}, {Leighly}, \& {Eracleous}}]{hcg+01}
{Halpern}, J.~P., {Camilo}, F., {Gotthelf}, E.~V., {Helfand}, D.~J., {Kramer},
  M., {Lyne}, A.~G., {Leighly}, K.~M., \& {Eracleous}, M. 2001, ApJ, 552, L125

\bibitem[{Halpern \& Ruderman(1993)}]{hr93}
Halpern, J.~P. \& Ruderman, M. 1993, ApJ, 415, 286

\bibitem[{Halpern \& Wang(1997)}]{hw97}
Halpern, J.~P. \& Wang, F. Y.-H. 1997, ApJ, 477, 905

\bibitem[{Hansen \& Phinney(1997)}]{hp97}
Hansen, B. \& Phinney, E.~S. 1997, MNRAS, 291, 569

\bibitem[{Harding \& Muslimov(2001)}]{hm01}
Harding, A.~K. \& Muslimov, A.~G. 2001, ApJ, 556, 987

\bibitem[{Harding \& Muslimov(2002)}]{hm02}
---. 2002, ApJ, 568, 862

\bibitem[{Harding {et~al.}(2002{\natexlab{a}})Harding, Muslimov, \&
  Zhang}]{hmz02}
Harding, A.~K., Muslimov, A.~G., \& Zhang, B. 2002{\natexlab{a}}, ApJ, 576, 366

\bibitem[{Harding {et~al.}(2002{\natexlab{b}})Harding, Strickman, Gwinn,
  Dodson, Moffet, \& McCulloch}]{hsg+02}
Harding, A.~K., Strickman, M.~S., Gwinn, C., Dodson, R., Moffet, D., \&
  McCulloch, P. 2002{\natexlab{b}}, ApJ, 576, 376

\bibitem[{Harnden {et~al.}(1985)Harnden, Grant, Seward, \& Kahn}]{hgsk85}
Harnden, F.~R., Grant, P.~D., Seward, F.~D., \& Kahn, S.~M. 1985, ApJ, 299, 828

\bibitem[{{Harrus} {et~al.}(2003){Harrus}, {Slane}, {Gaensler}, {Hughes},
  {Moffett}, \& {Dodson}}]{hsg+03}
{Harrus}, I., {Slane}, P., {Gaensler}, B., {Hughes}, J., {Moffett}, D., \&
  {Dodson}, R. 2003, in IAU Symposium 218, Young Neutron Stars and their
  Environment, ed. F.~Camilo \& B.~Gaensler (ASP)

%\bibitem[{Hartman {et~al.}(1999)Hartman, Bertsch, Bloom, Chen, Deines-Jones,
%  Esposito, Fichtel, Friedlander, Hunter, McDonald, Sreekumar, Thompson, Jones,
%  Lin, Michelson, Nolan, Tompkins, Kanbach, Mayer-Hasselwander, M\"ucke, Pohl,
%  Reimer, Kniffen, Schneid, von Montigny, Mukherjee, \& Dingus}]{hbb+99}
%Hartman, R.~C., Bertsch, D.~L., Bloom, S.~D., Chen, A.~W., Deines-Jones, P.,
%  Esposito, J.~A., Fichtel, C.~E., Friedlander, D.~P., Hunter, S.~D., McDonald,
%  L.~M., Sreekumar, P., Thompson, D.~J., Jones, B.~B., Lin, Y.~C., Michelson,
%  P.~F., Nolan, P.~L., Tompkins, W.~F., Kanbach, G., Mayer-Hasselwander, H.~A.,
%  M\"ucke, A., Pohl, M., Reimer, O., Kniffen, D.~A., Schneid, E.~J., von
%  Montigny, C., Mukherjee, R., \& Dingus, B.~L. 1999, ApJS, 123, 79

\bibitem[{Hartman {et~al.}(1999)}]{hbb+99}
Hartman, R.~C. et al. 1999, ApJS, 123, 79

\bibitem[{{Helfand} {et~al.}(2003{\natexlab{a}}){Helfand}, {Ag{\" u}eros}, \&
  {Gotthelf}}]{hag03}
{Helfand}, D.~J., {Ag{\" u}eros}, M.~A., \& {Gotthelf}, E.~V.
  2003{\natexlab{a}}, ApJ, 592, 941

\bibitem[{Helfand \& Becker(1984)}]{hb84}
Helfand, D.~J. \& Becker, R.~H. 1984, Nature, 307, 215

\bibitem[{Helfand \& Becker(1987)}]{hb87}
---. 1987, ApJ, 314, 203

\bibitem[{{Helfand} {et~al.}(1980){Helfand}, {Chanan}, \& {Novick}}]{hcn80}
{Helfand}, D.~J., {Chanan}, G.~A., \& {Novick}, R. 1980, Nature, 283, 337

\bibitem[{{Helfand} {et~al.}(2003{\natexlab{b}}){Helfand}, {Collins}, \&
  {Gotthelf}}]{hcg03}
{Helfand}, D.~J., {Collins}, B.~F., \& {Gotthelf}, E.~V. 2003{\natexlab{b}},
  ApJ, 582, 783

\bibitem[{{Helfand} {et~al.}(2001){Helfand}, {Gotthelf}, \& {Halpern}}]{hgh01}
{Helfand}, D.~J., {Gotthelf}, E.~V., \& {Halpern}, J.~P. 2001, ApJ, 556, 380

\bibitem[{{Hessels} {et~al.}(2003){Hessels}, {Roberts}, {Ransom}, {Kaspi},
  {Tam}, \& {Freire}}]{hrr+03}
{Hessels}, J.~W., {Roberts}, M.~S., {Ransom}, S.~M., {Kaspi}, V.~M., {Tam}, C.,
  \& {Freire}, P.~C. 2003, in IAU Symposium 218, Young Neutron Stars and their
  Environment, ed. F.~Camilo \& B.~Gaensler (ASP)

\bibitem[{{Hester} {et~al.}(2002){Hester}, {Mori}, {Burrows}, {Gallagher},
  {Graham}, {Halverson}, {Kader}, {Michel}, \& {Scowen}}]{hmb+02}
{Hester}, J.~J., {Mori}, K., {Burrows}, D., {Gallagher}, J.~S., {Graham},
  J.~R., {Halverson}, M., {Kader}, A., {Michel}, F.~C., \& {Scowen}, P. 2002,
  ApJ, 577, L49

%\bibitem[{{Hester} {et~al.}(1996){Hester}, {Stone}, {Scowen}, {Jun},
%  {Gallagher}, {Norman}, {Ballester}, {Burrows}, {Casertano}, {Clarke},
%  {Crisp}, {Griffiths}, {Hoessel}, {Holtzman}, {Krist}, {Mould}, {Sankrit},
%  {Stapelfeldt}, {Trauger}, {Watson}, \& {Westphal}}]{hss96b}
%{Hester}, J.~J., {Stone}, J.~M., {Scowen}, P.~A., {Jun}, B., {Gallagher},
%  J.~S., {Norman}, M.~L., {Ballester}, G.~E., {Burrows}, C.~J., {Casertano},
%  S., {Clarke}, J.~T., {Crisp}, D., {Griffiths}, R.~E., {Hoessel}, J.~G.,
%  {Holtzman}, J.~A., {Krist}, J., {Mould}, J.~R., {Sankrit}, R., {Stapelfeldt},
%  K.~R., {Trauger}, J.~T., {Watson}, A., \& {Westphal}, J.~A. 1996, ApJ, 456,
% 225

\bibitem[{{Hester} {et~al.}(1996)}]{hss96b}
Hester, J. J. et al. 1996, ApJ, 456, 225


\bibitem[{{Heyl} \& {Hernquist}(1997)}]{hh97a}
{Heyl}, J.~S. \& {Hernquist}, L. 1997, ApJ, 491, L95

\bibitem[{{Heyl} \& {Hernquist}(2001)}]{hh01}
---. 2001, MNRAS, 324, 292

\bibitem[{{Heyl} \& {Shaviv}(2002)}]{hs02}
{Heyl}, J.~S. \& {Shaviv}, N.~J. 2002, Phys. Rev. D, 66, 23002

\bibitem[{Heyl {et~al.}(2003)Heyl, Shaviv, \& Lloyd}]{hsl03}
Heyl, J.~S., Shaviv, N.~J., \& Lloyd, D. 2003, MNRAS, 342, 134

\bibitem[{{Hibschman} \& {Arons}(2001)}]{ha01}
{Hibschman}, J.~A. \& {Arons}, J. 2001, ApJ, 546, 382

\bibitem[{{Hirayama} {et~al.}(1999){Hirayama}, {Cominsky}, {Kaspi}, {Nagase},
  {Tavani}, {Kawai}, \& {Grove}}]{hck+99}
{Hirayama}, M., {Cominsky}, L.~R., {Kaspi}, V.~M., {Nagase}, F., {Tavani}, M.,
  {Kawai}, N., \& {Grove}, J.~E. 1999, ApJ, 521, 718

\bibitem[{Hirayama {et~al.}(1996)Hirayama, Nagase, Tavani, Kaspi, Kawia, \&
  Arons}]{hnt+96}
Hirayama, M., Nagase, F., Tavani, M., Kaspi, V.~M., Kawia, N., \& Arons, J.
  1996, PASJ, 48, 833

\bibitem[{Hirotani {et~al.}(2003)Hirotani, Harding, \& Shibata}]{hhs03}
Hirotani, K., Harding, A.~K., \& Shibata, S. 2003, ApJ, 591, 334

\bibitem[{Hirotani \& Shibata(2001)}]{hs01}
Hirotani, K. \& Shibata, S. 2001, MNRAS, 325, 1228

\bibitem[{Ho \& Lai(2001)}]{hl01a}
Ho, W. C.~G. \& Lai, D. 2001, MNRAS, 327, 1081

\bibitem[{{Ho} \& {Lai}(2003)}]{hl03}
{Ho}, W.~C.~G. \& {Lai}, D. 2003, MNRAS, 338, 233

\bibitem[{Ho \& Lai(2004)}]{hl04}
Ho, W. C.~G. \& Lai, D. 2004, ApJ, 607, 420

\bibitem[{{Ho} {et~al.}(2003){Ho}, {Lai}, {Potekhin}, \& {Chabrier}}]{hlpc03}
{Ho}, W.~C.~G., {Lai}, D., {Potekhin}, A.~Y., \& {Chabrier}, G. 2003, ApJ, 599,
  1293

\bibitem[{Hughes {et~al.}(2001)Hughes, Slane, Burrows, Garmire, Nousek, Olbert,
  \& Keohane}]{hsb+01}
Hughes, J.~P., Slane, P.~O., Burrows, D.~N., Garmire, G., Nousek, J.~A.,
  Olbert, C.~M., \& Keohane, J.~W. 2001, ApJ, 559, L153

\bibitem[{Hughes {et~al.}(2000)Hughes, Slane, \& Plucinsky}]{hsp00}
Hughes, J.~P., Slane, P.~O., \& Plucinsky, P. 2000, ApJ, 542, 386

\bibitem[{Johnston {et~al.}(1992)Johnston, Manchester, Lyne, Bailes, Kaspi,
  Qiao, \& D'Amico}]{jml+92}
Johnston, S., Manchester, R.~N., Lyne, A.~G., Bailes, M., Kaspi, V.~M., Qiao,
  G., \& D'Amico, N. 1992, ApJ, 387, L37

\bibitem[{Jones {et~al.}(2002)Jones, Stappers, \& Gaensler}]{jsg02}
Jones, D.~H., Stappers, B.~W., \& Gaensler, B.~M. 2002, A\&A, 389, L1

\bibitem[{Kaminker {et~al.}(2002)Kaminker, Yakovlev, \& Gnedin}]{kyg02}
Kaminker, A.~D., Yakovlev, D.~G., \& Gnedin, O.~Y. 2002, A\&A, 383, 1076

\bibitem[{Kanbach(2002)}]{kan02}
Kanbach, G. 2002, in Neutron Stars, Pulsars, and Supernova Remnants, ed.
  W.~Becker, H.~Lesch, \& J.~Tr\"umper (Garching bei M\"unchen:
  Max-Plank-Institut f\"ur extraterrestrische Physik), 91

\bibitem[{{Kanno}(1975)}]{kan75}
{Kanno}, S. 1975, PASJ, 27, 287

\bibitem[{Kaplan {et~al.}(2001)Kaplan, Kulkarni, \& Murray}]{kkm01}
Kaplan, D.~L., Kulkarni, S.~R., \& Murray, S.~S. 2001, ApJ, 558, 270

\bibitem[{{Kaplan} {et~al.}(2002){Kaplan}, {Kulkarni}, \& {van
  Kerkwijk}}]{kkv02}
{Kaplan}, D.~L., {Kulkarni}, S.~R., \& {van Kerkwijk}, M.~H. 2002, ApJ, 579,
  L29

\bibitem[{{Kaplan} {et~al.}(2003{\natexlab{a}}){Kaplan}, {Kulkarni}, \& {van
  Kerkwijk}}]{kkv03}
---. 2003{\natexlab{a}}, ApJ, 588, L33

\bibitem[{Kaplan {et~al.}(2002)Kaplan, van Kerkwijk, \& Anderson}]{kva02}
Kaplan, D.~L., van Kerkwijk, M.~H., \& Anderson, J. 2002, ApJ, 571, 447

\bibitem[{{Kaplan} {et~al.}(2003{\natexlab{b}}){Kaplan}, {van Kerkwijk},
  {Marshall}, {Jacoby}, {Kulkarni}, \& {Frail}}]{kvm+03}
{Kaplan}, D.~L., {van Kerkwijk}, M.~H., {Marshall}, H.~L., {Jacoby}, B.~A.,
  {Kulkarni}, S.~R., \& {Frail}, D.~A. 2003{\natexlab{b}}, ApJ, 590, 1008

\bibitem[{{Kargaltsev} {et~al.}(2002){Kargaltsev}, {Pavlov}, {Sanwal}, \&
  {Garmire}}]{kpsg02}
{Kargaltsev}, O., {Pavlov}, G.~G., {Sanwal}, D., \& {Garmire}, G.~P. 2002, ApJ,
  580, 1060

\bibitem[{Kaspi {et~al.}(1997)Kaspi, Bailes, Manchester, Stappers, Sandhu,
  Navarro, \& D'Amico}]{kbm+97}
Kaspi, V.~M., Bailes, M., Manchester, R.~N., Stappers, B.~W., Sandhu, J.~S.,
  Navarro, J., \& D'Amico, N. 1997, ApJ, 485, 820

\bibitem[{Kaspi {et~al.}(2001{\natexlab{a}})Kaspi, Gotthelf, Gaensle, \&
  Lyutikov"}]{kggl01}
Kaspi, V.~M., Gotthelf, E.~V., Gaensle, B.~M., \& Lyutikov", M.
  2001{\natexlab{a}}, ApJ, 562, L163

\bibitem[{Kaspi {et~al.}(1992)Kaspi, Manchester, Johnston, Lyne, \&
  D'Amico}]{kmj+92}
Kaspi, V.~M., Manchester, R.~N., Johnston, S., Lyne, A.~G., \& D'Amico, N.
  1992, ApJ, 399, L155

\bibitem[{Kaspi {et~al.}(1996)Kaspi, Manchester, Johnston, Lyne, \&
  D'Amico}]{kmj+96}
---. 1996, AJ, 111, 2028

\bibitem[{Kaspi {et~al.}(1994)Kaspi, Manchester, Siegman, Johnston, \&
  Lyne}]{kms+94}
Kaspi, V.~M., Manchester, R.~N., Siegman, B., Johnston, S., \& Lyne, A.~G.
  1994, ApJ, 422, L83

\bibitem[{Kaspi {et~al.}(2001{\natexlab{b}})Kaspi, Roberts, Vasisht, Gotthelf,
  Pivovaroff, \& Kawai}]{krv+01}
Kaspi, V.~M., Roberts, M. S.~E., Vasisht, G., Gotthelf, E.~V., Pivovaroff, M.,
  \& Kawai, N. 2001{\natexlab{b}}, ApJ, 560, 371

\bibitem[{Kaspi {et~al.}(1995)Kaspi, Tavani, Nagase, Hirayama, Hoshino, Aoki,
  Kawai, \& Arons}]{ktn+95}
Kaspi, V.~M., Tavani, M., Nagase, F., Hirayama, M., Hoshino, M., Aoki, T.,
  Kawai, N., \& Arons, J. 1995, ApJ, 453, 424

\bibitem[{Kellner(2002)}]{kel02}
Kellner, S. 2002, PhD thesis, Technische Universität München

\bibitem[{Kennel \& Coroniti(1984{\natexlab{a}})}]{kc84a}
Kennel, C.~F. \& Coroniti, F.~V. 1984{\natexlab{a}}, ApJ, 283, 694

\bibitem[{Kennel \& Coroniti(1984{\natexlab{b}})}]{kc84}
---. 1984{\natexlab{b}}, ApJ, 283, 710

\bibitem[{{Khangoulian} \& {Bogovalov}(2003)}]{kb03}
{Khangoulian}, D.~V. \& {Bogovalov}, S.~V. 2003, Astronomy Letters, 29, 495

%\bibitem[{Kifune {et~al.}(1993)Kifune, Hayashida, Tamura, Teshima, Tanimori,
%  Ogio, Kakimoto, Tsukagoshi, Yoshikoshi, Matsubara, Muraki, Mizumoto, Suda,
%  Hara, Fujii, Kabe, Watase, Fujimoto, Edwards, Patterson, Rowell, Roberts, \&
%  Thornton}]{kht+93}
%Kifune, T., Hayashida, N., Tamura, T., Teshima, M., Tanimori, T., Ogio, S.,
%  Kakimoto, F., Tsukagoshi, T., Yoshikoshi, T., Matsubara, Y., Muraki, Y.,
%  Mizumoto, Y., Suda, T., Hara, T., Fujii, H., Kabe, S., Watase, Y., Fujimoto,
%  M., Edwards, P.~G., Patterson, J.~R., Rowell, G.~P., Roberts, M.~D., \&
%  Thornton, G.~J.
%\newblock IAU circular 5905

\bibitem[{Kifune {et~al.}(1993)}]{kht+93}
Kifune, T. et al.
\newblock IAU circular 5905


\bibitem[{{Kirk} {et~al.}(2000){Kirk}, {Guthmann}, {Gallant}, \&
  {Achterberg}}]{kgga00}
{Kirk}, J.~G., {Guthmann}, A.~W., {Gallant}, Y.~A., \& {Achterberg}, A. 2000,
  ApJ, 542, 235

\bibitem[{{Kirk} \& {Skj{\ae}raasen}(2003)}]{ks03}
{Kirk}, J.~G. \& {Skj{\ae}raasen}, O. 2003, ApJ, 591, 366

\bibitem[{{Komissarov} \& {Lyubarsky}(2003)}]{kl03}
{Komissarov}, S.~S. \& {Lyubarsky}, Y.~E. 2003, MNRAS, 344, L93

%\bibitem[{{Kothes} \& {Reich}(2001)}]{kr01}
%{Kothes}, R. \& {Reich}, W. 2001, A\&A, 372, 627

\bibitem[{Krause-Polstorff \& Michel(1985)}]{km85}
Krause-Polstorff, J. \& Michel, F.~C. 1985, MNRAS, 213, 43P

\bibitem[{Kuiper {et~al.}(2001)Kuiper, Hermsen, Cusumano, Diehl, Schönfelder,
  Strong, Bennett, \& McConnell}]{khc+01}
Kuiper, L., Hermsen, W., Cusumano, G., Diehl, R., Schönfelder, V., Strong, A.,
  Bennett, K., \& McConnell, M.~L. 2001, A\&A, 378, 918

\bibitem[{{Kuiper} {et~al.}(1998){Kuiper}, {Hermsen}, {Verbunt}, \&
  {Belloni}}]{khvb98}
{Kuiper}, L., {Hermsen}, W., {Verbunt}, F., \& {Belloni}, T. 1998, A\&A, 336,
  545

\bibitem[{Kuiper {et~al.}(2000)Kuiper, Hermsen, Verbunt, Thompson, Stairs,
  Strickman, \& Cusumano}]{khv+00b}
Kuiper, L., Hermsen, W., Verbunt, F., Thompson, D.~J., Stairs, I. H.;~Lyne,
  A.~G., Strickman, M.~S., \& Cusumano, G. 2000, A\&A, 359, 615

\bibitem[{Kulkarni \& Hester(1988)}]{kh88}
Kulkarni, S.~R. \& Hester, J.~J. 1988, Nature, 335, 801

\bibitem[{Kulkarni {et~al.}(1992)Kulkarni, Phinney, Evans, \&
  Hasinger}]{kpeh92}
Kulkarni, S.~R., Phinney, E.~S., Evans, C.~R., \& Hasinger, G. 1992, Nature,
  359, 300

\bibitem[{Lai \& Ho(2003)}]{lh03b}
Lai, D. \& Ho, W. 2003, Phys. Rev. Lett., 91, 071101

\bibitem[{{Lai} \& {Ho}(2002)}]{lh02}
{Lai}, D. \& {Ho}, W.~C.~G. 2002, ApJ, 566, 373

\bibitem[{{Lai} \& {Ho}(2003)}]{lh03a}
---. 2003, ApJ, 588, 962

\bibitem[{Lampland(1921)}]{lam21}
Lampland, C.~O. 1921, PASP, 33, 79

\bibitem[{{Larson} \& {Link}(1999)}]{ll99}
{Larson}, M.~B. \& {Link}, B. 1999, ApJ, 521, 271

\bibitem[{Lattimer(1992)}]{lat92}
Lattimer, J.~M. 1992, in The Structure and Evolution of Neutron Stars, ed.
  D.~Pines, R.~Tmagaki, \& S.~Tsuruta (USA: Addison-Wesley)

\bibitem[{{Lattimer} \& {Prakash}(2001)}]{lp01}
{Lattimer}, J.~M. \& {Prakash}, M. 2001, ApJ, 550, 426

\bibitem[{{Lattimer} {et~al.}(1991){Lattimer}, {Prakash}, {Pethick}, \&
  {Haensel}}]{lpph91}
{Lattimer}, J.~M., {Prakash}, M., {Pethick}, C.~J., \& {Haensel}, P. 1991,
  Phys. Rev. Lett., 66, 2701

\bibitem[{{Lodenqual} {et~al.}(1974){Lodenqual}, {Canuto}, {Ruderman}, \&
  {Tsuruta}}]{lcrt74}
{Lodenqual}, J., {Canuto}, V., {Ruderman}, M., \& {Tsuruta}, S. 1974, ApJ, 190,
  141

\bibitem[{Lu {et~al.}(2002)Lu, Wang, Aschenbach, Durouchoux, \& Song}]{lwa+02}
Lu, F.~J., Wang, Q.~D., Aschenbach, B., Durouchoux, P., \& Song, L.~M. 2002,
  ApJ, 568, L49

\bibitem[{{Lu} {et~al.}(2003){Lu}, {Wang}, \& {Lang}}]{lwl03}
{Lu}, F.~J., {Wang}, Q.~D., \& {Lang}, C.~C. 2003, AJ, 126, 319

\bibitem[{Lyne \& Lorimer(1994)}]{ll94}
Lyne, A.~G. \& Lorimer, D.~R. 1994, Nature, 369, 127

\bibitem[{Lyne \& Manchester(1988)}]{lm88}
Lyne, A.~G. \& Manchester, R.~N. 1988, MNRAS, 234, 477

\bibitem[{Lyne {et~al.}(1998)Lyne, Manchester, Lorimer, Bailes, D'Amico,
  Tauris, Johnston, Bell, \& Nicastro}]{lml+98}
Lyne, A.~G., Manchester, R.~N., Lorimer, D.~R., Bailes, M., D'Amico, N.,
  Tauris, T.~M., Johnston, S., Bell, J.~F., \& Nicastro, L. 1998, MNRAS, 295,
  743

\bibitem[{Lyne {et~al.}(1996)Lyne, Pritchard, Graham-Smith, \& Camilo}]{lpgc96}
Lyne, A.~G., Pritchard, R.~S., Graham-Smith, F., \& Camilo, F. 1996, Nature,
  381, 497

\bibitem[{Lyne {et~al.}(1988)Lyne, Pritchard, \& Smith}]{lps88}
Lyne, A.~G., Pritchard, R.~S., \& Smith, F.~G. 1988, MNRAS, 233, 667

\bibitem[{Manchester(1996)}]{man96}
Manchester, R.~N. 1996, in Pulsars: Problems and Progress, {IAU} Colloquium
  160, ed. S.~Johnston, M.~A. Walker, \& M.~Bailes (San Francisco: ASP),
  193--196

\bibitem[{Manchester {et~al.}(1993)Manchester, Staveley-Smith, \&
  Kesteven}]{msk93}
Manchester, R.~N., Staveley-Smith, L., \& Kesteven, M.~J. 1993, ApJ, 411, 756

\bibitem[{Marshall {et~al.}(1998)Marshall, Gotthelf, Zhang, Middleditch, \&
  Wang}]{mgz+98}
Marshall, F.~E., Gotthelf, E.~V., Zhang, W., Middleditch, J., \& Wang, Q.~D.
  1998, ApJ, 499, L179

\bibitem[{Marshall \& Schulz(2002)}]{ms02}
Marshall, H.~L. \& Schulz, N.~S. 2002, ApJ, 574, 377

\bibitem[{{Matsui} {et~al.}(1988){Matsui}, {Long}, \& {Tuohy}}]{mlt88}
{Matsui}, Y., {Long}, K.~S., \& {Tuohy}, I.~R. 1988, ApJ, 329, 838

\bibitem[{{McGowan} {et~al.}(2003){McGowan}, {Kennea}, {Zane}, {C{\' o}rdova},
  {Cropper}, {Ho}, {Sasseen}, \& {Vestrand}}]{mkz+03}
{McGowan}, K.~E., {Kennea}, J.~A., {Zane}, S., {C{\' o}rdova}, F.~A.,
  {Cropper}, M., {Ho}, C., {Sasseen}, T., \& {Vestrand}. 2003, ApJ, 591, 380

\bibitem[{McLaughlin {et~al.}(2001)McLaughlin, Cordes, Deshpande, Gaensler,
  Hankins, Kaspi, \& Kern}]{mcd+01}
McLaughlin, M.~A., Cordes, J.~M., Deshpande, A.~A., Gaensler, B.~M., Hankins,
  T.~H., Kaspi, V.~M., \& Kern, J.~S. 2001, ApJ, 547, L41

\bibitem[{Melatos(1998)}]{mel98}
Melatos, A. 1998, Mem. della Soc. Ast. It., 69, 1009

\bibitem[{{Melrose}(2000)}]{mel00a}
{Melrose}, D.~B. 2000, in Pulsar Astronomy - 2000 and Beyond, {IAU} Colloquium
  177, ed. M.~Kramer, N.~Wex, \& R.~Wielebinski (San Francisco: ASP), 721

\bibitem[{Mereghetti {et~al.}(1996)Mereghetti, Bignami, \& Caraveo}]{mbc96}
Mereghetti, S., Bignami, G.~F., \& Caraveo, P.~A. 1996, ApJ, 464, 842

\bibitem[{{Mereghetti} {et~al.}(2002{\natexlab{a}}){Mereghetti}, {De Luca},
  {Caraveo}, {Becker}, {Mignani}, \& {Bignami}}]{mdc+02}
{Mereghetti}, S., {De Luca}, A., {Caraveo}, P.~A., {Becker}, W., {Mignani}, R.,
  \& {Bignami}, G.~F. 2002{\natexlab{a}}, ApJ, 581, 1280

\bibitem[{{Mereghetti} {et~al.}(2002{\natexlab{b}}){Mereghetti}, {Tiengo}, \&
  {Israel}}]{mti02}
{Mereghetti}, S., {Tiengo}, A., \& {Israel}, G.~L. 2002{\natexlab{b}}, ApJ,
  569, 275

\bibitem[{Michel(1969)}]{mic69}
Michel, F.~C. 1969, ApJ, 158, 727

\bibitem[{Morton(1964)}]{mor64}
Morton, D.~C. 1964, ApJ, 140, 460

\bibitem[{{Motch} {et~al.}(1999){Motch}, {Haberl}, {Zickgraf}, {Hasinger}, \&
  {Schwope}}]{mhz+99}
{Motch}, C., {Haberl}, F., {Zickgraf}, F.-J., {Hasinger}, G., \& {Schwope},
  A.~D. 1999, A\&A, 351, 177

\bibitem[{{Motch} {et~al.}(2003){Motch}, {Zavlin}, \& {Haberl}}]{mzh03}
{Motch}, C., {Zavlin}, V.~E., \& {Haberl}, F. 2003, A\&A, 408, 323

\bibitem[{{Mukherjee} {et~al.}(2000){Mukherjee}, {Gotthelf}, {Halpern}, \&
  {Tavani}}]{mght00}
{Mukherjee}, R., {Gotthelf}, E.~V., {Halpern}, J., \& {Tavani}, M. 2000, ApJ,
  542, 740

\bibitem[{Murray {et~al.}(2002)Murray, Slane, Seward, Ransom, \&
  Gaensler}]{mss+02}
Murray, S.~S., Slane, P.~O., Seward, F.~D., Ransom, S.~M., \& Gaensler, B.~M.
  2002, ApJ, 568, 226

\bibitem[{Muslimov \& Harding(2003)}]{mh03}
Muslimov, A.~G. \& Harding, A.~K. 2003, ApJ, 588, 430

\bibitem[{Muslimov \& Harding(2004)}]{mh04}
---. 2004, ApJ

\bibitem[{Muslimov \& Tsygan(1992)}]{mt92}
Muslimov, A.~G. \& Tsygan, A.~I. 1992, MNRAS, 255, 61

\bibitem[{{Ng} \& {Romani}(2004)}]{nr04}
{Ng}, C.~. \& {Romani}, R.~W. 2004, ApJ, 601, 479

\bibitem[{{Nicastro} {et~al.}(2004){Nicastro}, {Cusumano}, {L{\" o}hmer},
  {Kramer}, {Kuiper}, {Hermsen}, {Mineo}, \& {Becker}}]{ncl+04}
{Nicastro}, L., {Cusumano}, G., {L{\" o}hmer}, O., {Kramer}, M., {Kuiper}, L.,
  {Hermsen}, W., {Mineo}, T., \& {Becker}, W. 2004, A\&A, 413, 1065

\bibitem[{{Nolan} {et~al.}(2003){Nolan}, {Tompkins}, {Grenier}, \&
  {Michelson}}]{ntgm03}
{Nolan}, P.~L., {Tompkins}, W.~F., {Grenier}, I.~A., \& {Michelson}, P.~F.
  2003, ApJ, 597, 615

\bibitem[{\"Ogelman {et~al.}(1993)\"Ogelman, Finley, \& Zimmermann}]{ofz93}
\"Ogelman, H., Finley, J.~P., \& Zimmermann, H.~U. 1993, Nature, 361, 136

\bibitem[{{Olbert} {et~al.}(2001){Olbert}, {Clearfield}, {Williams}, {Keohane},
  \& {Frail}}]{ocw+01}
{Olbert}, C.~M., {Clearfield}, C., {Williams}, N., {Keohane}, J., \&
  {Frail}, D. 2001, ApJ, 554, L205

\bibitem[{{Olbert} {et~al.}(2003){Olbert}, {Keohane}, {Arnaud}, {Dyer},
  {Reynolds}, \& {Safi-Harb}}]{oka+03}
{Olbert}, C.~M., {Keohane}, J.~W., {Arnaud}, K.~A., {Dyer}, K.~K., {Reynolds},
  S.~P., \& {Safi-Harb}, S. 2003, ApJ, 592, L45

\bibitem[{Oort \& Walraven(1956)}]{ow56}
Oort, J. \& Walraven, T. 1956, Bull. Astr. Inst. Netherlands, 12, 285

\bibitem[{Ostriker \& Gunn(1969)}]{og69}
Ostriker, J.~P. \& Gunn, J.~E. 1969, ApJ, 157, 1395

\bibitem[{Ostriker {et~al.}(1970)Ostriker, Rees, \& Silk}]{ors70}
Ostriker, J.~P., Rees, M.~J., \& Silk, J. 1970, Astrophys. Lett., 6, 179

\bibitem[{\"Ozel(2001)}]{oze01}
\"Ozel, F. 2001, ApJ, 563, 276

\bibitem[{\"Ozel(2003)}]{oze03}
---. 2003, ApJ, 583, 402

\bibitem[{Pacini \& Salvati(1973)}]{ps73}
Pacini, F. \& Salvati, M. 1973, ApJ, 186, 249

\bibitem[{{Page}(1995)}]{pag95}
{Page}, D. 1995, ApJ, 442, 273

\bibitem[{{Page}(1998)}]{pag98}
{Page}, D. 1998, in The Many Faces of Neutron Stars. Edited by R. Buccheri, J.
  van Paradijs, and M. A. Alpar. Dordrecht ; Boston : Kluwer Academic
  Publishers, 1998., p.539, 539+

\bibitem[{Pavlov {et~al.}(2002{\natexlab{a}})Pavlov, Sanwal, Garmire, \&
  Zavlin}]{psgz02}
Pavlov, G.~G., Sanwal,D.,Garmire,G., \& Zavlin, V. 2002{\natexlab{a}},
  in Neutron Stars in Supernova Remnants, ed. P. Slane \& B. Gaensler
  (San Francisco: ASP), 247

\bibitem[{{Pavlov} {et~al.}(2001){Pavlov}, {Sanwal}, {K{\i}z{\i}ltan}, \&
  {Garmire}}]{pskg01}
{Pavlov}, G.~G., {Sanwal}, D., {K{\i}z{\i}ltan}, B., \& {Garmire}, G.~P. 2001,
  ApJ, 559, L131

\bibitem[{{Pavlov} \& {Shibanov}(1978)}]{ps78}
{Pavlov}, G.~G. \& {Shibanov}, I.~A. 1978, Astronomicheskii Zhurnal, 55, 373

\bibitem[{{Pavlov} {et~al.}(2003){Pavlov}, {Teter}, {Kargaltsev}, \&
  {Sanwal}}]{ptks03}
{Pavlov}, G.~G., {Teter}, M.~A., {Kargaltsev}, O., \& {Sanwal}, D. 2003, ApJ,
  591, 1157

\bibitem[{Pavlov \& Zavlin(2000)}]{pz00}
Pavlov, G.~G. \& Zavlin, V.~E. 2000, in Pulsar Astronomy - 2000 and Beyond,
  {IAU} Colloquium 177, ed. M.~Kramer, N.~Wex, \& R.~Wielebinski (San
  Francisco: ASP), 613--618

\bibitem[{Pavlov \& Zavlin(2003)}]{pz03}
Pavlov, G.~G. \& Zavlin, V.~E. 2003, in Proc. of the XXI Texas Symposium on
  Relativistic Astrophysics, ed. R.~Bandiera, R.~Maiolino, \& F.~Mannucci
  (World Sci. Publishing Co.), 319
%astro-ph/0305453

\bibitem[{Pavlov {et~al.}(2000)Pavlov, Zavlin, Aschenbach, Tr\"{u}mper, \&
  Sanwal}]{pza+00}
Pavlov, G.~G., Zavlin, V.~E., Aschenbach, B., Tr\"{u}mper, J., \& Sanwal, D.
  2000, ApJ, 531, L53

\bibitem[{Pavlov {et~al.}(2002{\natexlab{b}})Pavlov, Zavlin, \& Sanwal}]{pzs02}
Pavlov, G.~G., Zavlin, V.~E., \& Sanwal, D. 2002{\natexlab{b}}, in Heraeus
  Seminar on Neutron Stars, Pulsars, and Supernova Remnants, 
  MPE Report 278., ed. W.~Becker, H.~Lesch, \& J.~Tr\"umper, 273
%astro-ph/0206024

\bibitem[{Pavlov {et~al.}(2001)Pavlov, Zavlin, Sanwal, Burwitz, \&
  Garmire}]{pzs+01}
Pavlov, G.~G., Zavlin, V.~E., Sanwal, D., Burwitz, V., \& Garmire, G.~P. 2001,
  ApJ, 552, L129

\bibitem[{{Pavlov} {et~al.}(2002){Pavlov}, {Zavlin}, {Sanwal}, \& {Tr{\"
  u}mper}}]{pzst02}
{Pavlov}, G.~G., {Zavlin}, V.~E., {Sanwal}, D., \& {Tr{\" u}mper}, J. 2002,
  ApJ, 569, L95

\bibitem[{{Perna} {et~al.}(2003){Perna}, {Narayan}, {Rybicki}, {Stella}, \&
  {Treves}}]{pnr+03}
{Perna}, R., {Narayan}, R., {Rybicki}, G., {Stella}, L., \& {Treves}, A. 2003,
  ApJ, 594, 936

\bibitem[{Pethick(1992)}]{pet92}
Pethick, C.~J. 1992, Reviews of Modern Physics, 64, 1133

\bibitem[{Petre {et~al.}(1996)Petre, Becker, \& Winkler}]{pbw96}
Petre, R., Becker, C.~M., \& Winkler, P.~F. 1996, ApJ, 465, L43

\bibitem[{{Petre} {et~al.}(1982){Petre}, {Kriss}, {Winkler}, \&
  {Canizares}}]{pkwc82}
{Petre}, R., {Kriss}, G.~A., {Winkler}, P.~F., \& {Canizares}, C.~R. 1982, ApJ,
  258, 22

\bibitem[{{Petre} {et~al.}(2002){Petre}, {Kuntz}, \& {Shelton}}]{pks02}
{Petre}, R., {Kuntz}, K.~D., \& {Shelton}, R.~L. 2002, ApJ, 579, 404

\bibitem[{P\'etri {et~al.}(2002)P\'etri, Heyvaerts, \& Bonazzola}]{phb02}
P\'etri, J., Heyvaerts, J., \& Bonazzola, S. 2002, A\&A, 384, 414

\bibitem[{{Porquet} {et~al.}(2003){Porquet}, {Decourchelle}, \&
  {Warwick}}]{pdw03}
{Porquet}, D., {Decourchelle}, A., \& {Warwick}, R.~S. 2003, A\&A, 401, 197

\bibitem[{Potekhin {et~al.}(2003)Potekhin, Yakovlev, Chabrier, \&
  Gnedin}]{pycg03}
Potekhin, A., Yakovlev, D., Chabrier, G., \& Gnedin, O. 2003, ApJ, 594, 404

\bibitem[{{Potekhin} \& {Yakovlev}(2001)}]{py01}
{Potekhin}, A.~Y. \& {Yakovlev}, D.~G. 2001, A\&A, 374, 213

\bibitem[{Prakash {et~al.}(1988)Prakash, Ainsworth, \& Lattimer}]{pal88}
Prakash, A.~Y., Ainsworth, T.~L., \& Lattimer, J.~M. 1988, Phys. Rev. Lett.,
  61, 2518

\bibitem[{Prakash {et~al.}(1992)Prakash, Prakash, Lattimer, \&
  Pethick}]{pplp92}
Prakash, M., Prakash, M., Lattimer, J.~M., \& Pethick, C.~J. 1992, ApJ, 390,
  L77

\bibitem[{Predehl \& Kulkarni(1995)}]{pk95}
Predehl, P. \& Kulkarni, S.~R. 1995, A\&A, 294, L29

\bibitem[{Psaltis {et~al.}(2000)Psaltis, \"{O}zel, \& DeDeo}]{pod00}
Psaltis, D., \"{O}zel, F., \& DeDeo, S. 2000, ApJ, 544, 390

\bibitem[{Radhakrishnan \& Cooke(1969)}]{rc69a}
Radhakrishnan, V. \& Cooke, D.~J. 1969, Astrophys. Lett., 3, 225

\bibitem[{{Rajagopal} \& {Romani}(1996)}]{rr96a}
{Rajagopal}, M. \& {Romani}, R.~W. 1996, ApJ, 461, 327

\bibitem[{Rajagopal {et~al.}(1997)Rajagopal, Romani, \& Miller}]{rrm97}
Rajagopal, M., Romani, R.~W., \& Miller, M.~C. 1997, ApJ, 479, 347

\bibitem[{Rankin(1993)}]{ran93}
Rankin, J.~M. 1993, ApJ, 405, 285

\bibitem[{Ransom {et~al.}(2002)Ransom, Gaensler, \& Slane}]{rgs02}
Ransom, S.~M., Gaensler, B.~M., \& Slane, P.~O. 2002, ApJ, 570, L75

\bibitem[{Rees \& Gunn(1974)}]{rg74}
Rees, M.~J. \& Gunn, J.~E. 1974, MNRAS, 167, 1

\bibitem[{Reich {et~al.}(1984)Reich, F\"urst, \& Sofue}]{rfs84}
Reich, W., F\"urst, E., \& Sofue, Y. 1984, A\&A, 133, 4

\bibitem[{Reynolds \& Aller(1985)}]{ra85}
Reynolds, S.~P. \& Aller, H.~D. 1985, AJ, 90, 2312

\bibitem[{Reynolds \& Chevalier(1984)}]{rc84}
Reynolds, S.~P. \& Chevalier, R.~A. 1984, ApJ, 278, 630

\bibitem[{{Roberts} {et~al.}(2001){Roberts}, {Romani}, \& {Johnston}}]{rrj01}
{Roberts}, M.~S.~E., {Romani}, R.~W., \& {Johnston}, S. 2001, ApJ, 561, L187

\bibitem[{Roberts {et~al.}(1999)Roberts, Romani, Johnston, \& Green}]{rrjg99}
Roberts, M. S.~E., Romani, R.~W., Johnston, S., \& Green, A.~J. 1999, ApJ, 515,
  712

\bibitem[{Roberts {et~al.}(2001)Roberts, Romani, \& Kawai}]{rrk01}
Roberts, M. S.~E., Romani, R.~W., \& Kawai, N. 2001, ApJS, 133, 451

\bibitem[{Roberts {et~al.}(2003)Roberts, Tam, Kaspi, Lyutikov, Vasisht,
  Pivovaroff, Gotthelf, \& Kawai}]{rtk+03}
Roberts, M. S.~E., Tam, C.~R., Kaspi, V.~M., Lyutikov, M., Vasisht, G.,
  Pivovaroff, M., Gotthelf, E.~V., \& Kawai, N. 2003, ApJ, 588, 992

\bibitem[{Romani(1987)}]{rom87}
Romani, R.~W. 1987, ApJ, 313, 718

\bibitem[{Romani(1996)}]{rom96a}
---. 1996, ApJ, 470, 469

\bibitem[{Romani {et~al.}(1997)Romani, Cordes, \& Yadigaroglu}]{rcy97}
Romani, R.~W., Cordes, J.~M., \& Yadigaroglu, I.-A. 1997, ApJ, 484, L137

\bibitem[{Romani \& Yadigaroglu(1995)}]{ry95}
Romani, R.~W. \& Yadigaroglu, I.-A. 1995, ApJ, 438, 314

\bibitem[{Ruderman(2003)}]{rud03}
Ruderman, M. 2003, in X-ray and Gamma-ray Astrophysics of Galactic Sources, in
  press (astro-ph/0310777)

\bibitem[{Ruderman \& Sutherland(1975)}]{rs75}
Ruderman, M.~A. \& Sutherland, P.~G. 1975, ApJ, 196, 51

\bibitem[{Rutledge {et~al.}(2003)Rutledge, Fox, Bogosavljevic, \&
  Mahabal}]{rfbm03}
Rutledge, R.~E., Fox, D.~W., Bogosavljevic, M., \& Mahabal, A. 2003, ApJ,
  submitted (astro-ph/0302107)

\bibitem[{Ryan {et~al.}(2001)Ryan, Wagner, \& Starrfield}]{rws01}
Ryan, E., Wagner, R.~M., \& Starrfield, S.~G. 2001, ApJ, 548, 811

\bibitem[{Safi-Harb {et~al.}(1995)Safi-Harb, Ogelman, \& Finley}]{sof95}
Safi-Harb, S., Ogelman, H., \& Finley, J.~P. 1995, ApJ, 439, 722

\bibitem[{Saito(1998)}]{sai98}
Saito, Y. 1998, PhD thesis, University of Tokyo

\bibitem[{Sakurai {et~al.}(2001)Sakurai, Kawai, Torii, Negoro, Nagase, Shibata,
  \& Becker}]{skt+01}
Sakurai, I., Kawai,N.,Torii,K.,Negoro,H.,Nagase,F.,Shibata, S., \&
  Becker, W. 2001, PASJ, 53, 535

\bibitem[{{Sanwal} {et~al.}(2002{\natexlab{a}}){Sanwal}, {Garmire}, {Garmire},
  {Pavlov}, \& {Mignani}}]{sgg+02}
{Sanwal}, D., {Garmire}, G.~P., {Garmire}, A., {Pavlov}, G.~G., \& {Mignani},
  R. 2002{\natexlab{a}}, BAAS, 200, 7201

\bibitem[{{Sanwal} {et~al.}(2002{\natexlab{b}}){Sanwal}, {Pavlov}, {Zavlin}, \&
  {Teter}}]{spzt02}
{Sanwal}, D., {Pavlov}, G.~G., {Zavlin}, V.~E., \& {Teter}, M.~A.
  2002{\natexlab{b}}, ApJ, 574, L61

\bibitem[{Scargle(1969)}]{sca69}
Scargle, J.~D. 1969, ApJ, 156, 401

\bibitem[{{Schwope} {et~al.}(1999){Schwope}, {Hasinger}, {Schwarz}, {Haberl},
  \& {Schmidt}}]{shs+99}
{Schwope}, A.~D., {Hasinger}, G., {Schwarz}, R., {Haberl}, F., \& {Schmidt}, M.
  1999, A\&A, 341, L51

\bibitem[{Sedov(1959)}]{sed59}
Sedov, L.~I. 1959, Similarity and Dimensional Methods in Mechanics (New York:
  Academic Press)

\bibitem[{Shibanov \& Yakovlev(1996)}]{sy96}
Shibanov, Y.~A. \& Yakovlev, D.~G. 1996, A\&A, 309, 171

\bibitem[{{Shibata} {et~al.}(2003){Shibata}, {Tomatsuri}, {Shimanuki}, {Saito},
  \& {Mori}}]{sts+03}
{Shibata}, S., {Tomatsuri}, H., {Shimanuki}, M., {Saito}, K., \& {Mori}, K.
  2003, MNRAS, 346, 841

\bibitem[{Shull {et~al.}(1989)Shull, Fesen, \& Saken}]{sfs89}
Shull, J.~M., Fesen, R.~A., \& Saken, J.~M. 1989, ApJ, 346, 860

\bibitem[{Slane(1994)}]{sla94}
Slane, P. O. 1994, ApJ, 437, 458

\bibitem[{Slane {et~al.}(2000)Slane, Chen, Schulz, Seward, Hughes, \&
  Gaensler}]{scs+00}
Slane, P. O., Chen,Y.,Schulz,N.,Seward,F.,Hughes,J., \& Gaensler,
  B. 2000, ApJ, 533, L29

\bibitem[{{Slane} {et~al.}(2001){Slane}, {Hughes}, {Edgar}, {Plucinsky},
  {Miyata}, {Tsunemi}, \& {Aschenbach}}]{she+01}
{Slane}, P., {Hughes}, J.~P., {Edgar}, R.~J., {Plucinsky}, P.~P., {Miyata}, E.,
  {Tsunemi}, H., \& {Aschenbach}, B. 2001, ApJ, 548, 814

\bibitem[{{Slane} {et~al.}(2003){Slane}, {Zimmerman}, {Hughes}, {Seward},
  {Gaensler}, \& {Clarke}}]{szh+03}
{Slane}, P., {Zimmerman}, E.~R., {Hughes}, J.~P., {Seward}, F.~D., {Gaensler},
  B.~M., \& {Clarke}, M.~J. 2003, ApJ, 601, 1045

\bibitem[{Slane {et~al.}(2002)Slane, Helfand, \& Murray}]{shm02}
Slane, P.~O., Helfand, D.~J., \& Murray, S.~S. 2002, ApJ, 571, L45

\bibitem[{Spitkovsky \& Arons(2002)}]{sa02}
Spitkovsky, A. \& Arons, J. 2002, in Neutron Stars in Supernova Remnants, ed.
  P.~O. Slane \& B.~M. Gaensler (San Francisco: ASP), 81

\bibitem[{Spruit \& Phinney(1998)}]{sp98}
Spruit, H. \& Phinney, E.~S. 1998, Nature, 393, 139

\bibitem[{{Stappers} {et~al.}(2003){Stappers}, {Gaensler}, {Kaspi}, {van der
  Klis}, \& {Lewin}}]{sgk+03}
{Stappers}, B.~W., {Gaensler}, B.~M., {Kaspi}, V.~M., {van der Klis}, M., \&
  {Lewin}, W.~H.~G. 2003, Science, 299, 1372

\bibitem[{Staelin \& Reifenstein(1968)}]{sr68}
Staelin, D. H. \& Reifenstein, E. C. III, 1968, Science, 162, 1481

\bibitem[{{Tam} \& {Roberts}(2003)}]{tr03}
{Tam}, C. \& {Roberts}, M.~S.~E. 2003, ApJ, 598, L27

\bibitem[{Tam {et~al.}(2002)Tam, Roberts, \& Kaspi}]{trk02}
Tam, C., Roberts, M. S.~E., \& Kaspi, V.~M. 2002, ApJ, 572, 202

\bibitem[{Tananbaum(1999)}]{tan99}
Tananbaum, H.
\newblock IAU circular 7246

\bibitem[{Tavani \& Arons(1997)}]{ta97}
Tavani, M. \& Arons, J. 1997, ApJ, 477, 439

\bibitem[{Tavani {et~al.}(1994)Tavani, Arons, \& Kaspi}]{tak94}
Tavani, M., Arons, J., \& Kaspi, V.~M. 1994, ApJ, 433, L37

%\bibitem[{Tennant {et~al.}(2001)Tennant, Becker, Juda, Elsner, Kolodziejczak,
%  Murray, O'Dell, Paerels, Swartz, Shibazaki, \& Weisskopf}]{tbj+01}
%Tennant, A.~F., Becker, W., Juda, M., Elsner, R.~F., Kolodziejczak, J.~J.,
%  Murray, S.~S., O'Dell, S.~L., Paerels, F., Swartz, D.~A., Shibazaki, N., \&
%  Weisskopf, M.~C. 2001, ApJ, 554, L173

\bibitem[{Tennant {et~al.}(2001)}]{tbj+01}
Tennant, A. F. et al. 2001, ApJ, 554, L173

%\bibitem[{Thompson {et~al.}(1999)Thompson, Bailes, Bertsch, Cordes, D'Amico,
%  Esposito, Finley, Hartman, Hermsen, Kanbach, Kaspi, Kniffen, Kuiper, Lin,
%  Manchester, Matz, Mayer-Hasselwander, Michelson, Nolan, Ogelman, Pohl,
%  Ramanamurthy, Sreekumar, Reimer, Taylor, \& Ulmer}]{tbb+99}
%Thompson, D.~J., Bailes, M., Bertsch, D.~L., Cordes, J., D'Amico, N., Esposito,
%  J.~A., Finley, J., Hartman, R.~C., Hermsen, W., Kanbach, G., Kaspi, V.~M.,
%  Kniffen, D.~A., Kuiper, L., Lin, Y.~C., Manchester, R., Matz, S.~M.,
%  Mayer-Hasselwander, H.~A., Michelson, P.~F., Nolan, P.~L., Ogelman, H., Pohl,
%  M., Ramanamurthy, P.~V., Sreekumar, P., Reimer, O., Taylor, J.~H., \& Ulmer,
%  M. 1999, ApJ, 516, 297

\bibitem[{Thompson {et~al.}(1999)}]{tbb+99}
Thompson, D. J. et al. 1999, ApJ, 516, 297

\bibitem[{{Thorsett} {et~al.}(2002){Thorsett}, {Brisken}, \& {Goss}}]{tbg02}
{Thorsett}, S.~E., {Brisken}, W.~F., \& {Goss}, W.~M. 2002, ApJ, 573, L111

\bibitem[{Torii {et~al.}(1998)Torii, Kinugasa, Toneri, Asanuma, Tsunemi,
  Donati, Mitsuda, Gotthelf, \& Petre}]{tkt+98}
Torii, K., Kinugasa, K., Toneri, T., Asanuma, T., Tsunemi, H., Donati, T.,
  Mitsuda, K., Gotthelf, E.~V., \& Petre, R. 1998, ApJ, 494, L207

\bibitem[{Torii {et~al.}(1997)Torii, Tsunemi, Dotani, \& Mitsuda}]{ttd+97}
Torii, K., Tsunemi, H., Dotani, T., \& Mitsuda, K. 1997, ApJ, 489, 145

\bibitem[{Torii {et~al.}(1999)Torii, Tsunemi, Dotani, Mitsuda, Kawai, Kinugasa,
  Saito, \& Shibata}]{ttd+99}
Torii, K., Tsunemi, H., Dotani, T., Mitsuda, K., Kawai, N., Kinugasa, K.,
  Saito, Y., \& Shibata, S. 1999, ApJ, 523, L69

\bibitem[{Treves \& Colpi(1991)}]{tc91}
Treves, A. \& Colpi, M. 1991, A\&A, 241, 107

\bibitem[{{Treves} {et~al.}(2000){Treves}, {Turolla}, {Zane}, \&
  {Colpi}}]{ttzc00}
{Treves}, A., {Turolla}, R., {Zane}, S., \& {Colpi}, M. 2000, PASP, 112, 297

\bibitem[{Tsuruta(1964)}]{tsu64}
Tsuruta, S. 1964, PhD thesis, Columbia University

\bibitem[{{Tsuruta}(1986)}]{tsu86}
{Tsuruta}, S. 1986, Comments on Astrophysics, 11, 151

\bibitem[{Tsuruta(1998)}]{tsu98}
Tsuruta, S. 1998, Phys. Rep., 292, 1

\bibitem[{Tsuruta \& Cameron(1966)}]{tc66}
Tsuruta, S. \& Cameron, A. G.~W. 1966, Can. J. Phys., 44, 1863

\bibitem[{{Tsuruta} {et~al.}(2002){Tsuruta}, {Teter}, {Takatsuka}, {Tatsumi},
  \& {Tamagaki}}]{ttt+02}
{Tsuruta}, S., {Teter}, M.~A., {Takatsuka}, T., {Tatsumi}, T., \& {Tamagaki},
  R. 2002, ApJ, 571, L143

\bibitem[{Tuohy \& Garmire(1980)}]{tg80}
Tuohy, I. \& Garmire, G. 1980, ApJ, 239, 107

\bibitem[{Umeda {et~al.}(1993)Umeda, Shibazaki, Nomoto, \& Tsuruta}]{usnt93}
Umeda, H., Shibazaki, N., Nomoto, K., \& Tsuruta, S. 1993, ApJ, 408, 186

\bibitem[{Usov \& Melrose(1995)}]{um95}
Usov, V.~V. \& Melrose, D.~B. 1995, Aust. J. Phys., 48, 571

\bibitem[{{van der Swaluw}(2003)}]{van03}
{van der Swaluw}, E. 2003, A\&A, 404, 939

\bibitem[{{van der Swaluw} {et~al.}(2003{\natexlab{a}}){van der Swaluw},
  {Achterberg}, {Gallant}, {Downes}, \& {Keppens}}]{vag+03}
{van der Swaluw}, E., {Achterberg},A.,{Gallant},Y.,{Downes},T., \&
  {Keppens}, R. 2003{\natexlab{a}}, A\&A, 397, 913

\bibitem[{van~der Swaluw {et~al.}(2001)van~der Swaluw, Achterberg, Gallant, \&
  T\'{o}th}]{vagt01}
van~der Swaluw, E., Achterberg, A., Gallant, Y.~A., \& T\'{o}th, G. 2001, A\&A,
  380, 309

\bibitem[{{van der Swaluw} {et~al.}(2003{\natexlab{b}}){van der Swaluw},
  {Downes}, \& {Keegan}}]{vdk03}
{van der Swaluw}, E., {Downes}, T.~P., \& {Keegan}, R. 2003{\natexlab{b}}, A\&A

\bibitem[{van Kerkwijk {et~al.}(2004)van Kerkwijk, Kaplan, Durant, Kulkarni, \&
  Paerels}]{vkd+04}
van Kerkwijk, M.~H., Kaplan, D.~L., Durant, M., Kulkarni, S.~R., \& Paerels, F.
  2004, ApJ, 608, 432

\bibitem[{{van Kerkwijk} \& {Kulkarni}(2001)}]{vk01}
{van Kerkwijk}, M.~H. \& {Kulkarni}, S.~R. 2001, A\&A, 380, 221

\bibitem[{{van Riper} \& {Lamb}(1981)}]{vl81}
{van Riper}, K.~A. \& {Lamb}, D.~Q. 1981, ApJ, 244, L13

\bibitem[{van Riper {et~al.}(1995)van Riper, Link, \& Epstein}]{vle95}
van Riper, K.~A., Link, B., \& Epstein, R.~I. 1995, ApJ, 448, 294

%\bibitem[{Vasisht {et~al.}(1996)Vasisht, Aoki, Dotani, Kulkarni, \&
%  Nagase}]{vad+96}
%Vasisht, G., Aoki, T., Dotani, T., Kulkarni, S.~R., \& Nagase, F. 1996, ApJ,
%  456, 59

\bibitem[{Vasisht {et~al.}(1997)Vasisht, Kulkarni, Anderson, Hamilton, \&
  Kawai}]{vka+97}
Vasisht, G., Kulkarni, S.~R., Anderson, S.~B., Hamilton, T.~T., \& Kawai, N.
  1997, ApJ, 476, L43

\bibitem[{{Velusamy} {et~al.}(1992){Velusamy}, {Roshi}, \& {Venugopal}}]{vrv92}
{Velusamy}, T., {Roshi}, D., \& {Venugopal}, V.~R. 1992, MNRAS, 255, 210

\bibitem[{{Wallace} {et~al.}(1997){Wallace}, {Landecker}, \& {Taylor}}]{wlt97}
{Wallace}, B.~J., {Landecker}, T.~L., \& {Taylor}, A.~R. 1997, AJ, 114, 2068

\bibitem[{Walter(2001)}]{wal01}
Walter, F.~M. 2001, ApJ, 549, 433

\bibitem[{Walter \& Lattimer(2002)}]{wl02}
Walter, F.~M. \& Lattimer, J.~M. 2002, ApJ, 576, L145

\bibitem[{Walter \& Matthews(1997)}]{wm97}
Walter, F.~M. \& Matthews, L.~D. 1997, Nature, 389, 358

\bibitem[{{Walter} {et~al.}(1996){Walter}, {Wolk}, \& {Neuhauser}}]{wwn96}
{Walter}, F.~M., {Wolk}, S.~J., \& {Neuhauser}, R. 1996, Nature, 379, 233

\bibitem[{Wang {et~al.}(1998)Wang, Ruderman, Halpern, \& Zhu}]{wrhz98}
Wang, F. Y.-H., Ruderman, M., Halpern, J.~P., \& Zhu, T. 1998, ApJ, 498, 373

\bibitem[{Wang {et~al.}(2001)Wang, Gotthelf, Chu, \& Dickel}]{wgc+01}
Wang, Q.~D., Gotthelf, E.~V., Chu, Y.-H., \& Dickel, J.~R. 2001, ApJ, 559, 275

\bibitem[{{Wang} {et~al.}(2002){Wang}, {Lu}, \& {Lang}}]{wll02}
{Wang}, Q.~D., {Lu}, F., \& {Lang}, C.~C. 2002, ApJ, 581, 1148

\bibitem[{Webb {et~al.}(2004)Webb, Olive, Barret, Kramer, Cognard, \&
  Lohmer}]{wob+04}
Webb, N.~A., Olive, J.~F., Barret, D., Kramer, M., Cognard, I., \& Lohmer, O.
  2004, A\&A, 419, 269

\bibitem[{Weekes(1991)}]{wee91}
Weekes, T.~C. 1991, Space Sci. Rev., 59, 315

\bibitem[{{Weiler}(1978)}]{wei78}
{Weiler}, K.~W. 1978, in Proc. from the Workshop on Supernovae and
  Supernova Remnants, Vol.~49, 545--552

\bibitem[{Weiler \& Panagia(1980)}]{wp80}
Weiler, K.~W. \& Panagia, N. 1980, A\&A, 90, 269

\bibitem[{Weiler \& Shaver(1978)}]{ws78}
Weiler, K.~W. \& Shaver, P.~A. 1978, A\&A, 70, 389

%\bibitem[{{Weisskopf} {et~al.}(2000){Weisskopf}, {Hester}, {Tennant}, {Elsner},
%  {Schulz}, {Marshall}, {Karovska}, {Nichols}, {Swartz}, {Kolodziejczak}, \&
%  {O'Dell}}]{wht+00}
%{Weisskopf}, M.~C., {Hester}, J., {Tennant}, A., {Elsner}, R.,
%  {Schulz}, N., {Marshall}, H., {Karovska}, M., {Nichols}, J.,
%%  {Swartz}, D., {Kolodziejczak}, J., \& {O'Dell}, S. 2000, ApJ, 536,
%  L81

\bibitem[{Weisskopf {et~al.}(2000)}]{wht+00}
Weisskopf, M. C. et al. 2000, ApJ, 536, L81


\bibitem[{Weisskopf {et~al.}(2004)Weisskopf, O'Dell, Paerels, Elsner, Becker,
  Tennant, \& Schwarz}]{wop+04}
Weisskopf, M.~C., O'Dell, S.~L., Paerels, F., Elsner, R.~F., Becker, W.,
  Tennant, A.~F., \& Schwarz, D.~A. 2004, ApJ, 601, 1050

\bibitem[{Whiteoak \& Green(1996)}]{wg96}
Whiteoak, J. B.~Z. \& Green, A.~J. 1996, A\&AS, 118, 329,
%  (http://www.physics.usyd.edu.au/astrop/wg96cat/)

\bibitem[{Woltjer {et~al.}(1997)Woltjer, Salvati, Pacini, \& Bandiera}]{wspb97}
Woltjer, L., Salvati, M., Pacini, F., \& Bandiera, R. 1997, A\&A, 325, 295

\bibitem[{Yakovlev {et~al.}(2004)Yakovlev, Gnedin, Kaminder, Levenfish, \&
  Potekhin}]{ygk+04}
Yakovlev, D.~G., Gnedin, O.~Y., Kaminker, A.~D., Levenfish, K.~P., \& Potekhin,
  A.~Y. 2004, Adv. Space Res., 33, 523

\bibitem[{Yakovlev {et~al.}(2002)Yakovlev, Gnedin, Kaminker, \&
  Potekhin}]{ygkp02}
Yakovlev, D.~G., Gnedin, O.~Y., Kaminker, A.~D., \& Potekhin, A.~Y. 2002, in
  Proc. of the 270. Heraeus Seminar on Neutron Stars, Pulsars and
  Supernova Remnants, MPE Report 278., ed. W.~Becker, H.~Lesch, \& J.~Tr\"umper, Bad Honnef., 
in press (astro-ph/0111429)

\bibitem[{Yakovlev {et~al.}(2001)Yakovlev, Kaminker, \& Gnedin}]{ykg01}
Yakovlev, D.~G., Kaminker, A.~D., \& Gnedin, O.~Y. 2001, A\&A, 379, L5

\bibitem[{{Yakovlev} {et~al.}(2002){Yakovlev}, {Kaminker}, {Haensel}, \&
  {Gnedin}}]{ykhg02}
{Yakovlev}, D.~G., {Kaminker}, A.~D., {Haensel}, P., \& {Gnedin}, O.~Y. 2002,
  A\&A, 389, L24

\bibitem[{Yakovlev {et~al.}(1999)Yakovlev, Levenfish, \& Shibanov}]{yls99}
Yakovlev, D.~G., Levenfish, K.~P., \& Shibanov, Y.~A. 1999, Phys. Usp., 42,
  737 

%\bibitem[{{Yoshikoshi} {et~al.}(1997){Yoshikoshi}, {Kifune}, {Dazeley},
%  {Edwards}, {Hara}, {Hayami}, {Kakimoto}, {Konishi}, {Masaike}, {Matsubara},
%  {Matsuoka}, {Mizumoto}, {Mori}, {Muraishi}, {Muraki}, {Naito}, {Nishijima},
%  {Oda}, {Ogio}, {Ohsaki}, {Patterson}, {Roberts}, {Rowell}, {Sako},
%  {Sakurazawa}, {Susukita}, {Suzuki}, {Tamura}, {Tanimori}, {Thornton},
%  {Yanagita}, \& {Yoshida}}]{ykd+97}
%{Yoshikoshi}, T., {Kifune}, T., {Dazeley}, S.~A., {Edwards}, P.~G., {Hara}, T.,
%  {Hayami}, Y., {Kakimoto}, F., {Konishi}, T., {Masaike}, A., {Matsubara}, Y.,
%  {Matsuoka}, T., {Mizumoto}, Y., {Mori}, M., {Muraishi}, H., {Muraki}, Y.,
%  {Naito}, T., {Nishijima}, K., {Oda}, S., {Ogio}, S., {Ohsaki}, T.,
%  {Patterson}, J.~R., {Roberts}, M.~D., {Rowell}, G.~P., {Sako}, T.,
%  {Sakurazawa}, K., {Susukita}, R., {Suzuki}, A., {Tamura}, T., {Tanimori}, T.,
%  {Thornton}, G.~J., {Yanagita}, S., \& {Yoshida}, T. 1997, ApJ, 487, L65+

\bibitem[{Yoshikoshi {et~al.}(1997)}]{ykd+97}
Yoshikoshi, T. et al. 1997, ApJ, 487, L65

\bibitem[{Zane {et~al.}(2001)Zane, Turolla, Stella, \& Treves}]{ztst01}
Zane, S., Turolla, R., Stella, L., \& Treves, A. 2001, ApJ, 560, 384

\bibitem[{Zavlin \& Pavlov(2002)}]{zp02}
Zavlin, V.~E. \& Pavlov, G.~G. 2002, in Proc. of the 270. Heraeus Seminar
  on Neutron Stars, Pulsars and Supernova Remnants, ed. W.~Becker, H.~Lesch, \&
  J.~Tr\"umper (Bad Honnef), 273
%astro-ph/0206025

\bibitem[{{Zavlin} {et~al.}(2004){Zavlin}, {Pavlov}, \& Sanwal}]{zps04}
{Zavlin}, V.~E., {Pavlov}, G.~G., \& Sanwal, D. 2004, ApJ, 606, 444

\bibitem[{Zavlin {et~al.}(2002)Zavlin, Pavlov, Sanwal, Manchester, Trümper,
  Halpern, \& Becker}]{zps+02}
Zavlin, V.~E., Pavlov, G.~G., Sanwal, D., Manchester, R.~N., Trümper, J.,
  Halpern, J.~P., \& Becker, W. 2002, ApJ, 569, 894

\bibitem[{{Zavlin} {et~al.}(2000){Zavlin}, {Pavlov}, {Sanwal}, \& {Tr{\"
  u}mper}}]{zpst00}
{Zavlin}, V.~E., {Pavlov}, G.~G., {Sanwal}, D., \& {Tr{\" u}mper}, J. 2000,
  ApJ, 540, L25

\bibitem[{{Zavlin} {et~al.}(1996){Zavlin}, {Pavlov}, \& {Shibanov}}]{zps96}
{Zavlin}, V.~E., {Pavlov}, G.~G., \& {Shibanov}, Y.~A. 1996, A\&A, 315, 141

\bibitem[{{Zavlin} {et~al.}(1998){Zavlin}, {Pavlov}, \& {Trumper}}]{zpt98}
{Zavlin}, V.~E., {Pavlov}, G.~G., \& {Tr\"umper}, J. 1998, A\&A, 331, 821

\bibitem[{{Zavlin} {et~al.}(1999){Zavlin}, {Tr{\" u}mper}, \& {Pavlov}}]{zpt99}
{Zavlin}, V.~E., {Tr{\" u}mper}, J., \& {Pavlov}, G.~G. 1999, ApJ, 525, 959

\bibitem[{Zavlin {et~al.}(1999)Zavlin, Tr\"{u}mper, \& Pavlov}]{ztp99}
Zavlin, V.~E., Tr\"{u}mper, J., \& Pavlov, G.~G. 1999, ApJ, 525, 959

\bibitem[{Zhang \& Cheng(1997)}]{zc97}
Zhang, L. \& Cheng, K.~S. 1997, ApJ, 487, 370

\bibitem[{Zwicky(1938)}]{zwi38}
Zwicky, F. 1938, ApJ, 88, 522

\end{thebibliography}

\end{document}